\documentclass[preprint]{aastex7}

\newcommand{\revision}[1]{#1}

\shorttitle{TZTD Model}
\shortauthors{Grimble et al.}

\begin{document}

\title{The Two-Zone Temperature Distribution Model: Inferences on the Structure and Composition of Dusty Protoplanetary Disks}

\author[0000-0002-8899-8914]{William Grimble}
\email{wag3983@g.rit.edu}
\affiliation{Department of Physics and Astronomy, Vanderbilt University, Nashville, TN 37235, USA}
\affiliation{Frist Center for Autism and Innovation, Vanderbilt University, 2414 Highland Avenue, Suite 115, Nashville, TN 37212, USA}
\affiliation{Chester F. Carlson Center for Imaging Science, Rochester Institute of Technology, 54 Lomb Memorial Drive, Rochester, NY 14623, USA}

\author[0000-0002-3138-8250]{Joel Kastner}
\email{jhk@cis.rit.edu}
\affiliation{Chester F. Carlson Center for Imaging Science, Rochester Institute of Technology, 54 Lomb Memorial Drive, Rochester, NY 14623, USA}

\author[0000-0001-9855-8261]{B. Sargent}
\email{sargent@stsci.edu}
\affiliation{Space Telescope and Science Institute, 3700 San Martin Drive, Baltimore, MD 21218, USA}
\affiliation{Center for Astrophysical Sciences, The William H. Miller III Department of Physics and Astronomy, Johns Hopkins University, Baltimore, MD 21218, USA}

\author[0000-0002-3481-9052]{Keivan G.\ Stassun}
\email{keivan.stassun@vanderbilt.edu}
\affiliation{Department of Physics and Astronomy, Vanderbilt University, Nashville, TN 37235, USA}

\begin{abstract}
In order to understand the mineralogy and structure of protoplanetary disks, it is important to analyze them from both an empirical spectrum-based perspective and a radiative transfer image-based perspective. In a prior paper, we set forth an empirical mineralogy mid-IR spectral model that conveyed spatial information and worked in tandem with a radiative transfer model, which formed the EaRTH Disk Model. In this article, we take the empirical portion of that model, the TZTD model, and refine it with a newly derived protoplanetary disk thermal emission \revision{formulation which uses a temperature distribution without requiring discrete integration}; this simplified model uses an empirical relation between spatial distribution variables, which permits radiative transfer models to directly fit these spatial distribution variables more freely within the provided empirical constraints. We test this model against several $Spitzer~Space~Telescope$ Infrared Spectrograph (IRS) spectra, primarily transition disks, and discuss the mineralogical and structural implications of the fits, including the implications for grain growth and processing within the atmospheric zones of the disks. 
\end{abstract}

\section{Introduction}\label{intro}
Circumstellar disks orbiting young (age $\lesssim$10 Myr) stars are the birthplaces of planets. The gas/dust mass ratios in such protoplanetary disks may significantly depart from the ``canonical'' ISM value of $\sim$100. This dust, however, is observable in the mid-infrared (mid-IR) spectra of star/disk systems as a result of thermal emission from the dust in the system via characteristic features such as the signature prominent 10 $\micron$ peak and a lesser peak at $\sim$20 $\micron$, both indicative of silicate dust grains \citep{2011ppcd.book...14C}. Mid-IR spectroscopic instruments have been and continue to be used to analyze the contents of protoplanetary disks in this way; the $Spitzer~Space~Telescope$ Infra-Red Spectrograph ($Spitzer$ IRS) has recorded archival mid-IR spectral data for hundreds of classical T Tauri stars, and the James Webb Space Telescope (JWST) is now doing the same with vastly improved sensitivity and spectral resolution. This spectroscopic analysis has established an extreme mineralogical diversity within the regions of the disks directly observable via mid-IR spectroscopy, specifically the inner, planet-forming regions of the disk \revision{\citep[see, for example,][]{2010ApJ...721..431J}}. Silicates and other minerals are easily and reliably distinguishable within this regime and a plausible mixture of these grain species can reproduce the shape of the spectrum \citep{2011ppcd.book..114H}. 

Various empirical models have been used to analyze mid-IR spectra to study the presence of silicates in protoplanetary disks via the features in these spectra. One example is the two-temperature empirical model \citep{2009ApJ...690.1193S,2009ApJS..182..477S}, used by many authors \citep[e.g.][]{2016ApJ...830...71F,2015ApJ...810...62R,2022ApJ...933...54L}; however, the simplifying assumption in the model, that the disk consists of two isothermal zones, does not yield realistic descriptions of protoplanetary disk structures. Another example of a well-tested empirical model is the two-layer temperature distribution, which assumes that sections of the disk can be modeled based on the assumption that each section has a unique temperature distribution \citep{2009ApJ...695.1024J,2010ApJ...721..431J}. This model has likewise been used to model SEDs successfully \citep[e.g.][]{2023Natur.620..516P} and more realistically represents protoplanetary disk structure, but struggles to fit spectra over a wide range of mid-IR wavelengths, primarily due to the presumption of only one population of dust in the optically thin atmosphere.

In \citet{2024ApJ...970..137G}, we set forth a prototype of an empirical model we referred to as the two-zone temperature distribution; this model combined the two-temperature model and the two-zone temperature distribution to make use of the best features of both. The two-zone temperature model forms the basis of the empirical portion of our combined Empirical and Radiative Transfer Hybrid (EaRTH) Disk Model, in which we study protoplanetary disks from a mineralogical perspective using an empirical model and from a structural perspective using a radiative transfer model. In the two-zone distribution model, we assume two zones in line with the two-temperature model, but each zone has a unique temperature distribution, in line with the two-layer temperature distribution. \revision{The mineralogical output, consisting of the mixture (mass distributions) of dust species and their temperature distributions, becomes the mineralogical input for the radiative transfer model, which then fits the mid-IR SED to establish fundamental disk parameters, such as scale height and viscosity parameter, as constrained by the temperature distributions yielded by the mineralogical fitting.}
However, within this model, the disk is spatially constrained by the empirically fitted temperature distribution, which may not well represent the spatial structure of the disk; an empirical model that realistically spatially describes the structure of a protoplanetary disk while permitting freedom for radiative transfer analyses to fit these structures would fundamentally strengthen the EaRTH Disk Model and any model studying disks from different modeling perspectives.

In this article, we derive a new thermal emission formula for use with the two-zone temperature distribution model that extends the formulation used to describe the two-layer temperature distribution \citep{2009ApJ...695.1024J,2010ApJ...721..431J}. We then test this refined two-zone temperature distribution model on a set of protoplanetary disks to demonstrate the efficacy of this model. This data set includes MP Mus, the star/disk system that was the focus of the study in \citet{2024ApJ...970..137G} for comparison with our previous empirical mineralogical results, and a subset of transition disks from the study by \citet{2020ApJ...892..111F} that will be the focus of a forthcoming, more rigorous analysis using the EaRTH Disk Model. 

\section{The Two-Zone Temperature Distribution (TZTD) Model Formulation}\label{method}
\subsection{Derivation of Thermal Emission}

We begin this analysis by deriving a new formula to characterize the thermal emission of a protoplanetary disk. This derivation and formula are closely based on previous protoplanetary disk thermal emission models; such models begin with the following derived from radiative transfer formulas \citep{1990AJ.....99..924B,2001ApJ...560..957D},

\begin{equation}\label{basicflux}
F_{\nu}=\frac{2\pi\cos\theta}{d^2}\int_{r_{in}}^{r_{out}}(1-\exp(-\frac{\tau_{\nu}(r)}{\cos\theta}))B_{\nu}(T(r))rdr
\end{equation}

\noindent where $\theta$ is the disk inclination, $d$ is the distance between the disk and observer, $B_{\nu}$ is the Planck blackbody function, and the optical depth $\tau_{\nu}$ of a protoplanetary disk can be expressed as

\begin{equation}
\tau_{\nu}(r)=\Sigma(r)\kappa_{\nu},
\end{equation}

\noindent where $\Sigma(r)$ is the surface density and $\kappa_{\nu}$ is the opacity of the disk. Equation \ref{basicflux} can be simplified in different ways, such as assuming an optically thin disk \citep[see, for example,][]{2015EPJWC.10200007W}, in which case the equation becomes 

\begin{equation}\label{thin}
F_{\nu}=\frac{2\pi}{d^2}\int_{r_{in}}^{r_{out}}\Sigma(r)\kappa_{\nu}B_{\nu}(T(r))rdr;
\end{equation}

\noindent from this equation, the assumption of an isothermal disk can also be made to produce the simplified equation \citep{2016IAUS..314..128C}

\begin{equation}
F_{\nu}=\frac{M}{d^2}\kappa_{\nu}B_{\nu}(T).
\end{equation}

\noindent Many thermal emission models use this basis to model the emission from protoplanetary disks.

One such model is the two-temperature (2-T) model \citep{2009ApJ...690.1193S,2009ApJS..182..477S}, which assumes that the disk can be partitioned into two isothermal sections, a ``warm" and ``cool" zone. The model is expressed as

\begin{equation}
\label{mod_flux1}
\begin{array}{l}
F_{\nu}({\lambda}_{k})^{mod}=B_{\nu}({\lambda}_k,T_c)[{\Omega}_c+{\sum_i}{a_{c,i}}{\kappa}_i({\lambda}_k)]\\
\indent +B_{\nu}({\lambda}_k,T_w)[{\Omega}_w+{\sum_j}{a_{w,j}}{\kappa}_j({\lambda}_k)].
\end{array}
\end{equation}

\noindent Here the subscripts $c$ and $w$ denote the cool and warm disk flux densities, respectively, $\Omega$ is the blackbody solid angle characterizing the continuum under the dust features, and $a_{c,i}$ is the mass weight of a given disk dust component ($a_{c,i}$=$m_{c,i}/d^2$).

The two-layer temperature distribution (TLTD) model \citep{2009ApJ...695.1024J,2010ApJ...721..431J} is another thermal emission model, which assumes the disk is partitioned into a rim and midplane, both of which generate mid-IR continuum emission, as well as a dust atmosphere with a single dust population; each zone is characterized by a radial temperature distribution characterized using a temperature distribution exponent $q$, such that

\begin{equation}
T(r)=T_{max}(\frac{r}{r_{in}})^q.
\end{equation}

\noindent The model itself is parameterized by the equation

\begin{equation}
\label{mod_flux0}
\begin{array}{l}
F_{\nu}({\lambda}_{k})^{mod}=\frac{\pi R_*^2}{d^2}B_{\nu}({\lambda}_{k},T_*)+D1\int_{T_{rim,max}}^{T_{rim,min}}\frac{2\pi}{d^2}B_{\nu}({\lambda}_{k},T)T^{\frac{2-q_{rim}}{q_{rim}}}dT\\
\indent+D2\int_{T_{mid,max}}^{T_{mid,min}}\frac{2\pi}{d^2}B_{\nu}({\lambda}_{k},T)T^{\frac{2-q_{mid}}{q_{mid}}}dT\\
\indent+\Sigma_{i=1}^{N}\Sigma_{i=1}^{M}D_{i,j}\kappa_{i,j}({\lambda}_{k})\int_{T_{atm,max}}^{T_{atm,min}}\frac{2\pi}{d^2}B_{\nu}({\lambda}_{k},T)T^{\frac{2-q_{atm}}{q_{atm}}}dT+\Sigma_{i=1}^{NP}C_iI_i^{PAH}.
\end{array}
\end{equation}

\noindent The first term represents the approximate stellar photospheric emission, where $R_*$ is the stellar radius; the second term represents the puffed-up inner rim of the protoplanetary disk, where $D$ is the constant corresponding to the mass fraction of the component in question; the third term represents the disk midplane. These first three terms characterize the continuum underlying the dust features. The fourth term represents the optically thin dusty atmosphere, where $\kappa_{i,j}({\lambda}_{k})$ is the opacity of mineral species $i$ and grain size $j$, $N$ is the total number of species and $M$ is the total number of sizes for a given species; the fifth and final term represents the contribution of polycyclic aromatic hydrocarbons (PAHs), where $I_i^{PAH}$ is the intensity function corresponding to PAH feature $i$, $C_i$ is the constant corresponding to the feature's contribution to the total flux, and $NP$ is the total amount of PAH feature functions used in the fit.

To derive our own thermal emission model, we start with the flux expression in Equation \ref{thin} assuming an optically thin disk and a radially exponential surface density distribution, that is,

\begin{equation}\label{surfdens}
\Sigma(r)=\Sigma_{in}(\frac{r}{r_{in}})^p,
\end{equation}

\noindent such that

\begin{equation}\label{thin_new}
F_{\nu}=\frac{2\pi\kappa_{\nu}}{d^2}\frac{\Sigma_{in}}{r_{in}^p}\int_{r_{in}}^{r_{out}}B_{\nu}(T(r))r^{p+1}dr.
\end{equation}

Since we assume temperature is radially dependent, in the vein of \citet{2009ApJ...695.1024J,2010ApJ...721..431J}, we use a variable substitution, 

\begin{equation}
dT=qT_{max}\frac{r^{q-1}}{r_{in}^q}dr=\frac{qT}{r}dr,T=[T_{max},T_{min}],
\end{equation}

\noindent and then incorporate this substitution into Equation \revision{\ref{thin_new}}, resulting in the equation

\begin{equation}\label{extendedflux}
F_{\nu}=\frac{2\pi\kappa_{\nu}}{d^2}\frac{\Sigma_{in}}{r_{in}^p}\int_{T_{max}}^{T_{min}}B_{\nu}(T)\frac{r_{in}^{p+2}}{qT_{max}^{(p+2)/q}}T^{\frac{p+2}{q}-1}dT=\frac{\kappa_{\nu}2\pi\Sigma_{in}r_{in}^2}{d^2qT_{max}^{(p+2)/q}}\int_{T_{max}}^{T_{min}}B_{\nu}(T)T^{\frac{p+2}{q}-1}dT.
\end{equation}

With that substitution, the integral is solely dependent on radially-distributed temperature. To condense the constants, we use the formula for mass of a protoplanetary disk \citep{2020ApJ...892..111F},

\begin{equation}
\label{mass}
M=2\pi\int_{r_{in}}^{r_{out}}\Sigma(r)rdr.
\end{equation}

We solve the integral using Equation \ref{surfdens}, then rearrange the equation to solve for constant terms,

\begin{equation}
M=2\pi\frac{\Sigma_{in}}{r_{in}^p}\int_{r_{in}}^{r_{out}}r^{p+1}dr=2\pi\frac{\Sigma_{in}}{r_{in}^p}\frac{1}{p+2}(r_{out}^{p+2}-r_{in}^{p+2}),
\end{equation}

\begin{equation}
2\pi\Sigma_{in}=\frac{r_{in}^pM(p+2)}{r_{out}^{p+2}-r_{in}^{p+2}}=\frac{M(p+2)}{r_{out}^2(\frac{r_{out}}{r_{in}})^p-r_{in}^2}=\frac{M(p+2)}{r_{in}^2((\frac{r_{out}}{r_{in}})^{p+2}-1)},
\end{equation}

\begin{equation}
2\pi\Sigma_{in}r_{in}^2=\frac{M(p+2)}{(\frac{T_{min}}{T_{max}})^{\frac{p+2}{q}}-1}.
\end{equation}

We can directly apply this substitution to Equation \ref{extendedflux} and then simplify the result to obtain

\begin{equation}
F_{\nu}=\frac{M\kappa_{\nu}(2+p)}{d^2 q(T_{max}^{\frac{2+p}{q}}-T_{min}^{\frac{2+p}{q}})}\int_{T_{min}}^{T_{max}}B_{\nu}(T)T^{\frac{p+2}{q}-1}dT.
\end{equation}

Finally, we replace the respective surface density and temperature distribution variables $p$ and $q$ with a new variable $x$, defined by

\begin{equation}
\label{x}
x=\frac{2+p}{q}.
\end{equation}

We hereafter refer to $x$ simply as the distribution parameter since it empirically relates the surface density and temperature distribution variables. This distribution parameter, which is proportional to the surface density distribution and inversely proportional to the temperature distribution, is most likely negative. This is because the numerator of $x$, $2+p$, is positive with $p>-2$, and a lesser value of $p$, i.e. a steeper reduction of mass with respect to radius, is borderline unphysical; the denominator of $x$, $q$, must be negative, as a positive value of $q$ represents an increase of temperature with radius, which is likewise unphysical. With this definition, the equation for thermal emission from a protoplanetary disk assuming radially-dependent temperature and surface density becomes

\begin{equation}
\label{thermeq}
F_{\nu}=\frac{M\kappa_{\nu}x}{d^2(T_{max}^x-T_{min}^x)}\int_{T_{min}}^{T_{max}}B_{\nu}(T)T^{x-1}dT.
\end{equation}

We warn of the limitations of this model. First, it assumes the disk is vertically isothermal; if one considers the atmosphere/surface and midplane as separate zones, this is a reasonable simplification based on radiative transfer models. However, as this formula is based on the optically thin assumption, this does not truly account for optically thick thermal emission from the midplane in the disk; one can typically simply introduce continuum emission from the midplane to account for this. This model also does not account for scattered emission; however, for mid-IR wavelengths greater than $\sim 7.7~\micron$, emission from a star/disk system is almost entirely dominated by thermal emission \citep[see][]{2024ApJ...970..137G}, so we recommend a lower wavelength bound of approximately that value. We also reiterate that this model assumes a disk that is azimuthally symmetric.

\subsection{Exact Thermal Emission Assuming x=-1}

We take note of a special case where Equation \ref{thermeq} can be simplified further for even faster computation. We expand the blackbody function

\begin{equation}
B_{\nu}(T)=\frac{2h\nu^3}{c^2(\exp(\frac{h\nu}{kT})-1)}
\end{equation}

\noindent and include that in Equation \ref{thermeq},

\begin{equation}
\label{thermeq2}
F_{\nu}=\frac{M\kappa_{\nu}x}{d^2(T_{max}^x-T_{min}^x)}(\frac{2h\nu^3}{c^2})\int_{T_{min}}^{T_{max}}\frac{T^{x-1}}{\exp(\frac{h\nu}{kT})-1}dT.
\end{equation}

\noindent We rewrite the integral as

\begin{equation}
\int_{T_{min}}^{T_{max}}\frac{T^{x-1}}{\exp(\frac{h\nu}{kT})-1}dT=(\frac{h\nu}{k})^{x-1}\int_{T_{min}}^{T_{max}}\frac{(\frac{h\nu}{kT})^{1-x}}{\exp(\frac{h\nu}{kT})-1}dT,
\end{equation}

\noindent then use a substitution, $u=\frac{h\nu}{kT}$, and implement $x=-1$ to rewrite the integral as

\begin{equation}
(\frac{h\nu}{k})^{-1}\int_{\frac{h\nu}{kT_{max}}}^{\frac{h\nu}{kT_{min}}}\frac{u^{-1-(-1)}}{\exp(u)-1}du=(\frac{k}{h\nu})\int_{\frac{h\nu}{kT_{max}}}^{\frac{h\nu}{kT_{min}}}\frac{1}{\exp(u)-1}du,
\end{equation}

\noindent which, after simplification, can be expressed as

\begin{equation} 
T_{max}^{-1}-T_{min}^{-1}+(\frac{k}{h\nu})\ln(\frac{B_{\nu}(T_{max})}{B_{\nu}(T_{min})}).
\end{equation}

We can apply this simplification to Equation \ref{thermeq2}, substituting in $x=-1$ throughout the equation such that

\begin{equation}
\label{thermeq3}
\begin{array}{l}
F_{\nu}=\frac{M\kappa_{\nu}x}{d^2(T_{max}^x-T_{min}^x)}(\frac{2h\nu^3}{c^2})\int_{T_{min}}^{T_{max}}\frac{T^{x-1}}{\exp(\frac{h\nu}{kT})-1}dT\\
\indent =\frac{-M\kappa_{\nu}}{d^2(T_{max}^{-1}-T_{min}^{-1})}(\frac{2k\nu^2}{c^2})(\frac{h\nu}{k}(T_{max}^{-1}-T_{min}^{-1})+\ln(\frac{B_{\nu}(T_{max})}{B_{\nu}(T_{min})}))
\end{array}
\end{equation}

\noindent Finally, we distribute and rewrite some constants to express the thermal emission flux using

\begin{equation}
F_{\nu}=\frac{2M\kappa_{\nu}}{{\lambda^2}d^2}(\frac{k\ln(\frac{B_{\nu}(T_{max})}{B_{\nu}(T_{min})})}{T_{min}^{-1}-T_{max}^{-1}}-\frac{hc}{\lambda}).
\end{equation}

\revision{To investigate whether $x=-1$ is a reasonable prior assumption or merely an interesting special case, we note that we can relate the flaring exponent ($\beta$) to $q$ via the equation \citep{1997ApJ...490..368C,2001ApJ...560..957D},
\begin{equation}
q=2\beta-3.
\end{equation}
For $x=-1$, we then have
\begin{equation}
\beta=\frac{1-p}{2}.
\end{equation}

\noindent With the constraint $x=-1$, the flaring exponent varies from $\beta=1.25$ for $p=-1.5$ to $\beta=0.75$ for $p=-0.5$, a reasonable range of potential flaring exponents and surface density exponents \citep[see, for example,][]{2016A&A...586A.103W}. Furthermore, this constraint (indeed, $x$ in general assuming $p>-2$ and $q<0$) carries the implicit assumption that a disk with less flaring has a steeper radial temperature gradient and a flatter radial density gradient. Thus, the simplifying assumption of $x=-1$ appears physically realistic. While other values of $x$ are physically plausible and can lead to very different flux density results, our fitting tests indicate that the results are relatively insensitive to the choice of $x$; this is 
possibly a result of the linear dependence of $x$ on $p$, which itself has little effect on the model disk SED \citep{2016A&A...586A.103W}. }

Applying this formulation to the TZTD model equation yields
\begin{equation}
\begin{array}{l}
F_{\nu}({\lambda}_{k})^{mod}=(\frac{2}{\lambda_k^2})({a_{out}}{\kappa_{out}({\lambda}_k)}(\frac{k\ln(\frac{B_{\nu}({\lambda}_{k},T_{max,o})}{B_{\nu}({\lambda}_{k},T_{min})})}{T_{min}^{-1}-T_{max,o}^{-1}}-\frac{hc}{\lambda_k}) \\
\indent +[{\sum_i}{a_{c,i}}{\kappa}_i({\lambda}_k)](\frac{k\ln(\frac{B_{\nu}({\lambda}_{k},T_{max,c})}{B_{\nu}({\lambda}_{k},T_{max,o})})}{T_{max,o}^{-1}-T_{max,c}^{-1}}-\frac{hc}{\lambda_k}) \\
\indent +[{\sum_j}{a_{w,j}}{\kappa}_j({\lambda}_k)](\frac{k\ln(\frac{B_{\nu}({\lambda}_{k},T_{max,w})}{B_{\nu}({\lambda}_{k},T_{max,c})})}{T_{max,c}^{-1}-T_{max,w}^{-1}}-\frac{hc}{\lambda_k}).
\end{array}
\end{equation}

In this revised formulation, we have assumed the disk is optically thin, which is a reasonable assumption for the disk atmosphere; instead of using a general assumption of the optically thick regions of the disk such as the midplane, which would require a more complicated formulation, we fit the mass weight and opacity of a pre-set distribution corresponding to the assumption of a dust continuum, elaborated further in \S \ref{res}. Furthermore, as we elaborate on further in \S \ref{res}, an additional component is necessary to improve the fits, particularly at longer wavelengths, so we add an ``outer" zone with its own dust and temperature distribution and maximum temperature $T_{max,o}$. This simplification is useful for dust population analysis if faster results are needed and spatial distribution exponents are not a focus of the analysis.

We use this form of the TZTD model in this article and fit for $T_{max,w}$, $T_{max,c}$ (as a fraction of $T_{max,w}$), and $T_{max,o}$ (as a fraction of $T_{max,c}$) nonlinearly and the values of $a$ for each dust component linearly for each set of nonlinear variables, such that $a$ is nonnegative by iteratively setting the component with the minimum sub-zero flux value to zero until none are negative, in line with the two-temperature model \citep{2009ApJ...690.1193S,2009ApJS..182..477S} and our previous model \citep{2024ApJ...970..137G}. We make use of simplicial homology global optimization (SHGO) \citep{endres_simplicial_2018} via \verb|scipy.optimize.shgo|, as it is faster than other optimization algorithms and consistently converges to a single point; this optimization algorithm requires the nonlinear variables to be bounded, so the maximum temperatures are bounded between a minimum of 19 K, in line with \citet{2009ApJ...690.1193S,2009ApJS..182..477S}, and a maximum of 1500 K, the approximate dust evaporation temperature \citep{2001ApJ...560..957D}. The minimum temperature of the outer zone $T_{min}$ is set to 19 K. 

\section{Application of the TZTD Model to Transition Disk Observations}\label{res}
\subsection{Data Analysis}

We test the disk spectral energy distribution formulation described in \S \ref{method} by modeling archival $Spitzer$ IRS spectroscopy data collected for the star/disk system of MP Mus to compare the results with our previous empirical analysis in \citet{2024ApJ...970..137G}, as well as a subset of transition disks selected from the sample studied by \citet{2020ApJ...892..111F}. The targets we analyze are listed in Table \ref{targets}. We use $Spitzer$ IRS spectra for the analysis retrieved from CASSIS\footnote{The Combined Atlas of Sources with Spitzer IRS Spectra (CASSIS) is a product of the IRS instrument team, supported by NASA and JPL. CASSIS is supported by the "Programme National de Physique Stellaire" (PNPS) of CNRS/INSU co-funded by CEA and CNES and through the "Programme National Physique et Chimie du Milieu Interstellaire" (PCMI) of CNRS/INSU with INC/INP co-funded by CEA and CNES.}, version LR7 \citep{2011ApJS..196....8L}; indeed, the subset of targets we selected for this analysis consists of those with complete processed $Spitzer$ IRS spectral data across our desired mid-IR wavelength range of ~7.7 $\micron$ to ~37 $\micron$. We outlined our preprocessing of $Spitzer$ IRS data in \citet{2024ApJ...970..137G}. To briefly summarize, we adjust for mispointing by scaling whichever portion of the spectrum, short-low (SL) or long-low (LL) is lower to a factor that matches that of the higher value at the 14-$\micron$ boundary, adjust for extinction using the $J$ and $K_s$ magnitudes with the extinction curves by \citet{2009ApJ...693L..81M}, and set the flux uncertainty, defined as the RMS error from the archival data, to a minimum of 1$\%$ of the flux value at the given point. Additionally, in line with our recommendation from \S \ref{method}, we restrict the model fitting to wavelengths longer than 7.7 $\micron$ for each spectrum so disk thermal emission is the dominant source of flux in the spectrum.

{\catcode`\&=11
\gdef\citeg{\citet{2006A&A...460..695T,2002AJ....124.1670M}}}
{\catcode`\&=11
\gdef\citeh{\citep{2016A&A...595A...1G,2023A&A...674A...1G,2021A&A...649A...2L}}}
{\catcode`\&=11
\gdef\citei{\citet{2023A&A...673A..77R}}}
\begin{deluxetable*}{lcccccccc}
\tabletypesize{\scriptsize}
\deluxetablecaption{Transition disk sample\label{targets}}
\tablehead{
\colhead{AOR Key} & \colhead{Target} & \colhead{SpT$\tablenotemark{a}$} & \colhead{$T_{eff,*}\tablenotemark{b}$ (K) } & \colhead{$d\tablenotemark{c}$ (pc)} & \colhead{$R_*\tablenotemark{d}~(R_{\odot})$} & \colhead{$M_*\tablenotemark{e}~(M_{\odot})$} & \colhead{$J\tablenotemark{f}$ mag} & \colhead{$K_s\tablenotemark{f}$ mag}
}
\startdata
5198336 & MP Mus & K1 & 4920 & 97.9 $\pm$ 0.1 & 1.45 & 1.3 & 8.28 & 7.29\\
25348096 & CIDA 9 & M2 & 3490 & 175.1 $\pm$ 2.7 & 0.52 & 0.36 & 12.81 & 11.16\\
21875712 & CQ Tau & F2 & 6710 & 149.4 $\pm$ 1.3 & 1.83 & 1.63 & 7.93 & 6.17\\
26144000 & CS Cha & K2 & 4760 & 168.8 $\pm$ 1.9 & 1.77 & 1.4 & 9.11 & 8.20\\
16346624 & DM Tau & M2 & 3490 & 144.0 $\pm$ 0.5 & 1.26 & 0.39 & 10.44 & 9.52\\
12676352 & DoAr 44 & K2 & 4760 & 146.3 $\pm$ 0.5 & 1.45 & 1.4 & 9.23 & 7.61\\
16261888 & GG Tau AA Ab & K7 & 3970 & 116.4 $\pm$ 6.4 & 1.93 & 0.66 & 8.67 & 7.36\\
26141696 & GM Aur & K3 & 4550 & 158.1 $\pm$ 1.2 & 1.60 & 1.32 & 9.34 & 8.28\\
26141440 & IP Tau & M0 & 3770 & 129.4 $\pm$ 0.3 & 1.39 & 0.54 & 9.78 & 8.35\\
19666432 & J1604.3-2130 & K3 & 4550 & 145.3 $\pm$ 0.6 & 1.11 & 1.1 & 9.95 & 8.51\\
26140672 & LkCa 15 & K2 & 4760 & 157.2 $\pm$ 0.7 & 1.42 & 1.3 & 9.42 & 8.16\\
22144768 & PDS 70 & K7 & 3970 & 112.4 $\pm$ 0.2 & 1.24 & 0.8 & 9.55 & 8.54\\
5199104 & RX J1842.9-3532 & K2 & 4760 & 151.0 $\pm$ 0.4 & 1.32 & 1.14 & 9.50 & 8.17\\
5200640 & RX J1852.3-3700 & K2 & 4760 & 147.1 $\pm$ 0.5 & 1.14 & 1.05 & 9.77 & 9.01\\
26141184 & RY Tau & G2 & 5870 & 138.2 $\pm$ 3.9 & 2.78 & 2.25 & 7.16 & 5.40\\
12698880 & SR 21 & G4 & 5620 & 136.4 $\pm$ 0.6 & 1.38 & 2.12 & 8.75 & 6.72\\
21883136 & SR 24 S & K6 & 4020 & 100.4 $\pm$ 2.3 & 1.00 & 0.87 & 9.75 & 7.06\\
26140928 & UX Tau A & G8 & 5210 & 142.2 $\pm$ 0.7 & 1.67 & 1.4 & 8.62 & 7.55\\
21890304 & V1247 Ori & F0 & 7280 & 401.3 $\pm$ 3.2 & 2.93 & 1.82 & 8.88 & 7.41\\
12698368 & WSB 60 & M6 & 2600 & 135.0 $\pm$ 1.5 & 1.24 & 0.24 & 11.31 & 9.32\\
\enddata
\tablenotetext{a}{MP Mus spectral type given by \citeg, all others by \citet{2020ApJ...892..111F}}
\tablenotetext{b}{Based on spectral type using Table 6 in \citet{2013ApJS..208....9P}}
\tablenotetext{c}{Based on $Gaia$ DR3 parallax value and uncertainty \citeh}
\tablenotetext{d}{Based on bolometric luminosity and effective temperature}
\tablenotetext{e}{MP Mus mass given by \citei, all others by \citet{2020ApJ...892..111F}}
\tablenotetext{f}{2MASS data \citep{2006AJ....131.1163S}}
\end{deluxetable*}

{\catcode`\&=11
\gdef\citea{\cite{1995A&A...300..503D}}}
{\catcode`\&=11
\gdef\citeb{\cite{1994A&A...292..641J}}}
{\catcode`\&=11
\gdef\citec{\cite{2003A&A...408..193J}}}
{\catcode`\&=11
\gdef\cited{\cite{1998A&A...339..904J}}}
{\catcode`\&=11
\gdef\citee{\cite{2006A&A...451..357S}}}
{\catcode`\&=11
\gdef\citef{\cite{2013A&A...553A..81Z}}}

We use mostly the same set of opacity functions to fit the $Spitzer$ spectra as we detailed in \citet{2024ApJ...970..137G}, which we summarize in Table \ref{opacity}. We assume a distribution of hollow spheres (DHS) \citep{2003A&A...404...35M} with $V_{max}=0.9$ for each mineral, with grain-size power law exponent -3.5. We used the DHS algorithm as programmed in MCFOST \citep{2006A&A...459..797P,2009A&A...498..967P} to generate the opacity functions for each mineral for use in the TZTD model. However, we make use of an estimate of a dust continuum opacity function for the fit to account for continuum emission; this ``continuum opacity" is listed in Table \ref{opacity} and is calculated using the MCFOST implementation of Mie scattering, itself based on that by \citet{1983asls.book.....B}. Additionally, we include a similar opacity, but with the inclusion of smaller grains, to model the contribution from the outermost parts of the disk in the mid-IR SED. This function, listed in Table \ref{opacity}, is also generated assuming Mie scattering rather than using DHS; DHS is most appropriate for modelling the features of smaller, sub-micron grains, but a Mie scattering assumption is appropriate for these grains in the outer disk since they are at temperatures too low to contribute significantly to mid-IR spectral features beyond a Rayleigh-Jeans tail at the longest wavelengths in the spectrum. 

\begin{deluxetable*}{lccc}
\tabletypesize{\scriptsize}
\deluxetablecaption{Minerals and corresponding optical constants used to calculate opacities.\label{opacity}}
\tablehead{
\colhead{Name} & \colhead{Description of Optical Constants} & \colhead{Assumed Size($\micron$)} & \colhead{Reference}
}
\startdata
Outer Disk & Amorphous olivine MgFeSiO\ensuremath{_{4}} composition & 0.01-20000 & 1\\
Continuum (Cont) & Amorphous olivine MgFeSiO\ensuremath{_{4}} composition & 100-20000 & 1\\
Small Amorphous Pyroxene (SmAmPy) & Amorphous pyroxene of cosmic composition & 0.01-1 & 2\\
Large Amorphous Pyroxene (LgAmPy) & Amorphous pyroxene of cosmic composition & 1-5 & 2\\
Small Amorphous Polivene (SmAmPo) & Amorphous silicate of Mg\ensuremath{_{1.5}}SiO\ensuremath{_{3.5}} composition & 0.01-1 & 3\\
Large Amorphous Polivene (LgAmPo) & Amorphous silicate of Mg\ensuremath{_{1.5}}SiO\ensuremath{_{3.5}} composition & 1-5 & 3\\
Small Amorphous Olivine (SmAmOl) & Amorphous olivine MgFeSiO\ensuremath{_{4}} composition & 0.01-1 & 1\\
Large Amorphous Olivine (LgAmOl) & Amorphous olivine MgFeSiO\ensuremath{_{4}} composition & 1-5 & 1\\
Small Crystalline Enstatite (SmCryEnst) & Natural clinoenstatite MgSiO\ensuremath{_{3}} along crystallographic axes (p,s1,s2) & 0.01-1 & 4\\
Large Crystalline Enstatite (LgCryEnst) & Natural clinoenstatite MgSiO\ensuremath{_{3}} along crystallographic axes (p,s1,s2) & 1-2 & 4\\
Small Crystalline Forsterite (SmCryForst) & Forsterite Mg\ensuremath{_{2}}SiO\ensuremath{_{4}} along crystallographic axes (1u,2u,3u) & 0.01-1 & 5\\
Large Crystalline Forsterite (LgCryForst) & Forsterite Mg\ensuremath{_{2}}SiO\ensuremath{_{4}} along crystallographic axes (1u,2u,3u) & 1-2 & 5\\
Small Crystalline Silica (SmCrySil) & \ensuremath{\alpha}-quartz SiO\ensuremath{_{2}} at 955 K along crystallographic axes (\ensuremath{E_{para},E_{perp1},E_{perp2}}) & 0.01-1 & 6\\
Large Crystalline Silica (LgCrySil) & \ensuremath{\alpha}-quartz SiO\ensuremath{_{2}} at 955 K along crystallographic axes (\ensuremath{E_{para},E_{perp1},E_{perp2}}) & 1-2 & 6\\
\enddata
\tablerefs{(1) \citea; (2) \citeb; (3) \citec; (4) \cited; (5) \citee; (6) \citef}
\end{deluxetable*}

We summarize the results of our analysis in Table \ref{res1}; here, we list the maximum zone temperatures, along with the reduced $\chi^2$ value of each fit. We list the numerical mineralogical results in Table \ref{minres} in Appendix \ref{appe}. Figure \ref{res_plot} demonstrates a compilation of the plots of the fit results for the disks in Table \ref{targets}; detailed results are in Appendix \ref{appe}. As in \citet{2024ApJ...970..137G}, the uncertainties we demonstrate are derived using a process described by \citet{2009ApJS..182..477S}, in which we estimate the parameter value shift such that the reduced $\chi^2$ value increases by 1, using a Taylor expansion of the reduced $\chi^2$ formula for the linear variables and an iterative brute force method for the nonlinear variables; this yields uncertainty estimates that tend to be more conservative than Monte Carlo-based uncertainty estimations \citep{1976ApJ...210..642A,2009ApJS..182..477S}.

\begin{figure*}[htb!]
    \gridline{\fig{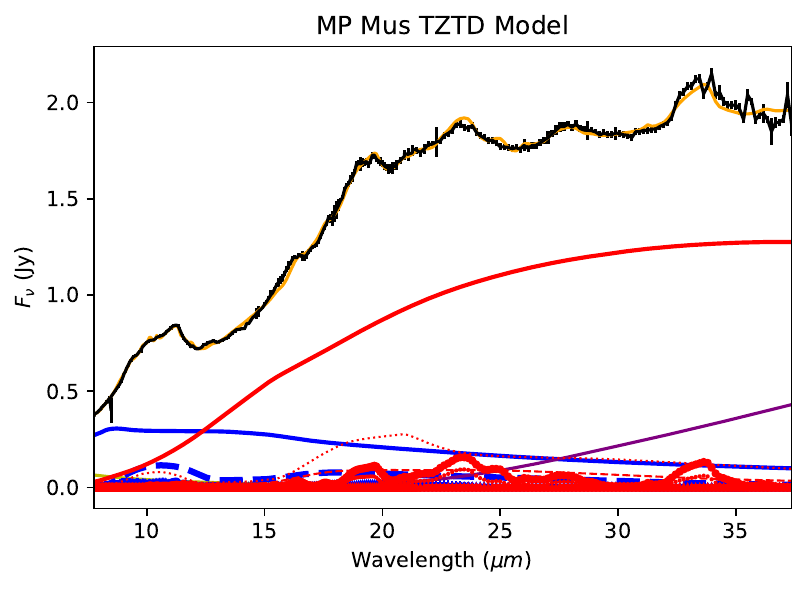}{0.3\textwidth}{}
              \fig{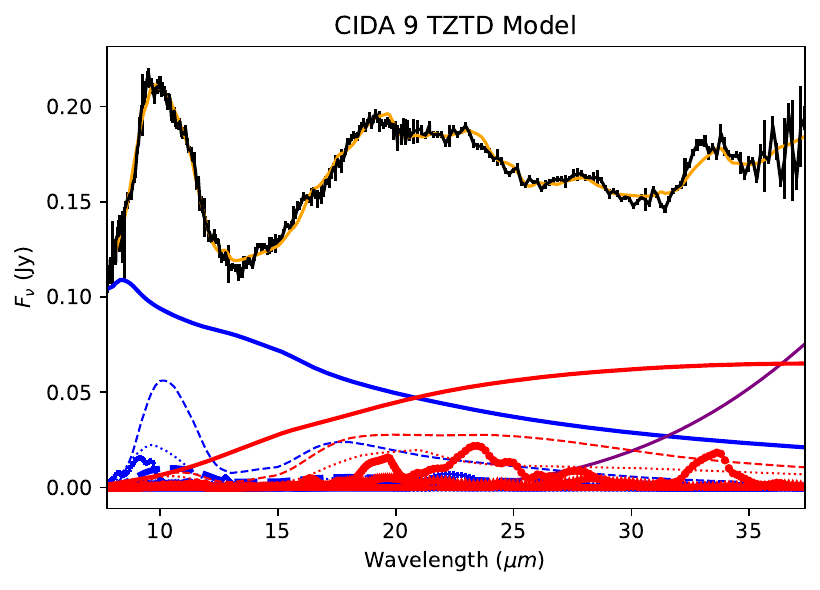}{0.3\textwidth}{}
              \fig{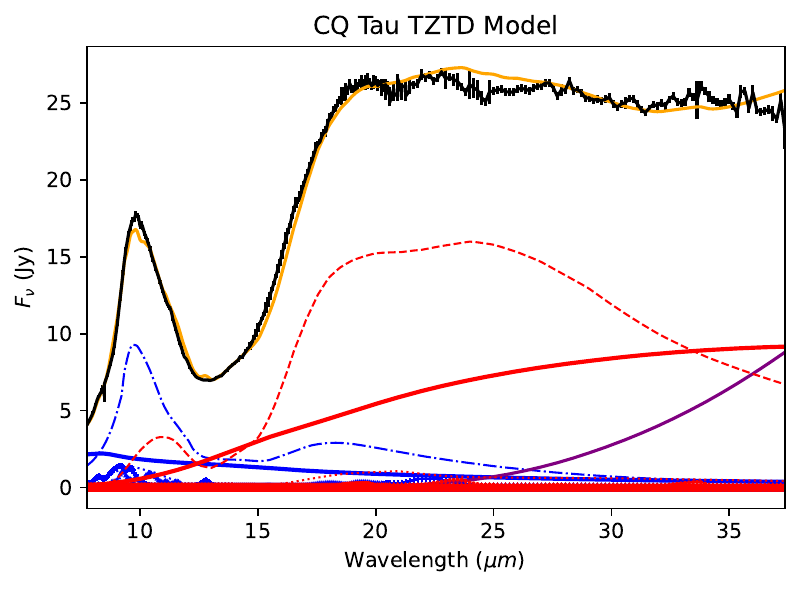}{0.3\textwidth}{}
              }
    \gridline{\fig{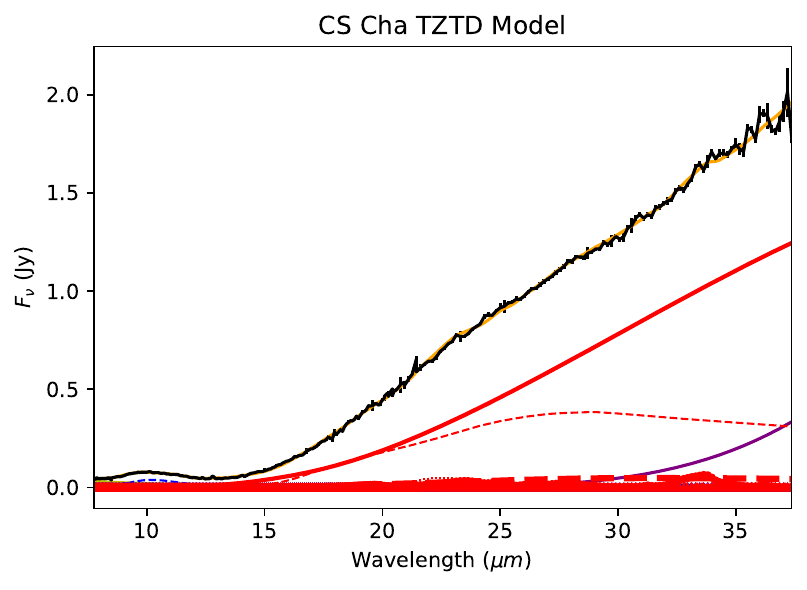}{0.3\textwidth}{}
              \fig{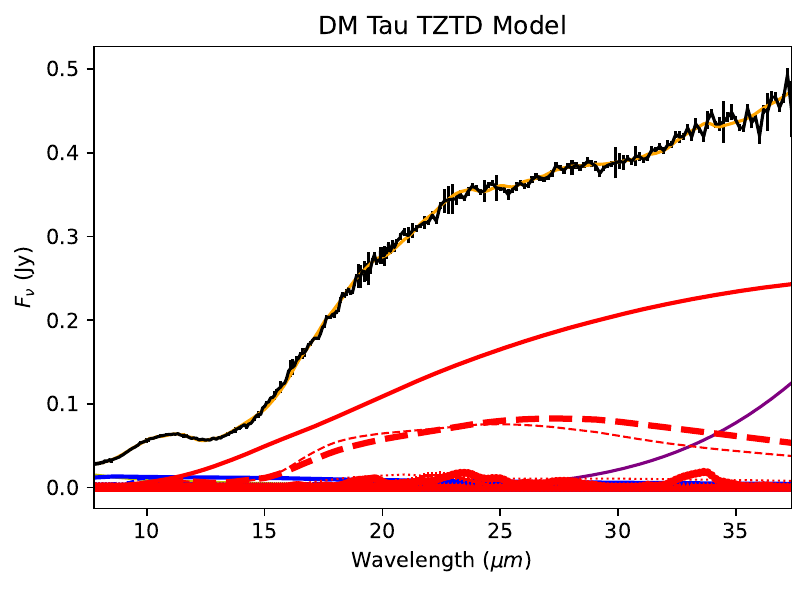}{0.3\textwidth}{}
              \fig{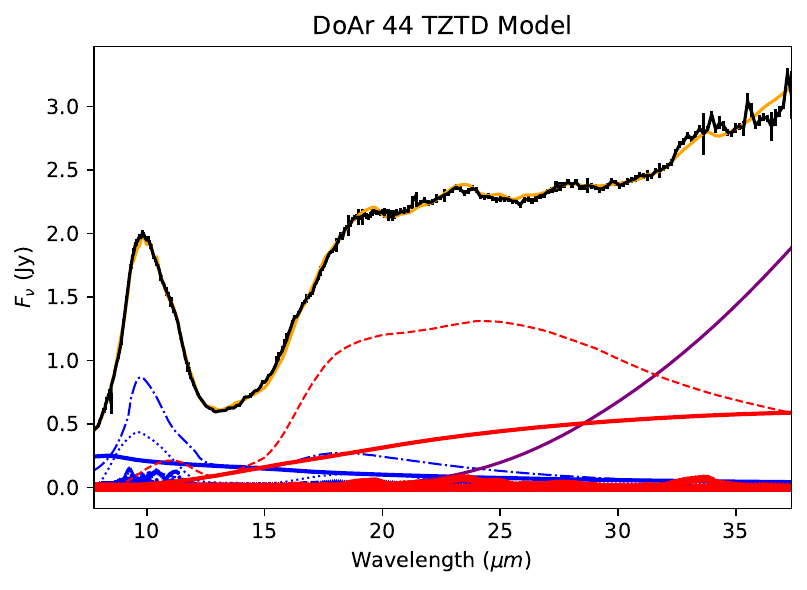}{0.3\textwidth}{}
              }
    \gridline{\fig{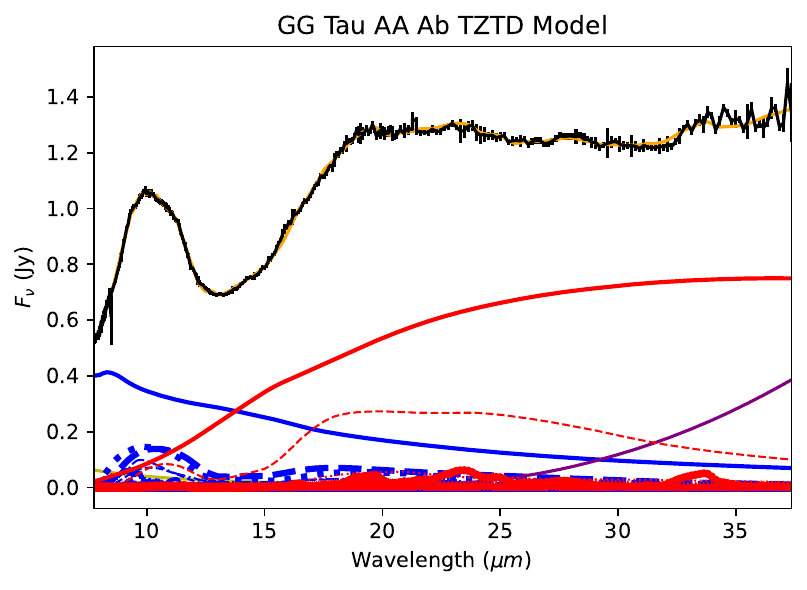}{0.3\textwidth}{}
              \fig{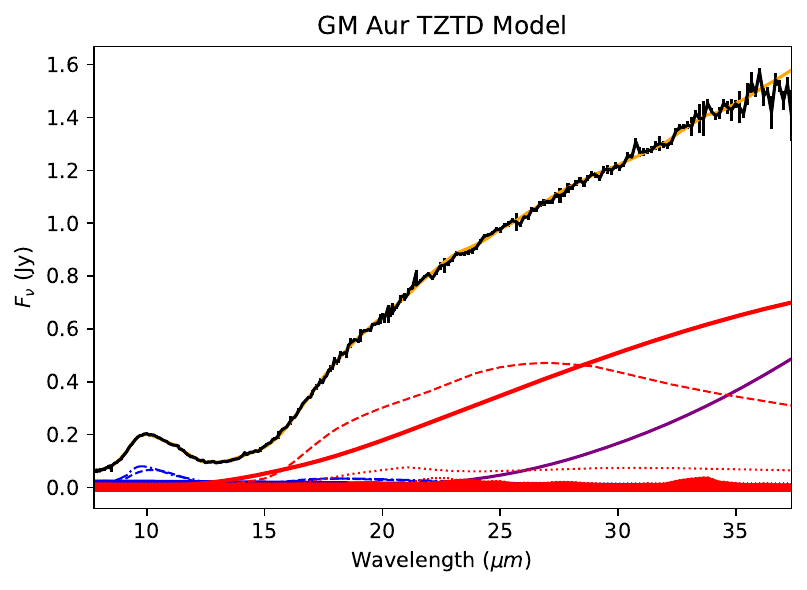}{0.3\textwidth}{}
              \fig{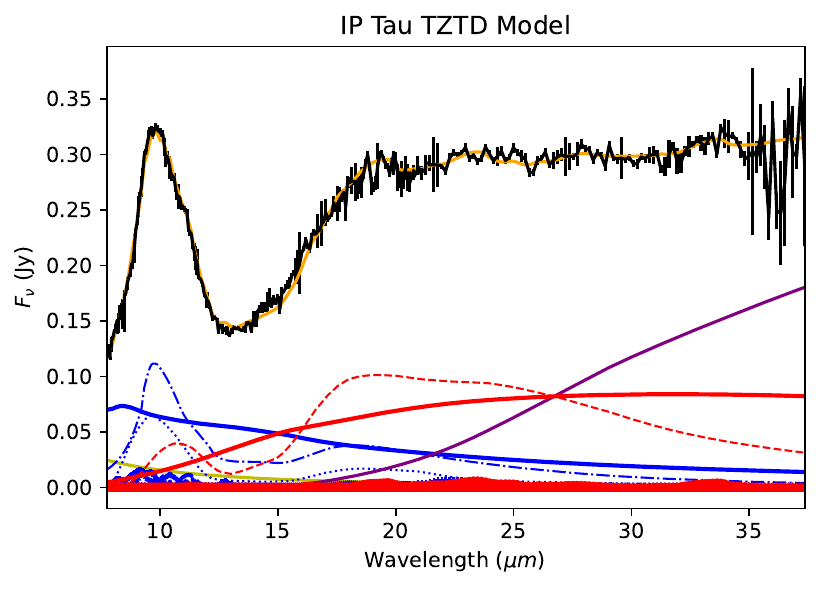}{0.3\textwidth}{}
              }
    \gridline{\fig{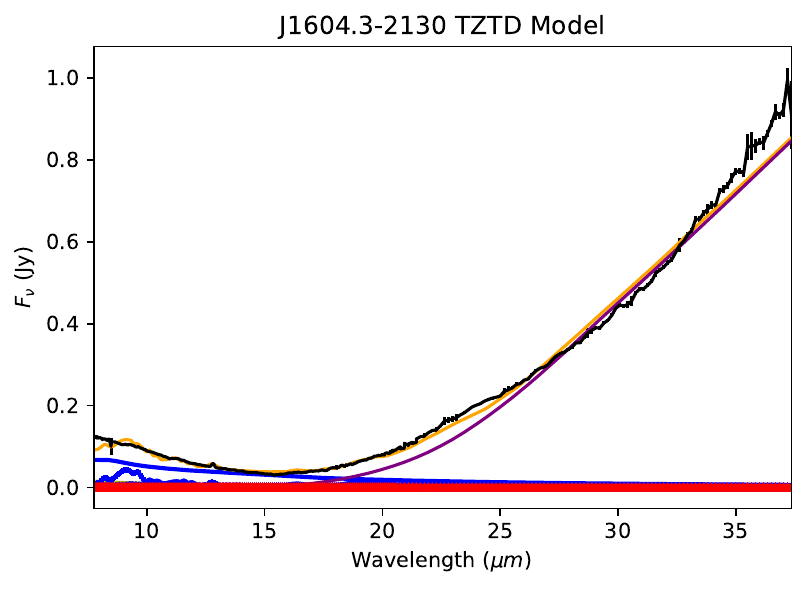}{0.3\textwidth}{}
              \fig{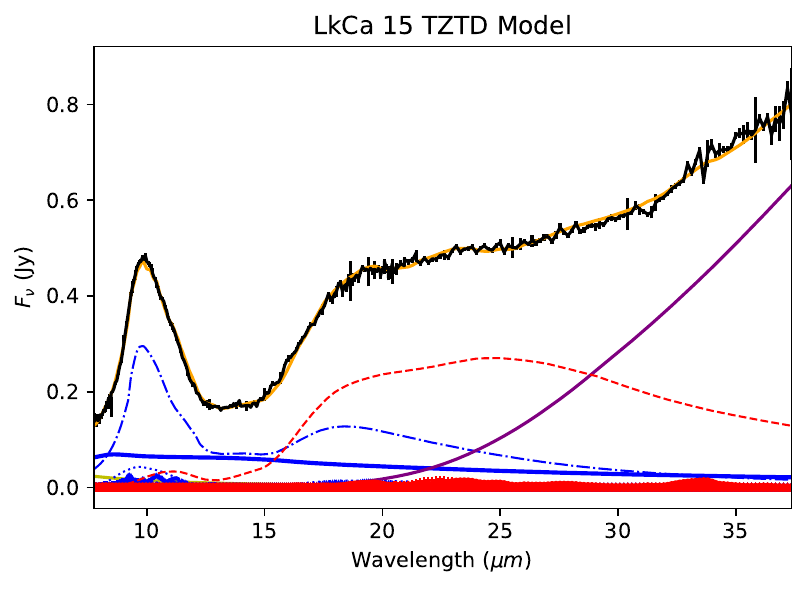}{0.3\textwidth}{}
              \fig{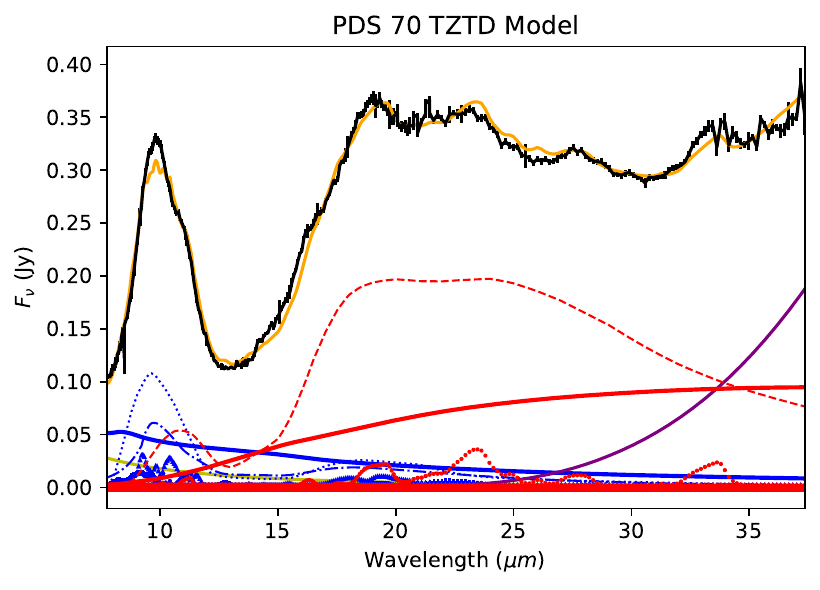}{0.3\textwidth}{}
              }
    \caption{Model fit plots of TZTD empirical mineralogical analysis of $Spitzer$ IRS spectra of targets indicated in Table \ref{targets}; Red: Cool disk component constituents, Blue: Warm disk component constituents; see Figure Set A1 in Appendix \ref{appe} for legend of dust components \label{res_plot}}
\end{figure*}

\begin{figure*}[htb!]
    \gridline{\fig{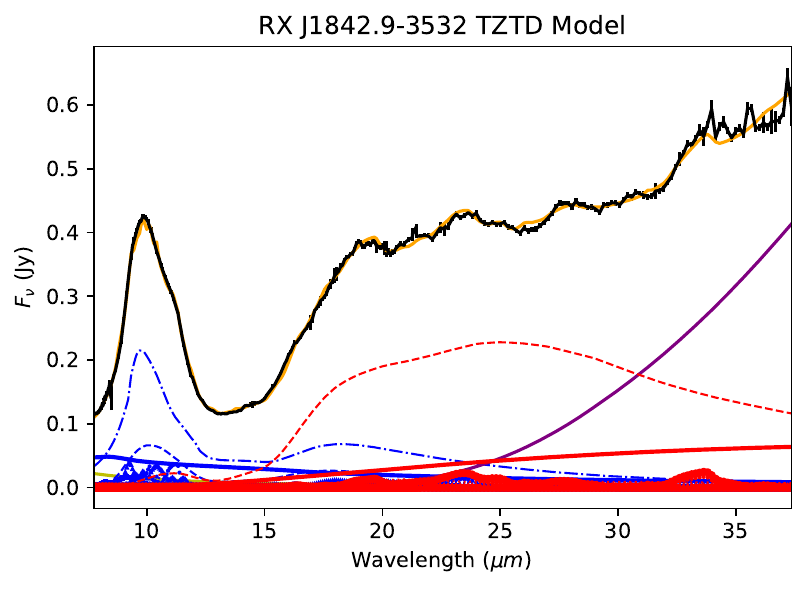}{0.3\textwidth}{}
              \fig{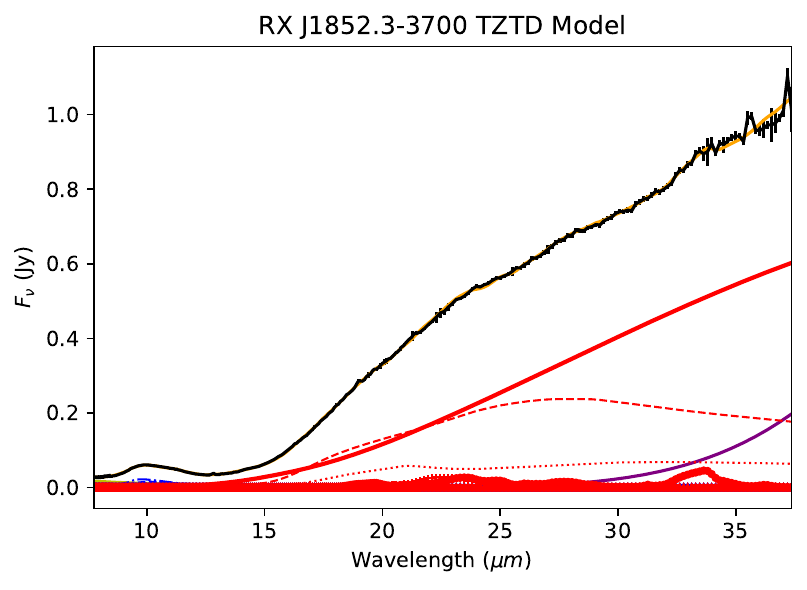}{0.3\textwidth}{}
              }
    \gridline{\fig{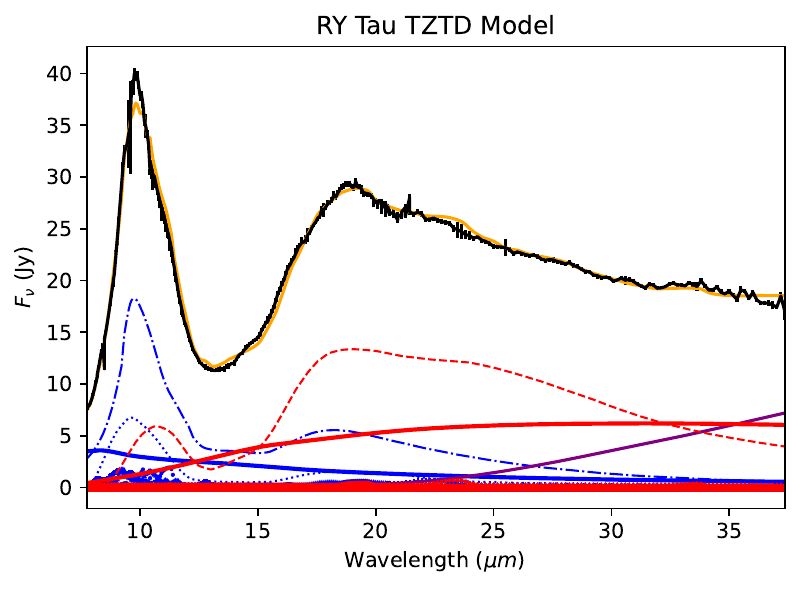}{0.3\textwidth}{}
              \fig{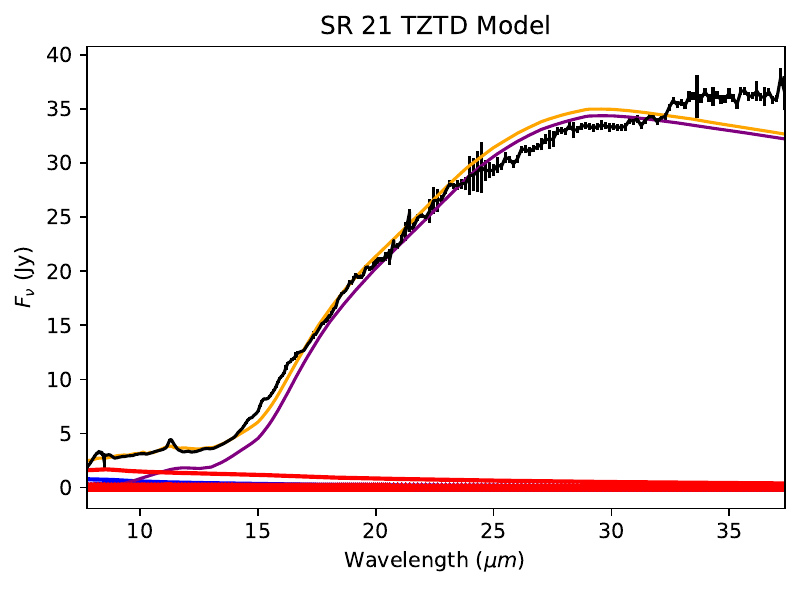}{0.3\textwidth}{}
              \fig{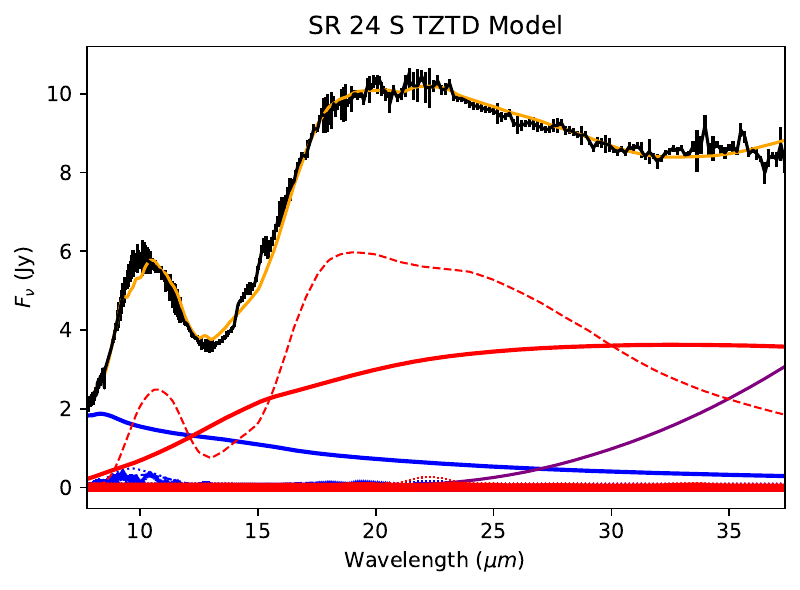}{0.3\textwidth}{}
              }
    \gridline{\fig{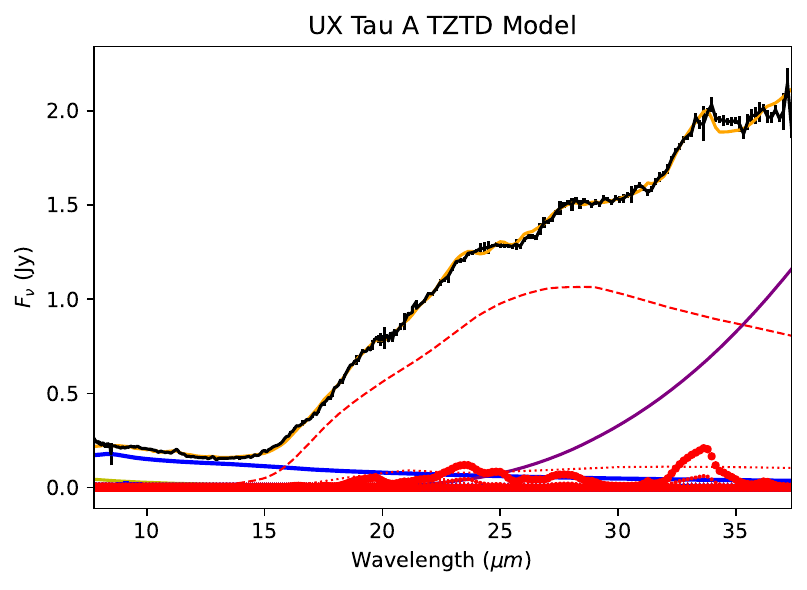}{0.3\textwidth}{}
              \fig{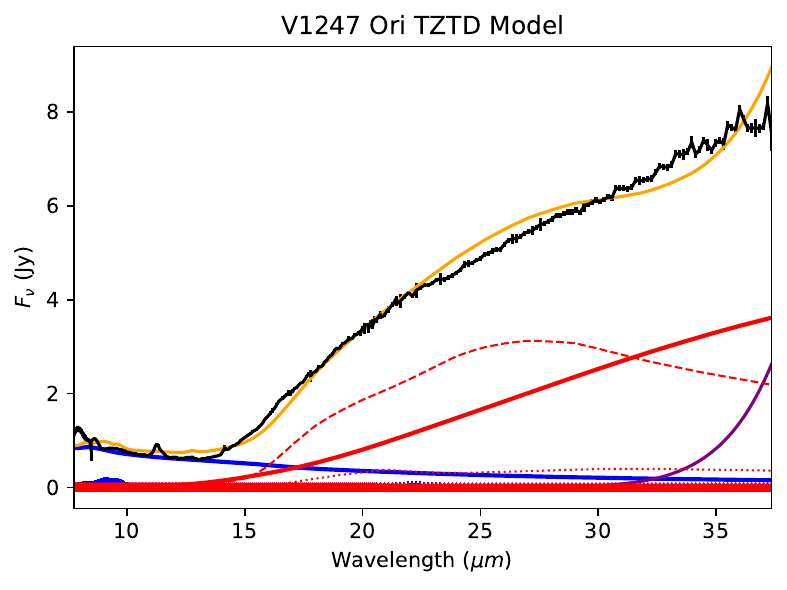}{0.3\textwidth}{}
              \fig{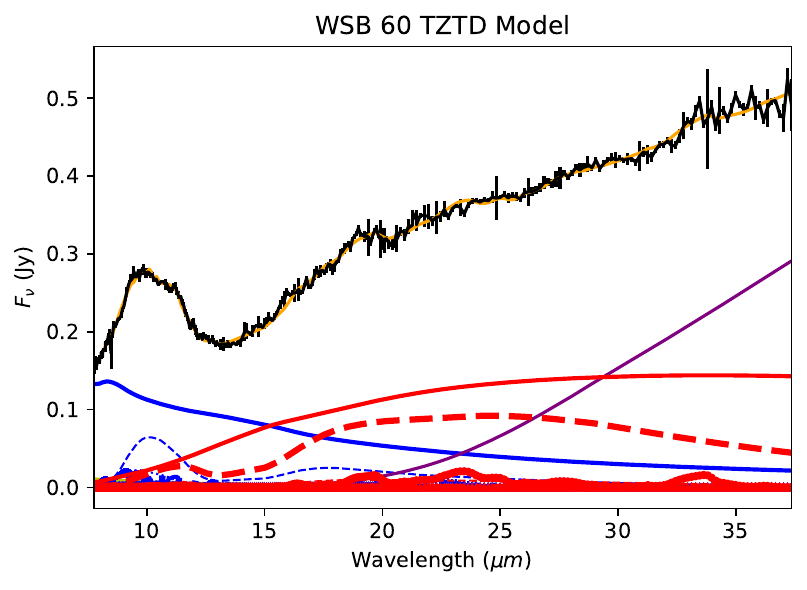}{0.3\textwidth}{}
              }
    \figurenum{1}
    \caption{continued.}
\end{figure*}

\begin{deluxetable*}{lcccc}
\tabletypesize{\scriptsize}
\deluxetablecaption{TZTD Model Results\label{res1}}
\tablehead{
\colhead{Target} & \colhead{$T_{max,w}$ (K)} & \colhead{$T_{max,c}$ (K)} & \colhead{$T_{max,o}$ (K)} & \colhead{$\chi^2$/d.o.f}
}
\startdata
MP Mus & $750^{+230}_{-100}$ & $210^{+90}_{-70}$ & $70^{+40}_{-40}$ & 1.7\\
CIDA 9 & $1200^{+750}_{-300}$ & $220^{+180}_{-140}$ & $50^{+40}_{-40}$ & 4.42\\
CQ Tau & $1500^{+90}_{-60}$ & $200^{+10}_{-10}$ & $50^{+10}_{-10}$ & 9.64\\
CS Cha & $920^{+70}_{-20}$ & $110^{+10}_{-10}$ & $40^{+30}_{-30}$ & 6.8\\
DM Tau & $800^{+50}_{-50}$ & $170^{+10}_{-10}$ & $40^{+30}_{-20}$ & 2.65\\
DoAr 44 & $1500^{+300}_{-140}$ & $190^{+40}_{-40}$ & $50^{+20}_{-10}$ & 3.98\\
GG Tau AA Ab & $1400^{+710}_{-140}$ & $220^{+150}_{-110}$ & $50^{+40}_{-40}$ & 1.28\\
GM Aur & $870^{+120}_{-40}$ & $130^{+20}_{-20}$ & $50^{+20}_{-20}$ & 3.07\\
IP Tau & $1170^{+360}_{-310}$ & $250^{+90}_{-110}$ & $80^{+40}_{-40}$ & 4.49\\
J1604.3-2130 & $1110^{+30}_{-10}$ & $1100^{+40}_{-30}$ & $70^{+0}_{-0}$ & 56.4\\
LkCa 15 & $830^{+140}_{-100}$ & $170^{+30}_{-50}$ & $60^{+20}_{-20}$ & 4.98\\
PDS 70 & $1500^{+120}_{-70}$ & $210^{+20}_{-20}$ & $40^{+10}_{-10}$ & 12.8\\
RX J1842.9-3532 & $1500^{+510}_{-150}$ & $160^{+70}_{-60}$ & $60^{+20}_{-20}$ & 4.22\\
RX J1852.3-3700 & $940^{+50}_{-20}$ & $120^{+10}_{-10}$ & $30^{+20}_{-20}$ & 3.53\\
RY Tau & $1500^{+510}_{-200}$ & $260^{+110}_{-90}$ & $70^{+30}_{-30}$ & 7.56\\
SR 21 & $1500^{+30}_{-0}$ & $1130^{+30}_{-30}$ & $150^{+0}_{-0}$ & 44.8\\
SR 24 S & $1500^{+150}_{-130}$ & $250^{+30}_{-30}$ & $50^{+20}_{-10}$ & 5.83\\
UX Tau A & $1330^{+150}_{-40}$ & $110^{+10}_{-10}$ & $50^{+10}_{-10}$ & 9.56\\
V1247 Ori & $1500^{+70}_{-20}$ & $120^{+10}_{-10}$ & $20^{+40}_{-10}$ & 74.2\\
WSB 60 & $1480^{+460}_{-550}$ & $240^{+120}_{-110}$ & $70^{+40}_{-40}$ & 2.85\\
\enddata
\tablecomments{Temperatures rounded to nearest value of ten; error values of 0 indicate actual values less than 5}{}
\end{deluxetable*}

\subsection{Goodness of Fits}\label{goodness}

Based on the results of the analysis, we see that a majority of the fits from the TZTD model result in reduced $\chi^2$ values less than 5. We note that the parameters corresponding to the warm disk, particularly the maximum temperatures, have larger uncertainties, while the cool disk parameters have smaller uncertainties, indicating a high degree of precision with these parameters. 

\revision{This revised TZTD model produces a similar, but improved, fit to the MP Mus spectrum relative to the former model --- i.e., the version demonstrated in \citet{2024ApJ...970..137G} --- with a $\chi^2$ value of $\sim$1.7. For a fair comparison, when the fit from the previous study is performed over the same wavelength range (i.e. $>7.7~\micron$ as opposed to the full wavelength range from the previous study) and with a consistent set of opacities\footnote{The previous paper \citep{2024ApJ...970..137G} erroneously stated a value $V_{max}=0.9$ was used for the fits, whereas the adopted value was actually $V_{max}=0.8$; the fit $\chi^2$ values would reduce by order $\sim 10\%$ for $V_{max}=0.9$.}, the reduced $\chi^2$ drops from $\sim$2.6 to $\sim$1.8 with minor differences for a few fit parameters (e.g., $T_{max,w}$ and $q_c$), which is close to the result from the new model presented in this article, but still larger. Furthermore, the new version of the model fits three nonlinear variables, while the original version fitted five. In addition to requiring fewer nonlinear variables than the original TZTD model, the updated model fitting runs approximately 12 times faster, thanks to the new optimized formula described here and the consequent ability to leverage the faster SHGO optimization algorithm.} 

\revision{This revised model returns a similar cool zone temperature (210 K) to that determined via the original \citep{2024ApJ...970..137G} model (240 K), but returns a far lower maximum warm zone temperature (750 K) relative to the original (1500 K). This lower maximum temperature --- while perhaps partially due to our imposition of a wavelength limit of $>$7.7 \micron\ in the fitting (see \S \ref{fits}) --- is indicative of a larger inner cavity than found by the original model. The lower warm-component temperature found here is thus more closely in line with the results of \citet{2025NatAs.tmp..142R}, who found evidence for a planet-carved clearing of radius $\sim$1--3 au within the inner regions of the MP Mus disk.} 

\revision{The revised model also returns a somewhat different dust composition relative to the original model  (see the dust composition plot in Figure Set A1 for MP Mus, as well as the values in Table \ref{minres}, in Appendix \ref{appe}; the complete figure set of 20 images is available in the online journal). In the warm disk, large olivine remains the dominant mineral in the revised model, as was the case in the original model; small forsterite, large enstatite, and large silica are still detected, but in larger amounts. Small enstatite, absent from the original model, is now detected, in trace amounts, in the revised model. Large pyroxene, prominent in the original modeling, is no longer detected; instead, small olivine is now detected, with noticeable abundance. In the cool disk, small pyroxene is still dominant, moreso than in the original model, while the previous small amount of small enstatite is no longer detected; small and large forsterite are still present, similarly prominent, and largely responsible for the crystalline bumps in the mid-IR SED. We emphasize the presence of potential degeneracy within empirical dust component fits, particularly within the warm disk, which is notably cooler (albeit with a larger temperature uncertainty, as noted) and has a very different dust composition relative to the original model in \citet{2024ApJ...970..137G}. However, as noted, this fit still demonstrates improvement over that in \citet{2024ApJ...970..137G}.}

Four targets had reduced $\chi^2$ values greater than 10, and therefore the TZTD model struggled to fit them well. Three of these targets, those being J1604.3-2130, SR 21, and V1247 Ori, have very weak defining silicate features at lower wavelengths, with each of them dramatically increasing in flux at longer wavelengths. In the fits for each of these targets, the TZTD model determined their atmospheres to consist entirely of crystalline minerals, which is likely inaccurate, and characterized the outer disk to be contributing to the majority of the flux even at lower wavelengths (approximately $15-20~\micron$), which is likely accurate; furthermore, the fits removed the warm continuum in the case of SR 21 and the cool continuum in the case of the other three targets. This suggests a dearth of hot dust, or even dust at temperatures where mineralogy can be probed; in these cases, the ALMA-probed disk cavities are likely truly devoid or nearly devoid of dust. A similar issue occurs with the warm atmosphere of UX Tau A, but this fit produces a reduced $\chi^2$ value just below 10.

As for the remaining target with a high reduced $\chi^2$, PDS 70, the model simply struggled to model its complex features. 
\revision{ \citet{2024A&A...691A.148J} also found that modeling the PDS 70 disk assuming DHS dust grains provided results inferior to those using Gaussian random variable (GRF) grains. In certain cases, it may therefore behoove future models to consider different dust grain models in the case where the model's ``default" grain set struggles}. A similar problem occurs with the CQ Tau disk, where the reduced $\chi^2$ value of the fit is just below 10; the model finds the cool disk to consist mainly of cool small amorphous olivine, but this creates a flux excess in the 23-27 $\micron$ range.

\subsection{Fitted Temperatures and Mineralogies}\label{fits}

The maximum temperatures of the warm disks in many cases are very close to, if not at, the dust evaporation temperature. \revision{Furthermore, the uncertainties in these warm disk component temperatures, particularly the upper bounds, tend to be larger than for the cool disk temperatures.
From a physical standpoint, these high warm disk temperatures likely indicate} that, for these disks, the radius of the inner rim is effectively set by the dust sublimation point --- hence, the central star's luminosity --- rather than a large inner cavity opened by planet formation or disk photoevaporation processes. 
\revision{The relatively large errors in the warm components also likely reflect our imposition of a wavelength lower limit of 7.7 $\micron$, which makes the modeling less sensitive to emission from dust at temperatures approaching sublimation.}

Most of the disks in this analysis are transition disks, with published ALMA mm/sub-mm continuum images \citep[for MP Mus image, see][for other disks in this analysis]{2023A&A...673A..77R,2020ApJ...892..111F} indicating prevalent central cavities, i.e. hot dust indicated to be present by mid-IR spectral fitting is not visible in the mm/sub-mm regime. We note that the transition disks discussed here can have optically thick emission within and beyond the gaps visible within the disk; see Table \ref{minres}, where we find evidence of continuum emission within these disks, which likely comes from this optically thick emission. 

The cool disks tend to have very cool maximum temperatures, close to room temperature. These cool maximum temperatures indicate that the innermost parts of the cool disks, while distant from the inner rim, tend to be within a few au of the host star. Therefore, the emission probed by $Spitzer$ IRS likely arises from dust within the cavities in the cental regions of these disks, despite the seeming absence of dust within the ALMA images. This assertion may be tested via au-resolution JWST mid-IR imaging or ALMA sub-mm imaging at the highest possible resolution.

Outside of continuum emission, we find evidence of large amorphous minerals within the cooler atmospheric regions of these disks in only three targets, CS Cha (in which large olivine is detected in a trace amount), DM Tau (in which large and small olivine are detected in similar amounts), and WSB 60 (in in which large olivine dominates the cool disk); in the warm disks, however, we find some evidence of large amorphous grains. Only 3 disks showed evidence for large pyroxene in the warm disk (DM Tau, GG Tau AA Ab, and WSB 60), 3 disks for large olivine in the warm disk (MP Mus, CIDA 9, and GG Tau AA Ab), and none of the disks contained any large polivene, nor did any of the cool disk models contain small polivene; this means the only polivene detected in any of these disks is small polivene in the warm zone, and there are several detections of such. As opposed to amorphous silicates, large crystalline silicates appear in many of the cool disks, particularly forsterite, with silica and enstatite comparatively rarely found, suggesting grain processing if not growth in the cool disks \citep{2009ApJS..182..477S}. 

\revision{Note that though it is theoretically possible to extract the mass directly from the mass weights of the components, units corresponding to column/surface density are used here, in line with previous mid-IR modeling \citep[e.g.,][]{2009ApJ...690.1193S,2009ApJS..182..477S}. In general, however, we find disk dust masses that are typically lower than those obtained via ALMA data analysis. For example, the mass weight components of MP Mus indicate a dust mass of $\sim1.5\times10^{-5}~M_{\odot}$, whereas analysis of ALMA continuum imaging yields a dust mass an order of magnitude larger \citep{2023A&A...673A..77R}. The discrepancy arises because the bulk of the dust disk, which can be mapped by ALMA, is too cold to significantly thermally radiate in the mid-IR. }

\subsection{Comparisons of Mineralogical Results with Previous Work}

\revision{While a one-to-one comparison of mineralogical results and goodness of fit with previous modeling of these disks is difficult, given the different model parameterizations, we summarize here key points of agreement and disagreement. }

\subsubsection{MP Mus}
As we indicate in \S \ref{goodness}, both in this study and our previous one \citep{2024ApJ...970..137G}, we find pyroxene to be dominant in the cool disk and olivine in the warm disk of MP Mus, with significant amounts of forsterite in the cool disk and large crystalline minerals in the warm disk. The study by \citet{2008ApJ...683..479B} is in general agreement, having found the disk to be dominated by olivine at shorter wavelengths and pyroxene at longer wavelengths with small amounts of crystalline minerals throughout.

\subsubsection{DM Tau}
We find the disk of DM Tau to be dominated by both small and large olivine in the cool zone in our model along with a significant amount of small pyroxene; in the warm zone, we find a dominance of small olivine along with a significant amount of large pyroxene and large forsterite. \citet{2016ApJ...831..167M} found that DM Tau consists dominantly of olivine with a significant amount of pyroxene in the outer wall of the disk, which generally agrees with the distribution of minerals in our model. \citet{2009ApJS..182..477S} found that the warm disk of DM Tau consists most significantly of large amorphous silicates with some small olivine and trace ($<10\%$ each) amounts of forsterite, where we find more significant amounts of small olivine and forsterite with no large olivine; they found the cool disk of DM Tau to contain mostly small olivine, agreeing with our model though we also find significant large olivine and small pyroxene, as well as significant amounts of forsterite and silica, of which we find trace amounts of larger grains of those species. \citet{2011ApJ...734...51O} found small amorphous grain dominance and substantial silica in the cool disk, where we only find trace amounts of crystalline minerals, though our models agree in finding a dominant mixture of small and large amorphous grains; however, in the warm disk, they found significant amounts of both enstatite and silica but trace amounts of forsterite, where our TZTD model finds significant forsterite but trace amounts of silica and no enstatite, and where we find a mixture of small and large amorphous grains, they found a dominance of large grains.

\subsubsection{DoAr 44}
In the disk of DoAr 44, we find the cool disk to consist almost entirely of small olivine and the warm disk to be dominated by small polivene. \citet{2006A&A...456..535S} found a dominance of small olivine in the disk, in agreement with our model, though they also found large pyroxene in the disk, which we do not. \citet{2010A&A...520A..39O} found dominantly small amorphous materials in the warm disk zone and large amorphous materials in the cool disk zone, with trace amounts of crystalline minerals, including a small but noticeable amount of small forsterite in the cool disk, which is partially in agreement with our analysis, though we do not find large amorphous minerals in the cool disk.

\subsubsection{GG Tau AA Ab}
In the GG Tau disk, we find a dominance of large pyroxene and olivine in the warm disk, as well as a significant amount of small olivine and polivene; we also find that the cool disk is almost entirely small olivine with some small pyroxene and large forsterite and enstatite. \citet{2006A&A...456..535S} found some of each type of mineral in the disk except for silica, where we find trace amounts of large silica in the warm disk; they also found a dominance of small olivine with a lesser dominance of large pyroxene, the former of which lines up well with our cool disk model.
\citet{2006ApJ...646.1024H} found that the GG Tau disk is dominated by olivine, both small and large, with a significant amount of large pyroxene as well, and a trace amount of crystalline silicates, in general agreement with both zones in our model, though we find more of a mix of minerals in the warm disk and a significant amount of small pyroxene in the cool disk rather than large. \citet{2009ApJS..182..477S} found mostly large pyroxene, with significant amounts of small and large olivine and trace amounts of other materials, where we find a mix of large amorphous minerals along with some small amorphous minerals; in the cool disk, they found mostly small pyroxene with a significant amount of forsterite and trace amounts of other minerals, where we find the cool disk to consist mostly of small olivine but with a significant amount of small pyroxene and a small but noticeable amount of large forsterite along with trace amounts of other crystalline minerals. The model used by \citet{2011ApJ...734...51O} found a significant amount of silica along with trace amounts of other crystals and a dominance of large amorphous grains in the warm disk, where we find similar results, except with less crystalline minerals and a significant amount of small amorphous silicates along with the large grains.

\subsubsection{GM Aur}
In the warm disk of GM Aur, we find a dominance of small polivene with a significant amount of small olivine, and in the cool disk, we find a dominance of small olivine with a significant amount of small pyroxene; thus, throughout the whole disk, we find a large amount of small olivine. \citet{2006A&A...456..535S} found a dominance of both small and large olivine with trace amounts of other minerals in the disk, where we find no large olivine and a dominance of small polivene in the warm disk but a dominance of small olivine in the cool disk. \citet{2016ApJ...831..167M} found that GM Aur consists entirely of olivine and pyroxene in equal parts in the sublimation and outer walls, which is similar to our cool disk model though we find olivine to dominate. 
\citet{2009ApJS..182..477S} found the warm disk to consist almost entirely of small olivine, and while we find small polivene to be dominant in that zone, we also find a large amount of small olivine in agreement with the model; they found the cool disk to be dominated by small amorphous minerals, mostly pyroxene but also olivine, and our model agrees, but with the most dominant of the two reversed. \citet{2011ApJ...734...51O} found a significant amount of silica in both disk zones along with both enstatite and forsterite in the warm disk, where we only find trace amounts of crystalline minerals in both disks. In the warm disk, they found a dominance of small grains, in agreement with the TZTD model, but with a significant amount of large grains where the TZTD model finds none; in the cool disk, they found a dominance of large amorphous grains, but we find only small amorphous grains there. 

\subsubsection{IP Tau}
The TZTD model finds the disk of IP Tau to consist nearly entirely of small olivine in the cool disk and a dominance of small polivene in the warm disk along with significant amounts of small pyroxene.
The study by \citet{2009ApJS..182..477S} found the warm disk of IP Tau to consist almost entirely of small amorphous minerals, mostly olivine with some pyroxene, which lines up with ours, except we only find pyroxene in the warm disk and find polivene to be dominant; in the cool disk, they found a dominance of large pyroxene with a significant amount of small olivine, while we find small olivine dominates, with no pyroxene and a trace but noticeable amount of small and large forsterite. \citet{2011ApJ...734...51O} found a dominance of amorphous materials of various sizes with trace amounts of crystalline silicates in the warm disk, in agreement with our model, though we find only small amorphous silicates; however, in the cool disk, they found large amorphous grain dominance and a substantial amount of silica, which the TZTD model does not detect. 

\subsubsection{LkCa 15}
In the warm disk of LkCa 15, we find a dominance of small polivene in the warm disk with a noticeable amount of small pyroxene and trace amounts of crystalline minerals; in the cool disk, we find a dominance of small olivine with trace amounts of crystalline minerals. \citet{2006A&A...456..535S} found a dominance of both small and large olivine with trace amounts of other minerals, where we find a dominance of small polivene in the warm disk and a notable presence of small pyroxene in the warm disk but no large olivine in either disk, despite an agreement on the dominance of small olivine in the cool disk. \citet{2016ApJ...831..167M} found that in both the sublimation and outer walls of LkCa 15, the constituency is entirely olivine, which is close and in general agreement to our cool disk model. \citet{2009ApJS..182..477S} found that small olivine dominates the warm disk of LkCa 15, with a noticeable amount of large pyroxene, though we find small polivene to be dominant with small pyroxene also significant; they found the cool disk to consist dominantly of small pyroxene with noticeable amounts of forsterite and silica, though we find a dominance of small olivine. The model used by \citet{2011ApJ...734...51O} with LkCa 15 is in agreement with our model in the warm disk of LkCa 15, where only trace amounts of crystalline minerals were found, although they indicate a significant amount of large amorphous grains where we find none; however, in the cool disk, they found a highly significant amount of silica, but the TZTD model finds the cool disk to consist almost entirely of small olivine, where the previous model found a mixture of small and large amorphous grains with more large grains.

\subsubsection{RX J1842.9-3532}
In the disk of RX J1842.9-3532, we find a dominance of small amorphous minerals with trace but noticeable amounts of crystalline minerals; in particular, we find a significant amount of small olivine and a dominance of small polivene in the warm disk, and a significant amount of small pyroxene with a dominance of small olivine in the cool disk. The analysis by \citet{2008ApJ...683..479B} found the RX J1842.9-3532 disk to be dominated by olivine at shorter wavelengths and pyroxene at longer wavelengths, with trace amounts of crystalline minerals throughout; while the TZTD model finds polivene to be dominant in the warm disk, the cool disk dominance by olivine and the significant presence of olivine in the warm disk results in a general dominance of olivine in the disk, in agreement with the given model.

\subsubsection{RX J1852.3-3700}
We find a similar composition in the disk of RX J1852.3-3700 to that of RX J1842.9-3532, with a significant amount of small olivine and a dominance of small polivene in the warm disk, and a significant amount of small pyroxene with a dominance of small olivine in the cool disk, along with trace amounts of crystalline minerals in the disk. \citet{2008ApJ...683..479B} found the RX J1852.3-3700 disk to be almost entirely olivine, with only trace amounts of other minerals and a significant amount of pyroxene at longer wavelengths; as with the previous case, though we find significant amounts of other minerals, there is general agreement with the dominance of olivine in the disk.

\subsubsection{RY Tau}
In our model of RY Tau, we find the cool disk to consist almost entirely of small olivine with a trace amount of small forsterite; we find the warm disk to be dominated by small polivene, with a significant amount of small pyroxene and trace amounts of other minerals. \citet{2009A&A...508..247G} found a dominance of small pyroxene and large olivine in the RY Tau disk, along with the presence of both small and large forsterite, small enstatite, and small silica; our model is in agreement with the presence of small pyroxene, small forsterite, small enstatite, and large silica, but in our model, large forsterite and large olivine are entirely absent, with our model instead detecting a dominance of small olivine in the cool disk and small polivene in the warm disk. \citet{2006ApJ...646.1024H} found the disk to consist mostly of olivine, with some large pyroxene and trace amounts of crystalline minerals; in our model, we find small amorphous minerals of multiple species in both disks as the dominant constituency, but no large amorphous minerals at all, though we find some crystalline minerals. \citet{2011ApJ...734...51O} found significant amounts of enstatite and silica in the warm disk alongside trace amounts of forsterite and a dominance of large amorphous materials, but TZTD detects a smaller amount of enstatite and silica and a dominance of small amorphous materials.

\subsubsection{Further Comparisons}
\citet{2011ApJ...728...49E} targeted several of the same disks we did in our analysis, (CS Cha, DM Tau, GM Aur, IP Tau, LkCa 15, RY Tau, and UX Tau A) so we briefly summarize their results. Within the outer wall and disk zones assigned in their version of the model, in most cases, amorphous minerals made up more than $90\%$ of the dust; this tends to be consistent with the disk zones in our analysis. The exceptions to the dominance of amorphous minerals in the study are LkCa 15, which was found to have $11\%$ crystalline silica, and UX Tau A, which was found to have $10\%$ forsterite, $4\%$ enstatite, and $5\%$ silica. In the case of LkCa 15, our model finds only trace amounts of silica in both disk zones; in the case of UX Tau A, crystalline minerals entirely dominated the warm disk in our model while the cool disk has trace amounts of crystalline minerals, so the combination of both of our zones is ultimately the best comparison in this case. \citet{2011ApJ...728...49E} also modeled the optically thin regions of some of these disks (CS Cha, GM Aur, IP Tau, LkCa 15 within 1 au and from 1-15 au within its gap separately, and RY Tau), and in all cases, amorphous minerals were found to be dominant by our previous definition, consistent with our disks. In the case of LkCa 15, within 1 au, trace amounts of forsterite were found by the previous model, consistent with both of our disk zones; however, no crystalline minerals were found by the previous model in the optically thin region beyond that, which is inconsistent with both zones as we modeled them, as we find trace amounts of crystalline minerals in both. This same inconsistency applies between our findings of the mineralogy of RY Tau.

In a comprehensive search for crystallinity in disks using $Spitzer$ IRS data, \citet{2016AJ....151..146C} found crystallinity percentages and lower limits for several targets from our analysis; we compare those to the crystallinity percentages of the warm and cool disk zones in our model excluding continuum in Table \ref{crystallinity} in Appendix \ref{appe}. We note that in each case besides CQ Tau, GM Aur, LkCa 15, RY Tau, and SR 21, the crystallinity of the warm disk is between the lower limit and estimated crystallinity of the search by \citet{2016AJ....151..146C}; in the former four cases, the measured crystallinity is lower, and it is higher in the latter case. In the cool disks, in all cases except for MP Mus, GG Tau AA Ab, SR 21, and WSB 60, the TZTD-estimated crystallinity is lower than the imposed lower limits; in the cases of each disk except for SR 21, where it is above the estimated value, the crystallinity is above the lower limit but below the estimated value. Note that the TZTD model finds SR 21, a disk the model struggled to properly fit, to be entirely crystalline.

\section{Discussion}\label{disc}
\subsection{Model Assessment}
Based on the fact that many of our reduced $\chi^2$ values are low, we conclude that the TZTD model in its current state is able to produce robust fits to spectra of protoplanetary disks. In particular, the marked improvement of the fit to the MP Mus spectrum despite the use of fewer parameters highlights the promise of this version of the TZTD model to efficiently generate sets of disk structural and mineralogical model parameters that well reproduce mid-IR spectral data. This is aided by the model's simplicity, maintaining speed and few nonlinear parameters, while also capturing the temperature distribution inherent in protoplanetary disks. 

We make use of approximations of optically thick dust by using distributions including large dust while using the assumption of an optically thin disk to aid in this simplicity. However, while the assumption of an optically thick portion would be more accurate, it would also be more complex and slow down the model. Furthermore, it would require the model to fit for exact spatial parameters, such as the inner radius of the disk via the solid angle of the optically thick emission, as well as the value of $q$, further constraining the model spatially; since empirical mid-IR spectral analysis poorly constrains the disk spatially, this tends to return values that disagree with radiative transfer models, i.e. rendering an inaccurate view of the disk. Therefore, with our assumption, we trade greater accuracy for simplicity and leave spatial analysis to radiative transfer models.

While the model well fit many disks, three out of the four that were poorly fit were likely the result of lacking a significant amount of dust at temperatures such that thermal emission would largely be in the mid-IR regime. Since these disks lacked many distinct spectral features, a potential improvement could be made by removing the warm component from these disks and permitting a low maximum temperature to accommodate the increase in flux at longer wavelengths. More observation data of such disks as these, particularly at longer wavelengths, would also aid in mineralogy analysis.

\subsection{Future Directions}
We plan on developing the EaRTH Disk Model further and analyzing the transition disks studied in this article on the radiative transfer side of the model using MCFOST \citep{2006A&A...459..797P,2009A&A...498..967P}. Once these data are further analyzed and published, we seek to broaden the utility of the TZTD model and, by extension, the EaRTH Disk Model by analyzing spectra from newer spectroscopic surveys, such as those conducted by JWST using instruments such as the Mid-Infrared Instrument (MIRI) \citep{2015PASP..127..584R,2015PASP..127..595W}. 

Looking forward, we also see potential for using the EaRTH Disk Model with spectra from the PRobe far-Infrared Mission for Astrophysics (PRIMA) \citep{2022AAS...24030407B}. In particular, Band 1 of PRIMA's Far-Infrared Enhanced Survey Spectrometer (FIRESS) would record spectra from 24-43 $\micron$ ($R \sim 95-150$), overlapping with our current spectra. As we have demonstrated in our analysis, the overlapping region (24-37 $\micron$) is very informative on the mineralogy of protoplanetary disks in conjunction with our model, contributing to evidence of a significant amount of cool crystalline silicates in most of the disks we analyzed (see Figure \ref{res_plot}); further FIRESS observations of these or similar disks could confirm our finding, as well as allow expanded analysis using the longer wavelength results from Band 1 observations. This could strongly benefit analysis of disks lacking in warm dust or with a particular dominance of cool dust, such as those disks the TZTD model struggled to fit.

\section{Conclusions}\label{conc}
We have presented a new formulation for mid-IR thermal emission of a protoplanetary disk, which relies on a radial distribution of temperature and mass, but does not directly fit or assume either, only an empirical relation between the two. An assumption on this relation can be made for a faster model and can be done to omit spatial information analysis in regards to a mineralogy analysis; such analysis is best left to radiative transfer models. We assume a warm and cool zone within a protoplanetary disk, each with its own thermal emission distribution, as well as an outer disk with a generalized distribution to approximate its effects on the mid-IR SED. We derived this formula and used it to fit multiple $Spitzer$ IRS spectra to demonstrate the efficacy of this TZTD model.

We demonstrate the capabilities of the TZTD model by fitting a variety of $Spitzer$ IRS spectra to a large range of mineral opacities. In its present formulation, the model results in robust fits with these spectra, and these fits are highly informative. The fits reveal that cool disks have large crystalline minerals, mainly forsterite, indicative of grain processing. In a forthcoming paper, the set of transition disks analyzed here and the mineralogical parametrization from this model will be analyzed and fitted with a radiative transfer model, i.e. the EaRTH Disk Model, to improve said model's ability to reproduce mid-IR spectra and mm/sub-mm images in a consistent parametrization.

Empirical mineralogical modeling that returns some results that can be used to infer certain spatial information about the protoplanetary disk, such as a range of temperatures and an empirical relation between unknown distribution exponents, while not directly outputting such spatial information, is ideal for analyzing protoplanetary disks. The empirical information provided effectively constrains radiative transfer models while permitting them the freedom to fit directly for more specific spatial information. This modeling works well for $Spitzer$ spectra and will soon work with lower-resolution JWST spectra and lower-wavelength PRIMA spectra\revision{, i.e. $\sim24-43~\micron$}.

\section*{Acknowledgements}\label{ackn}

We thank the anonymous referee for their careful review and helpful comments which helped to greatly improve this article.

This research was supported by NASA Astrophysics Data Analysis Program grant 80NSSC22K0625 to RIT.

We acknowledge fellowship funding support from the Frist Center for Autism $\&$ Innovation at Vanderbilt University.

This research has made use of the NASA/ IPAC Infrared Science Archive, which is operated by
the Jet Propulsion Laboratory, California Institute of Technology, under contract with the National
Aeronautics and Space Administration.

This work is based in part on observations made with the $Spitzer~Space~Telescope$, obtained
from the NASA/ IPAC Infrared Science Archive, both of which are operated by the Jet Propulsion
Laboratory, California Institute of Technology under a contract with the National Aeronautics and
Space Administration.

This research makes use of data products from the Two Micron All Sky Survey, which is a joint project of the University of Massachusetts and the Infrared Processing and Analysis Center/California Institute of Technology, funded by the National Aeronautics and Space Administration and the National Science Foundation.

This work has made use of data from the European Space Agency (ESA) mission
{\it Gaia} (\url{https://www.cosmos.esa.int/gaia}), processed by the {\it Gaia}
Data Processing and Analysis Consortium (DPAC,
\url{https://www.cosmos.esa.int/web/gaia/dpac/consortium}). Funding for the DPAC
has been provided by national institutions, in particular the institutions
participating in the {\it Gaia} Multilateral Agreement.

This research has made use of the SIMBAD database,
operated at CDS, Strasbourg, France. The original description of the SIMBAD service was published in \citet{2000A&AS..143....9W}.

This research has made use of the VizieR catalogue access tool, CDS,
Strasbourg, France (DOI : 10.26093/cds/vizier). The original description of the VizieR service was published in \citet{2000A&AS..143...23O}.

The online documentation for SciPy~\citep{2020SciPy-NMeth} was referenced for the utility of some Python functions, as well as sample code to see how the functions work, particularly \verb'scipy.interpolate.interp1d' for data interpolation. The program made significant use of NumPy~\citep{oliphant2006guide,2011CSE....13b..22V,2020Natur.585..357H} and AstroPy \citep{2013A&A...558A..33A}. Graph plotting was done using Matplotlib~\citep{Hunter:2007}.

This research has made use of Wolfram Alpha to assist in the derivation of the model.

This research made use of the Heidelberg - Jena - St.Petersburg - Database of Optical Constants (HJPDOC) for the retrieval of silicate optical constants \citep{1999A&AS..136..405H,2003JQSRT..79..765J}.

\bibliography{biblio}
\bibliographystyle{aasjournalv7}

\restartappendixnumbering

\appendix{\section{Data Results}\label{appe}

\begin{deluxetable*}{lccc}
\tabletypesize{\tiny}
\deluxetablecaption{TZTD Mineralogy Results\label{minres}}
\tablehead{
\colhead{Target} & \colhead{wCont ($g~cm^{-2}$)} & \colhead{cCont ($g~cm^{-2}$)} & \colhead{Outer ($g~cm^{-2}$)}
}
\startdata
MP Mus & (1.9 $\pm$ 0.0596) $\times~10^{-16}$ & (3.03 $\pm$ 0.0658) $\times~10^{-14}$ & (2.88 $\pm$ 0.458) $\times~10^{-13}$\\
CIDA 9 & (2.99 $\pm$ 0.0798) $\times~10^{-17}$ & (2.54 $\pm$ 0.122) $\times~10^{-15}$ & (6.61 $\pm$ 0.955) $\times~10^{-13}$\\
CQ Tau & (5.08 $\pm$ 0.318) $\times~10^{-16}$ & (3.68 $\pm$ 0.168) $\times~10^{-13}$ & (3.49 $\pm$ 0.433) $\times~10^{-11}$\\
CS Cha & (1.19 $\pm$ 0.118) $\times~10^{-17}$ & (3.46 $\pm$ 0.102) $\times~10^{-13}$ & (2.51 $\pm$ 0.797) $\times~10^{-11}$\\
DM Tau & (9.96 $\pm$ 0.676) $\times~10^{-18}$ & (1.93 $\pm$ 0.0592) $\times~10^{-14}$ & (4.85 $\pm$ 1.01) $\times~10^{-12}$\\
DoAr 44 & (6.19 $\pm$ 0.384) $\times~10^{-17}$ & (2.42 $\pm$ 0.171) $\times~10^{-14}$ & (4.58 $\pm$ 0.289) $\times~10^{-12}$\\
GG Tau AA Ab & (8.89 $\pm$ 0.296) $\times~10^{-17}$ & (2.48 $\pm$ 0.0684) $\times~10^{-14}$ & (1.69 $\pm$ 0.284) $\times~10^{-12}$\\
GM Aur & (1.97 $\pm$ 0.166) $\times~10^{-17}$ & (6.31 $\pm$ 0.255) $\times~10^{-14}$ & (1.38 $\pm$ 0.189) $\times~10^{-12}$\\
IP Tau & (1.81 $\pm$ 0.0882) $\times~10^{-17}$ & (1.22 $\pm$ 0.073) $\times~10^{-15}$ & (4.45 $\pm$ 0.327) $\times~10^{-14}$\\
J1604.3-2130 & (2.91 $\pm$ 0.0672) $\times~10^{-18}$ & (0.0 $\pm$ 7.23) $\times~10^{-19}$ & (4.53 $\pm$ 0.0971) $\times~10^{-13}$\\
LkCa 15 & (4.62 $\pm$ 0.265) $\times~10^{-17}$ & (0.0 $\pm$ 4.38) $\times~10^{-16}$ & (6.55 $\pm$ 0.291) $\times~10^{-13}$\\
PDS 70 & (1.1 $\pm$ 0.0639) $\times~10^{-17}$ & (4.22 $\pm$ 0.235) $\times~10^{-15}$ & (3.35 $\pm$ 0.351) $\times~10^{-12}$\\
RX J1842.9-3532 & (1.38 $\pm$ 0.0855) $\times~10^{-17}$ & (3.23 $\pm$ 0.459) $\times~10^{-15}$ & (8.92 $\pm$ 0.487) $\times~10^{-13}$\\
RX J1852.3-3700 & (6.13 $\pm$ 0.605) $\times~10^{-18}$ & (1.44 $\pm$ 0.0418) $\times~10^{-13}$ & (2.88 $\pm$ 0.899) $\times~10^{-11}$\\
RY Tau & (5.96 $\pm$ 0.481) $\times~10^{-16}$ & (1.11 $\pm$ 0.0559) $\times~10^{-13}$ & (5.17 $\pm$ 0.493) $\times~10^{-12}$\\
SR 21 & (2.39 $\pm$ 0.215) $\times~10^{-17}$ & (7.86 $\pm$ 0.277) $\times~10^{-16}$ & (1.09 $\pm$ 0.0157) $\times~10^{-12}$\\
SR 24 S & (3.21 $\pm$ 0.147) $\times~10^{-16}$ & (9.5 $\pm$ 0.383) $\times~10^{-14}$ & (1.14 $\pm$ 0.165) $\times~10^{-11}$\\
UX Tau A & (8.87 $\pm$ 0.202) $\times~10^{-17}$ & (0.0 $\pm$ 6.77) $\times~10^{-15}$ & (6.82 $\pm$ 0.575) $\times~10^{-12}$\\
V1247 Ori & (3.33 $\pm$ 0.0888) $\times~10^{-16}$ & (1.78 $\pm$ 0.0771) $\times~10^{-12}$ & (8.95 $\pm$ 2.28) $\times~10^{-7}$\\
WSB 60 & (2.54 $\pm$ 0.0837) $\times~10^{-17}$ & (2.76 $\pm$ 0.121) $\times~10^{-15}$ & (1.62 $\pm$ 0.0986) $\times~10^{-13}$\\
\enddata
\end{deluxetable*}

\begin{deluxetable*}{lccccccc}
\tabletypesize{\tiny}
\tablenum{A1}
\deluxetablecaption{(continued.)}
\tablehead{
\colhead{Target} & \colhead{wSmAmPy ($g~cm^{-2}$)} & \colhead{wLgAmPy ($g~cm^{-2}$)} & \colhead{wSmAmOl ($g~cm^{-2}$)} & \colhead{wLgAmOl ($g~cm^{-2}$)} & \colhead{wSmAmPo ($g~cm^{-2}$)} & \colhead{wLgAmPo ($g~cm^{-2}$)}
}
\startdata
MP Mus & (0.0 $\pm$ 4.38) $\times~10^{-21}$ & (0.0 $\pm$ 6.39) $\times~10^{-21}$ & (7.55 $\pm$ 5.23) $\times~10^{-21}$ & (4.53 $\pm$ 0.685) $\times~10^{-20}$ & (0.0 $\pm$ 7.01) $\times~10^{-21}$ & (0.0 $\pm$ 8.1) $\times~10^{-21}$\\
CIDA 9 & (2.37 $\pm$ 0.683) $\times~10^{-21}$ & (0.0 $\pm$ 8.92) $\times~10^{-22}$ & (7.05 $\pm$ 0.679) $\times~10^{-21}$ & (2.04 $\pm$ 0.831) $\times~10^{-21}$ & (0.0 $\pm$ 9.73) $\times~10^{-22}$ & (0.0 $\pm$ 1.04) $\times~10^{-21}$\\
CQ Tau & (1.22 $\pm$ 0.35) $\times~10^{-19}$ & (0.0 $\pm$ 3.59) $\times~10^{-20}$ & (0.0 $\pm$ 3.74) $\times~10^{-20}$ & (0.0 $\pm$ 3.31) $\times~10^{-20}$ & (1.34 $\pm$ 0.0525) $\times~10^{-18}$ & (0.0 $\pm$ 4.63) $\times~10^{-20}$\\
CS Cha & (0.0 $\pm$ 9.56) $\times~10^{-22}$ & (0.0 $\pm$ 1.25) $\times~10^{-21}$ & (1.6 $\pm$ 0.103) $\times~10^{-20}$ & (0.0 $\pm$ 1.25) $\times~10^{-21}$ & (0.0 $\pm$ 1.42) $\times~10^{-21}$ & (0.0 $\pm$ 1.54) $\times~10^{-21}$\\
DM Tau & (0.0 $\pm$ 5.57) $\times~10^{-22}$ & (2.21 $\pm$ 0.848) $\times~10^{-21}$ & (3.35 $\pm$ 0.739) $\times~10^{-21}$ & (0.0 $\pm$ 9.9) $\times~10^{-22}$ & (0.0 $\pm$ 9.28) $\times~10^{-22}$ & (0.0 $\pm$ 1.08) $\times~10^{-21}$\\
DoAr 44 & (4.3 $\pm$ 0.441) $\times~10^{-20}$ & (0.0 $\pm$ 5.19) $\times~10^{-21}$ & (1.39 $\pm$ 0.493) $\times~10^{-20}$ & (0.0 $\pm$ 5.01) $\times~10^{-21}$ & (1.35 $\pm$ 0.0625) $\times~10^{-19}$ & (0.0 $\pm$ 6.01) $\times~10^{-21}$\\
GG Tau AA Ab & (0.0 $\pm$ 2.51) $\times~10^{-21}$ & (2.23 $\pm$ 0.353) $\times~10^{-20}$ & (7.22 $\pm$ 2.78) $\times~10^{-21}$ & (2.26 $\pm$ 0.355) $\times~10^{-20}$ & (1.33 $\pm$ 0.378) $\times~10^{-20}$ & (0.0 $\pm$ 4.25) $\times~10^{-21}$\\
GM Aur & (8.62 $\pm$ 1550) $\times~10^{-24}$ & (0.0 $\pm$ 2.14) $\times~10^{-21}$ & (2.44 $\pm$ 0.183) $\times~10^{-20}$ & (0.0 $\pm$ 2.21) $\times~10^{-21}$ & (3.99 $\pm$ 0.237) $\times~10^{-20}$ & (0.0 $\pm$ 2.56) $\times~10^{-21}$\\
IP Tau & (6.16 $\pm$ 0.882) $\times~10^{-21}$ & (0.0 $\pm$ 1.03) $\times~10^{-21}$ & (0.0 $\pm$ 9.2) $\times~10^{-22}$ & (0.0 $\pm$ 9.94) $\times~10^{-22}$ & (1.67 $\pm$ 0.125) $\times~10^{-20}$ & (0.0 $\pm$ 1.24) $\times~10^{-21}$\\
J1604.3-2130 & (0.0 $\pm$ 4.1) $\times~10^{-23}$ & (0.0 $\pm$ 6.71) $\times~10^{-23}$ & (0.0 $\pm$ 2.86) $\times~10^{-23}$ & (0.0 $\pm$ 3.6) $\times~10^{-23}$ & (0.0 $\pm$ 5.54) $\times~10^{-23}$ & (0.0 $\pm$ 6.94) $\times~10^{-23}$\\
LkCa 15 & (1.04 $\pm$ 0.287) $\times~10^{-20}$ & (0.0 $\pm$ 3.04) $\times~10^{-21}$ & (0.0 $\pm$ 2.72) $\times~10^{-21}$ & (0.0 $\pm$ 2.71) $\times~10^{-21}$ & (1.12 $\pm$ 0.0377) $\times~10^{-19}$ & (0.0 $\pm$ 3.59) $\times~10^{-21}$\\
PDS 70 & (8.97 $\pm$ 0.687) $\times~10^{-21}$ & (0.0 $\pm$ 6.23) $\times~10^{-22}$ & (0.0 $\pm$ 6.75) $\times~10^{-22}$ & (0.0 $\pm$ 5.81) $\times~10^{-22}$ & (8.04 $\pm$ 0.956) $\times~10^{-21}$ & (0.0 $\pm$ 8.14) $\times~10^{-22}$\\
RX J1842.9-3532 & (2.18 $\pm$ 1.0) $\times~10^{-21}$ & (0.0 $\pm$ 1.14) $\times~10^{-21}$ & (8.98 $\pm$ 1.09) $\times~10^{-21}$ & (0.0 $\pm$ 1.08) $\times~10^{-21}$ & (3.85 $\pm$ 0.139) $\times~10^{-20}$ & (0.0 $\pm$ 1.32) $\times~10^{-21}$\\
RX J1852.3-3700 & (0.0 $\pm$ 5.2) $\times~10^{-22}$ & (0.0 $\pm$ 7.2) $\times~10^{-22}$ & (5.63 $\pm$ 0.602) $\times~10^{-21}$ & (0.0 $\pm$ 7.24) $\times~10^{-22}$ & (1.09 $\pm$ 0.0802) $\times~10^{-20}$ & (0.0 $\pm$ 8.73) $\times~10^{-22}$\\
RY Tau & (4.49 $\pm$ 0.617) $\times~10^{-19}$ & (0.0 $\pm$ 6.76) $\times~10^{-20}$ & (5.55 $\pm$ 7.2) $\times~10^{-20}$ & (0.0 $\pm$ 6.81) $\times~10^{-20}$ & (1.92 $\pm$ 0.088) $\times~10^{-18}$ & (0.0 $\pm$ 8.03) $\times~10^{-20}$\\
SR 21 & (0.0 $\pm$ 9.95) $\times~10^{-22}$ & (0.0 $\pm$ 1.69) $\times~10^{-21}$ & (0.0 $\pm$ 9.04) $\times~10^{-22}$ & (0.0 $\pm$ 1.26) $\times~10^{-21}$ & (0.0 $\pm$ 1.34) $\times~10^{-21}$ & (0.0 $\pm$ 1.73) $\times~10^{-21}$\\
SR 24 S & (3.34 $\pm$ 1.84) $\times~10^{-20}$ & (0.0 $\pm$ 2.21) $\times~10^{-20}$ & (2.63 $\pm$ 2.65) $\times~10^{-20}$ & (0.0 $\pm$ 2.72) $\times~10^{-20}$ & (0.0 $\pm$ 2.9) $\times~10^{-20}$ & (0.0 $\pm$ 2.78) $\times~10^{-20}$\\
UX Tau A & (0.0 $\pm$ 1.28) $\times~10^{-21}$ & (0.0 $\pm$ 1.49) $\times~10^{-21}$ & (0.0 $\pm$ 1.26) $\times~10^{-21}$ & (0.0 $\pm$ 1.34) $\times~10^{-21}$ & (0.0 $\pm$ 1.94) $\times~10^{-21}$ & (0.0 $\pm$ 1.96) $\times~10^{-21}$\\
V1247 Ori & (0.0 $\pm$ 1.99) $\times~10^{-21}$ & (0.0 $\pm$ 2.49) $\times~10^{-21}$ & (0.0 $\pm$ 1.79) $\times~10^{-21}$ & (0.0 $\pm$ 2.04) $\times~10^{-21}$ & (0.0 $\pm$ 3.01) $\times~10^{-21}$ & (0.0 $\pm$ 3.34) $\times~10^{-21}$\\
WSB 60 & (1.66 $\pm$ 0.686) $\times~10^{-21}$ & (1.64 $\pm$ 0.969) $\times~10^{-21}$ & (5.74 $\pm$ 0.763) $\times~10^{-21}$ & (0.0 $\pm$ 9.86) $\times~10^{-22}$ & (0.0 $\pm$ 1.05) $\times~10^{-21}$ & (0.0 $\pm$ 1.18) $\times~10^{-21}$\\
\enddata
\end{deluxetable*}

\begin{deluxetable*}{lcccccc}
\tabletypesize{\tiny}
\deluxetablecaption{(continued)}
\tablenum{A1}
\tablehead{
\colhead{Target} & \colhead{wSmCryForst ($g~cm^{-2}$)} & \colhead{wSmCryEnst ($g~cm^{-2}$)} & \colhead{wSmCrySil ($g~cm^{-2}$)} & \colhead{wLgCryForst ($g~cm^{-2}$)} & \colhead{wLgCryEnst ($g~cm^{-2}$)} & \colhead{wLgCrySil ($g~cm^{-2}$)}
}
\startdata
MP Mus & (3.98 $\pm$ 3.79) $\times~10^{-21}$ & (5.41 $\pm$ 38.6) $\times~10^{-22}$ & (0.0 $\pm$ 1.27) $\times~10^{-21}$ & (0.0 $\pm$ 8.19) $\times~10^{-21}$ & (5.64 $\pm$ 5.2) $\times~10^{-21}$ & (4.41 $\pm$ 2.95) $\times~10^{-21}$\\
CIDA 9 & (1.99 $\pm$ 4.51) $\times~10^{-22}$ & (1.28 $\pm$ 54.6) $\times~10^{-23}$ & (0.0 $\pm$ 2.2) $\times~10^{-22}$ & (0.0 $\pm$ 8.58) $\times~10^{-22}$ & (2.12 $\pm$ 6.78) $\times~10^{-22}$ & (9.42 $\pm$ 4.67) $\times~10^{-22}$\\
CQ Tau & (5.09 $\pm$ 28.8) $\times~10^{-21}$ & (5.87 $\pm$ 3.22) $\times~10^{-20}$ & (0.0 $\pm$ 6.65) $\times~10^{-21}$ & (0.0 $\pm$ 4.04) $\times~10^{-20}$ & (0.0 $\pm$ 3.83) $\times~10^{-20}$ & (7.2 $\pm$ 1.88) $\times~10^{-20}$\\
CS Cha & (4.04 $\pm$ 6.51) $\times~10^{-22}$ & (5.38 $\pm$ 87.0) $\times~10^{-23}$ & (0.0 $\pm$ 3.36) $\times~10^{-22}$ & (1.09 $\pm$ 1.59) $\times~10^{-21}$ & (0.0 $\pm$ 1.07) $\times~10^{-21}$ & (2.17 $\pm$ 0.704) $\times~10^{-21}$\\
DM Tau & (8.16 $\pm$ 56.3) $\times~10^{-23}$ & (0.0 $\pm$ 4.82) $\times~10^{-22}$ & (0.0 $\pm$ 1.39) $\times~10^{-22}$ & (6.87 $\pm$ 14.0) $\times~10^{-22}$ & (0.0 $\pm$ 7.04) $\times~10^{-22}$ & (2.12 $\pm$ 3.21) $\times~10^{-22}$\\
DoAr 44 & (6.12 $\pm$ 3.71) $\times~10^{-21}$ & (8.71 $\pm$ 3.93) $\times~10^{-21}$ & (0.0 $\pm$ 1.03) $\times~10^{-21}$ & (0.0 $\pm$ 6.58) $\times~10^{-21}$ & (1.24 $\pm$ 5.09) $\times~10^{-21}$ & (3.11 $\pm$ 2.42) $\times~10^{-21}$\\
GG Tau AA Ab & (1.26 $\pm$ 1.99) $\times~10^{-21}$ & (1.31 $\pm$ 2.1) $\times~10^{-21}$ & (0.0 $\pm$ 7.72) $\times~10^{-22}$ & (0.0 $\pm$ 4.23) $\times~10^{-21}$ & (0.0 $\pm$ 2.78) $\times~10^{-21}$ & (2.85 $\pm$ 1.62) $\times~10^{-21}$\\
GM Aur & (8.73 $\pm$ 12.5) $\times~10^{-22}$ & (0.0 $\pm$ 1.35) $\times~10^{-21}$ & (0.0 $\pm$ 3.73) $\times~10^{-22}$ & (0.0 $\pm$ 3.19) $\times~10^{-21}$ & (0.0 $\pm$ 1.75) $\times~10^{-21}$ & (1.47 $\pm$ 0.906) $\times~10^{-21}$\\
IP Tau & (5.22 $\pm$ 5.97) $\times~10^{-22}$ & (7.51 $\pm$ 7.81) $\times~10^{-22}$ & (0.0 $\pm$ 2.73) $\times~10^{-22}$ & (0.0 $\pm$ 1.19) $\times~10^{-21}$ & (0.0 $\pm$ 9.21) $\times~10^{-22}$ & (8.42 $\pm$ 5.35) $\times~10^{-22}$\\
J1604.3-2130 & (6.87 $\pm$ 3.66) $\times~10^{-23}$ & (0.0 $\pm$ 3.43) $\times~10^{-23}$ & (0.0 $\pm$ 9.51) $\times~10^{-24}$ & (4.1 $\pm$ 0.828) $\times~10^{-22}$ & (5.57 $\pm$ 5.46) $\times~10^{-23}$ & (4.4 $\pm$ 0.416) $\times~10^{-22}$\\
LkCa 15 & (2.07 $\pm$ 1.97) $\times~10^{-21}$ & (3.85 $\pm$ 2.42) $\times~10^{-21}$ & (0.0 $\pm$ 8.68) $\times~10^{-22}$ & (0.0 $\pm$ 3.11) $\times~10^{-21}$ & (0.0 $\pm$ 2.8) $\times~10^{-21}$ & (1.99 $\pm$ 1.86) $\times~10^{-21}$\\
PDS 70 & (5.58 $\pm$ 5.26) $\times~10^{-22}$ & (1.6 $\pm$ 0.581) $\times~10^{-21}$ & (0.0 $\pm$ 1.79) $\times~10^{-22}$ & (0.0 $\pm$ 6.77) $\times~10^{-22}$ & (0.0 $\pm$ 7.3) $\times~10^{-22}$ & (6.57 $\pm$ 4.17) $\times~10^{-22}$\\
RX J1842.9-3532 & (2.07 $\pm$ 0.816) $\times~10^{-21}$ & (2.67 $\pm$ 0.896) $\times~10^{-21}$ & (0.0 $\pm$ 2.27) $\times~10^{-22}$ & (0.0 $\pm$ 1.37) $\times~10^{-21}$ & (3.01 $\pm$ 11.3) $\times~10^{-22}$ & (1.04 $\pm$ 0.542) $\times~10^{-21}$\\
RX J1852.3-3700 & (3.94 $\pm$ 4.33) $\times~10^{-22}$ & (1.7 $\pm$ 4.54) $\times~10^{-22}$ & (0.0 $\pm$ 1.32) $\times~10^{-22}$ & (0.0 $\pm$ 9.92) $\times~10^{-22}$ & (0.0 $\pm$ 6.15) $\times~10^{-22}$ & (2.87 $\pm$ 3.34) $\times~10^{-22}$\\
RY Tau & (6.42 $\pm$ 6.07) $\times~10^{-20}$ & (7.21 $\pm$ 5.68) $\times~10^{-20}$ & (0.0 $\pm$ 1.46) $\times~10^{-20}$ & (0.0 $\pm$ 8.15) $\times~10^{-20}$ & (0.0 $\pm$ 6.83) $\times~10^{-20}$ & (6.01 $\pm$ 3.05) $\times~10^{-20}$\\
SR 21 & (0.0 $\pm$ 1.27) $\times~10^{-21}$ & (0.0 $\pm$ 5.8) $\times~10^{-22}$ & (0.0 $\pm$ 3.13) $\times~10^{-22}$ & (0.0 $\pm$ 1.91) $\times~10^{-21}$ & (0.0 $\pm$ 8.9) $\times~10^{-22}$ & (3.17 $\pm$ 0.879) $\times~10^{-21}$\\
SR 24 S & (0.0 $\pm$ 1.96) $\times~10^{-20}$ & (1.63 $\pm$ 2.24) $\times~10^{-20}$ & (3.32 $\pm$ 4.65) $\times~10^{-21}$ & (0.0 $\pm$ 3.57) $\times~10^{-20}$ & (0.0 $\pm$ 2.42) $\times~10^{-20}$ & (1.14 $\pm$ 0.833) $\times~10^{-20}$\\
UX Tau A & (1.22 $\pm$ 0.966) $\times~10^{-21}$ & (4.73 $\pm$ 11.2) $\times~10^{-22}$ & (0.0 $\pm$ 3.57) $\times~10^{-22}$ & (0.0 $\pm$ 2.03) $\times~10^{-21}$ & (0.0 $\pm$ 1.47) $\times~10^{-21}$ & (1.78 $\pm$ 1.07) $\times~10^{-21}$\\
V1247 Ori & (0.0 $\pm$ 2.71) $\times~10^{-21}$ & (0.0 $\pm$ 1.65) $\times~10^{-21}$ & (0.0 $\pm$ 1.67) $\times~10^{-21}$ & (0.0 $\pm$ 4.62) $\times~10^{-21}$ & (0.0 $\pm$ 2.09) $\times~10^{-21}$ & (1.34 $\pm$ 0.503) $\times~10^{-20}$\\
WSB 60 & (5.12 $\pm$ 5.88) $\times~10^{-22}$ & (0.0 $\pm$ 5.61) $\times~10^{-22}$ & (0.0 $\pm$ 1.94) $\times~10^{-22}$ & (0.0 $\pm$ 1.17) $\times~10^{-21}$ & (2.43 $\pm$ 7.59) $\times~10^{-22}$ & (8.96 $\pm$ 4.42) $\times~10^{-22}$\\
\enddata
\end{deluxetable*}

\begin{deluxetable*}{lcccccc}
\tabletypesize{\tiny}
\deluxetablecaption{(continued)}
\tablenum{A1}
\tablehead{
\colhead{Target} & \colhead{cSmAmPy ($g~cm^{-2}$)} & \colhead{cLgAmPy ($g~cm^{-2}$)} & \colhead{cSmAmOl ($g~cm^{-2}$)} & \colhead{cLgAmOl ($g~cm^{-2}$)} & \colhead{cSmAmPo ($g~cm^{-2}$)} & \colhead{cLgAmPo ($g~cm^{-2}$)}
}
\startdata
MP Mus & (6.67 $\pm$ 0.779) $\times~10^{-18}$ & (0.0 $\pm$ 7.94) $\times~10^{-19}$ & (1.91 $\pm$ 0.567) $\times~10^{-18}$ & (0.0 $\pm$ 5.88) $\times~10^{-19}$ & (0.0 $\pm$ 8.58) $\times~10^{-19}$ & (0.0 $\pm$ 8.38) $\times~10^{-19}$\\
CIDA 9 & (7.77 $\pm$ 1.53) $\times~10^{-19}$ & (0.0 $\pm$ 1.47) $\times~10^{-19}$ & (9.5 $\pm$ 1.06) $\times~10^{-19}$ & (0.0 $\pm$ 1.05) $\times~10^{-19}$ & (0.0 $\pm$ 1.63) $\times~10^{-19}$ & (0.0 $\pm$ 1.54) $\times~10^{-19}$\\
CQ Tau & (4.93 $\pm$ 2.41) $\times~10^{-17}$ & (0.0 $\pm$ 1.93) $\times~10^{-17}$ & (6.22 $\pm$ 0.167) $\times~10^{-16}$ & (0.0 $\pm$ 1.44) $\times~10^{-17}$ & (0.0 $\pm$ 2.39) $\times~10^{-17}$ & (0.0 $\pm$ 2.05) $\times~10^{-17}$\\
CS Cha & (1.56 $\pm$ 1.36) $\times~10^{-17}$ & (0.0 $\pm$ 1.31) $\times~10^{-17}$ & (1.99 $\pm$ 0.0942) $\times~10^{-16}$ & (2.01 $\pm$ 0.923) $\times~10^{-17}$ & (0.0 $\pm$ 1.45) $\times~10^{-17}$ & (0.0 $\pm$ 1.37) $\times~10^{-17}$\\
DM Tau & (1.84 $\pm$ 0.761) $\times~10^{-18}$ & (0.0 $\pm$ 7.6) $\times~10^{-19}$ & (6.91 $\pm$ 0.534) $\times~10^{-18}$ & (7.01 $\pm$ 0.546) $\times~10^{-18}$ & (0.0 $\pm$ 8.15) $\times~10^{-19}$ & (0.0 $\pm$ 7.88) $\times~10^{-19}$\\
DoAr 44 & (0.0 $\pm$ 2.28) $\times~10^{-18}$ & (0.0 $\pm$ 2.12) $\times~10^{-18}$ & (5.55 $\pm$ 0.158) $\times~10^{-17}$ & (0.0 $\pm$ 1.53) $\times~10^{-18}$ & (0.0 $\pm$ 2.34) $\times~10^{-18}$ & (0.0 $\pm$ 2.18) $\times~10^{-18}$\\
GG Tau AA Ab & (1.85 $\pm$ 0.826) $\times~10^{-18}$ & (0.0 $\pm$ 8.28) $\times~10^{-19}$ & (7.62 $\pm$ 0.589) $\times~10^{-18}$ & (0.0 $\pm$ 6.09) $\times~10^{-19}$ & (0.0 $\pm$ 9.04) $\times~10^{-19}$ & (0.0 $\pm$ 8.8) $\times~10^{-19}$\\
GM Aur & (1.69 $\pm$ 0.34) $\times~10^{-17}$ & (0.0 $\pm$ 3.25) $\times~10^{-18}$ & (6.45 $\pm$ 0.234) $\times~10^{-17}$ & (0.0 $\pm$ 2.32) $\times~10^{-18}$ & (0.0 $\pm$ 3.57) $\times~10^{-18}$ & (0.0 $\pm$ 3.38) $\times~10^{-18}$\\
IP Tau & (0.0 $\pm$ 8.2) $\times~10^{-20}$ & (0.0 $\pm$ 7.96) $\times~10^{-20}$ & (1.07 $\pm$ 0.063) $\times~10^{-18}$ & (0.0 $\pm$ 6.32) $\times~10^{-20}$ & (0.0 $\pm$ 9.21) $\times~10^{-20}$ & (0.0 $\pm$ 8.79) $\times~10^{-20}$\\
J1604.3-2130 & (0.0 $\pm$ 7.98) $\times~10^{-22}$ & (0.0 $\pm$ 1.15) $\times~10^{-21}$ & (0.0 $\pm$ 4.8) $\times~10^{-22}$ & (0.0 $\pm$ 5.83) $\times~10^{-22}$ & (0.0 $\pm$ 8.77) $\times~10^{-22}$ & (0.0 $\pm$ 1.01) $\times~10^{-21}$\\
LkCa 15 & (0.0 $\pm$ 5.68) $\times~10^{-19}$ & (0.0 $\pm$ 5.29) $\times~10^{-19}$ & (1.14 $\pm$ 0.0392) $\times~10^{-17}$ & (0.0 $\pm$ 3.76) $\times~10^{-19}$ & (0.0 $\pm$ 5.97) $\times~10^{-19}$ & (0.0 $\pm$ 5.55) $\times~10^{-19}$\\
PDS 70 & (0.0 $\pm$ 3.16) $\times~10^{-19}$ & (0.0 $\pm$ 2.63) $\times~10^{-19}$ & (7.87 $\pm$ 0.218) $\times~10^{-18}$ & (0.0 $\pm$ 1.93) $\times~10^{-19}$ & (0.0 $\pm$ 3.18) $\times~10^{-19}$ & (0.0 $\pm$ 2.78) $\times~10^{-19}$\\
RX J1842.9-3532 & (9.89 $\pm$ 6.18) $\times~10^{-19}$ & (0.0 $\pm$ 5.77) $\times~10^{-19}$ & (1.36 $\pm$ 0.0421) $\times~10^{-17}$ & (0.0 $\pm$ 4.07) $\times~10^{-19}$ & (0.0 $\pm$ 6.35) $\times~10^{-19}$ & (0.0 $\pm$ 5.92) $\times~10^{-19}$\\
RX J1852.3-3700 & (4.4 $\pm$ 0.568) $\times~10^{-17}$ & (0.0 $\pm$ 5.46) $\times~10^{-18}$ & (9.71 $\pm$ 0.388) $\times~10^{-17}$ & (0.0 $\pm$ 3.85) $\times~10^{-18}$ & (0.0 $\pm$ 5.97) $\times~10^{-18}$ & (0.0 $\pm$ 5.66) $\times~10^{-18}$\\
RY Tau & (0.0 $\pm$ 7.26) $\times~10^{-18}$ & (0.0 $\pm$ 6.58) $\times~10^{-18}$ & (1.67 $\pm$ 0.0536) $\times~10^{-16}$ & (0.0 $\pm$ 5.03) $\times~10^{-18}$ & (0.0 $\pm$ 7.73) $\times~10^{-18}$ & (0.0 $\pm$ 7.07) $\times~10^{-18}$\\
SR 21 & (0.0 $\pm$ 1.19) $\times~10^{-20}$ & (0.0 $\pm$ 1.94) $\times~10^{-20}$ & (0.0 $\pm$ 1.13) $\times~10^{-20}$ & (0.0 $\pm$ 1.52) $\times~10^{-20}$ & (0.0 $\pm$ 1.64) $\times~10^{-20}$ & (0.0 $\pm$ 2.08) $\times~10^{-20}$\\
SR 24 S & (0.0 $\pm$ 4.82) $\times~10^{-18}$ & (0.0 $\pm$ 4.56) $\times~10^{-18}$ & (1.13 $\pm$ 0.0351) $\times~10^{-16}$ & (0.0 $\pm$ 3.4) $\times~10^{-18}$ & (0.0 $\pm$ 5.16) $\times~10^{-18}$ & (0.0 $\pm$ 4.81) $\times~10^{-18}$\\
UX Tau A & (4.41 $\pm$ 0.939) $\times~10^{-17}$ & (0.0 $\pm$ 8.65) $\times~10^{-18}$ & (2.72 $\pm$ 0.0642) $\times~10^{-16}$ & (0.0 $\pm$ 6.06) $\times~10^{-18}$ & (0.0 $\pm$ 9.75) $\times~10^{-18}$ & (0.0 $\pm$ 8.99) $\times~10^{-18}$\\
V1247 Ori & (5.13 $\pm$ 1.07) $\times~10^{-16}$ & (0.0 $\pm$ 8.49) $\times~10^{-17}$ & (2.49 $\pm$ 0.0735) $\times~10^{-15}$ & (0.0 $\pm$ 6.16) $\times~10^{-17}$ & (0.0 $\pm$ 9.67) $\times~10^{-17}$ & (0.0 $\pm$ 8.16) $\times~10^{-17}$\\
WSB 60 & (0.0 $\pm$ 1.35) $\times~10^{-19}$ & (0.0 $\pm$ 1.4) $\times~10^{-19}$ & (1.42 $\pm$ 0.985) $\times~10^{-19}$ & (1.43 $\pm$ 0.105) $\times~10^{-18}$ & (0.0 $\pm$ 1.51) $\times~10^{-19}$ & (0.0 $\pm$ 1.5) $\times~10^{-19}$\\
\enddata
\end{deluxetable*}

\begin{deluxetable*}{lcccccc}
\tabletypesize{\tiny}
\deluxetablecaption{(continued)}
\tablenum{A1}
\tablehead{
\colhead{Target} & \colhead{cSmCryForst ($g~cm^{-2}$)} & \colhead{cSmCryEnst ($g~cm^{-2}$)} & \colhead{cSmCrySil ($g~cm^{-2}$)} & \colhead{cLgCryForst ($g~cm^{-2}$)} & \colhead{cLgCryEnst ($g~cm^{-2}$)} & \colhead{cLgCrySil ($g~cm^{-2}$)}
}
\startdata
MP Mus & (5.21 $\pm$ 3.77) $\times~10^{-19}$ & (0.0 $\pm$ 4.96) $\times~10^{-19}$ & (0.0 $\pm$ 3.81) $\times~10^{-19}$ & (1.26 $\pm$ 0.476) $\times~10^{-18}$ & (0.0 $\pm$ 5.37) $\times~10^{-19}$ & (0.0 $\pm$ 4.68) $\times~10^{-19}$\\
CIDA 9 & (0.0 $\pm$ 6.88) $\times~10^{-20}$ & (0.0 $\pm$ 9.33) $\times~10^{-20}$ & (3.64 $\pm$ 72.0) $\times~10^{-21}$ & (2.87 $\pm$ 0.847) $\times~10^{-19}$ & (1.14 $\pm$ 0.984) $\times~10^{-19}$ & (0.0 $\pm$ 7.91) $\times~10^{-20}$\\
CQ Tau & (5.26 $\pm$ 11.7) $\times~10^{-18}$ & (0.0 $\pm$ 1.5) $\times~10^{-17}$ & (0.0 $\pm$ 6.67) $\times~10^{-18}$ & (0.0 $\pm$ 1.26) $\times~10^{-17}$ & (0.0 $\pm$ 1.43) $\times~10^{-17}$ & (0.0 $\pm$ 7.89) $\times~10^{-18}$\\
CS Cha & (0.0 $\pm$ 5.68) $\times~10^{-18}$ & (0.0 $\pm$ 8.13) $\times~10^{-18}$ & (9.1 $\pm$ 7.43) $\times~10^{-18}$ & (8.57 $\pm$ 7.05) $\times~10^{-18}$ & (0.0 $\pm$ 8.61) $\times~10^{-18}$ & (0.0 $\pm$ 8.24) $\times~10^{-18}$\\
DM Tau & (0.0 $\pm$ 3.74) $\times~10^{-19}$ & (0.0 $\pm$ 4.66) $\times~10^{-19}$ & (0.0 $\pm$ 4.11) $\times~10^{-19}$ & (6.25 $\pm$ 4.71) $\times~10^{-19}$ & (0.0 $\pm$ 5.07) $\times~10^{-19}$ & (4.2 $\pm$ 4.95) $\times~10^{-19}$\\
DoAr 44 & (1.01 $\pm$ 1.07) $\times~10^{-18}$ & (0.0 $\pm$ 1.3) $\times~10^{-18}$ & (4.91 $\pm$ 11.4) $\times~10^{-19}$ & (1.19 $\pm$ 1.28) $\times~10^{-18}$ & (0.0 $\pm$ 1.41) $\times~10^{-18}$ & (0.0 $\pm$ 1.23) $\times~10^{-18}$\\
GG Tau AA Ab & (0.0 $\pm$ 4.13) $\times~10^{-19}$ & (0.0 $\pm$ 5.12) $\times~10^{-19}$ & (0.0 $\pm$ 3.75) $\times~10^{-19}$ & (6.69 $\pm$ 5.08) $\times~10^{-19}$ & (2.84 $\pm$ 5.48) $\times~10^{-19}$ & (0.0 $\pm$ 4.45) $\times~10^{-19}$\\
GM Aur & (0.0 $\pm$ 1.52) $\times~10^{-18}$ & (0.0 $\pm$ 1.96) $\times~10^{-18}$ & (1.42 $\pm$ 1.65) $\times~10^{-18}$ & (1.1 $\pm$ 1.85) $\times~10^{-18}$ & (0.0 $\pm$ 2.1) $\times~10^{-18}$ & (0.0 $\pm$ 1.86) $\times~10^{-18}$\\
IP Tau & (2.44 $\pm$ 3.84) $\times~10^{-20}$ & (0.0 $\pm$ 5.03) $\times~10^{-20}$ & (0.0 $\pm$ 3.5) $\times~10^{-20}$ & (2.63 $\pm$ 4.88) $\times~10^{-20}$ & (0.0 $\pm$ 5.4) $\times~10^{-20}$ & (0.0 $\pm$ 3.98) $\times~10^{-20}$\\
J1604.3-2130 & (0.0 $\pm$ 6.82) $\times~10^{-22}$ & (0.0 $\pm$ 6.84) $\times~10^{-22}$ & (0.0 $\pm$ 2.29) $\times~10^{-22}$ & (0.0 $\pm$ 1.34) $\times~10^{-21}$ & (0.0 $\pm$ 9.99) $\times~10^{-22}$ & (0.0 $\pm$ 9.51) $\times~10^{-22}$\\
LkCa 15 & (0.0 $\pm$ 2.45) $\times~10^{-19}$ & (0.0 $\pm$ 3.07) $\times~10^{-19}$ & (2.66 $\pm$ 26.1) $\times~10^{-20}$ & (2.09 $\pm$ 2.99) $\times~10^{-19}$ & (0.0 $\pm$ 3.23) $\times~10^{-19}$ & (2.08 $\pm$ 2.96) $\times~10^{-19}$\\
PDS 70 & (3.79 $\pm$ 1.52) $\times~10^{-19}$ & (0.0 $\pm$ 1.82) $\times~10^{-19}$ & (0.0 $\pm$ 1.37) $\times~10^{-19}$ & (0.0 $\pm$ 1.67) $\times~10^{-19}$ & (0.0 $\pm$ 1.9) $\times~10^{-19}$ & (0.0 $\pm$ 1.44) $\times~10^{-19}$\\
RX J1842.9-3532 & (3.2 $\pm$ 2.84) $\times~10^{-19}$ & (0.0 $\pm$ 3.49) $\times~10^{-19}$ & (9.37 $\pm$ 30.6) $\times~10^{-20}$ & (5.28 $\pm$ 3.38) $\times~10^{-19}$ & (0.0 $\pm$ 3.75) $\times~10^{-19}$ & (0.0 $\pm$ 3.4) $\times~10^{-19}$\\
RX J1852.3-3700 & (0.0 $\pm$ 2.67) $\times~10^{-18}$ & (0.0 $\pm$ 3.4) $\times~10^{-18}$ & (0.0 $\pm$ 3.11) $\times~10^{-18}$ & (5.11 $\pm$ 3.23) $\times~10^{-18}$ & (0.0 $\pm$ 3.58) $\times~10^{-18}$ & (4.36 $\pm$ 3.58) $\times~10^{-18}$\\
RY Tau & (2.57 $\pm$ 3.81) $\times~10^{-18}$ & (0.0 $\pm$ 4.26) $\times~10^{-18}$ & (0.0 $\pm$ 2.6) $\times~10^{-18}$ & (0.0 $\pm$ 4.38) $\times~10^{-18}$ & (0.0 $\pm$ 4.43) $\times~10^{-18}$ & (0.0 $\pm$ 3.01) $\times~10^{-18}$\\
SR 21 & (2.54 $\pm$ 1.56) $\times~10^{-20}$ & (0.0 $\pm$ 7.41) $\times~10^{-21}$ & (0.0 $\pm$ 3.98) $\times~10^{-21}$ & (0.0 $\pm$ 2.22) $\times~10^{-20}$ & (0.0 $\pm$ 1.12) $\times~10^{-20}$ & (0.0 $\pm$ 1.05) $\times~10^{-20}$\\
SR 24 S & (0.0 $\pm$ 2.29) $\times~10^{-18}$ & (0.0 $\pm$ 2.88) $\times~10^{-18}$ & (1.23 $\pm$ 2.75) $\times~10^{-18}$ & (3.38 $\pm$ 28.5) $\times~10^{-19}$ & (0.0 $\pm$ 3.05) $\times~10^{-18}$ & (0.0 $\pm$ 3.08) $\times~10^{-18}$\\
UX Tau A & (3.75 $\pm$ 4.23) $\times~10^{-18}$ & (0.0 $\pm$ 5.71) $\times~10^{-18}$ & (3.59 $\pm$ 4.52) $\times~10^{-18}$ & (1.45 $\pm$ 0.501) $\times~10^{-17}$ & (0.0 $\pm$ 6.05) $\times~10^{-18}$ & (0.0 $\pm$ 4.98) $\times~10^{-18}$\\
V1247 Ori & (3.63 $\pm$ 46.8) $\times~10^{-18}$ & (0.0 $\pm$ 5.17) $\times~10^{-17}$ & (0.0 $\pm$ 2.68) $\times~10^{-17}$ & (0.0 $\pm$ 5.4) $\times~10^{-17}$ & (0.0 $\pm$ 4.89) $\times~10^{-17}$ & (0.0 $\pm$ 2.62) $\times~10^{-17}$\\
WSB 60 & (0.0 $\pm$ 6.35) $\times~10^{-20}$ & (0.0 $\pm$ 8.26) $\times~10^{-20}$ & (0.0 $\pm$ 6.83) $\times~10^{-20}$ & (1.23 $\pm$ 0.814) $\times~10^{-19}$ & (0.0 $\pm$ 9.14) $\times~10^{-20}$ & (0.0 $\pm$ 7.98) $\times~10^{-20}$\\
\enddata
\end{deluxetable*}

\begin{deluxetable*}{lcccc}
\tablenum{A2}
\deluxetablecaption{Crystallinity Percentages, previous results$^1$ vs. TZTD Results\label{crystallinity}}
\tablehead{
\colhead{Target} & \colhead{Crystallinity$^1$} & \colhead{Lower Limit$^1$} & \colhead{Crystallinity (TZTD, Warm Disk)} & \colhead{Crystallinity (TZTD, Cool Disk)}
}
\startdata
MP Mus & 37.5 & 5.2 & 21.6 & 17.2 \\
CQ Tau & 77.4 & 13.1 & 8.5 & 0.8 \\
DoAr 44 & 33.2 & 7.8 & 9.1 & 4.6 \\
GG Tau AA Ab & 26.2 & 4.6 & 7.6 & 9.1 \\
GM Aur & 47.4 & 7.1 & 3.5 & 3.0 \\
IP Tau & 33.0 & 5.4 & 8.5 & 4.5 \\
LkCa 15 & 35.3 & 8.0 & 6.1 & 3.8 \\ 
PDS 70 & 26.2 & 7.6 & 14.2 & 4.6 \\ 
RX J1842.9-3532 & 35.0 & 8.5 & 10.9 & 6.1 \\
RY Tau & 38.6 & 8.1 & 7.5 & 1.5 \\ 
SR 21 & 45.6 & 5.7 & 100.0 & 100.0 \\
SR 24 S & 41.4 & 5.9 & 34.2 & 1.4 \\ 
WSB 60 & 48.4 & 4.9 & 15.4 & 7.3 \\ 
\enddata
\tablerefs{(1) \citet{2016AJ....151..146C}}
\end{deluxetable*}

\figsetstart
\figsetnum{A1}
\figsettitle{Dust Composition Plots}

\figsetgrpstart
\figsetgrpnum{A1.1}
\figsetgrptitle{MP Mus}
\figsetplot{\gridline{(a) \fig{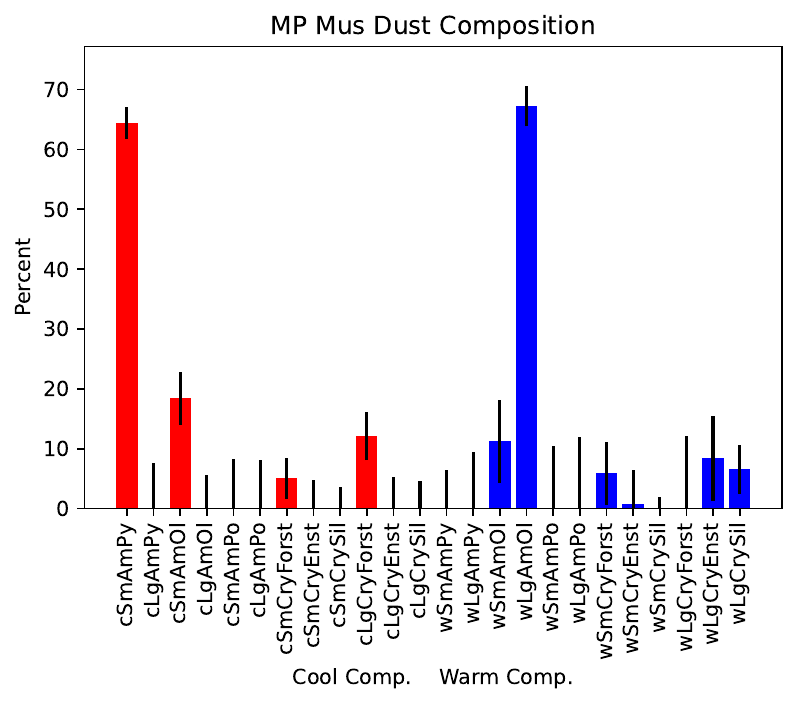}{0.5\textwidth}{}
              (b) \fig{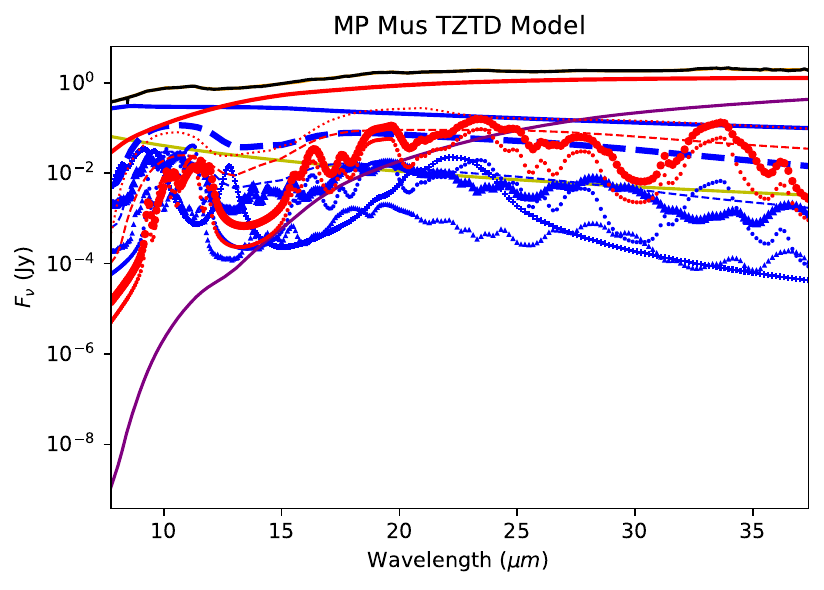}{0.5\textwidth}{}
              }
    \gridline{(c) \fig{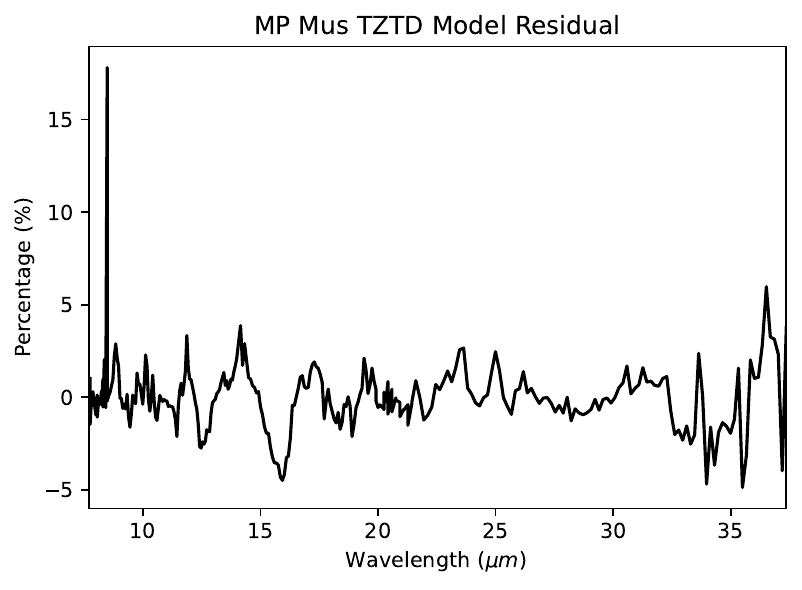}{0.5\textwidth}{}
              (d) \fig{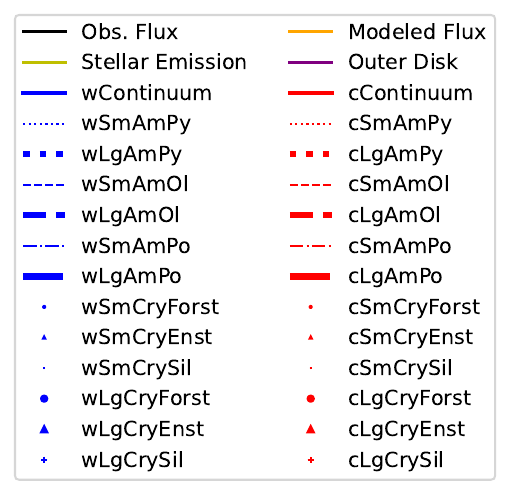}{0.45\textwidth}{}
              }}
\figsetgrpnote{Results of TZTD empirical mineralogical analysis of MP Mus $Spitzer$ IRS spectrum; (a) Dust composition of optically thin portions of protoplanetary disk, using values from Table \ref{minres}, (b) Logarithmic scaling of corresponding model fit plot in Figure \ref{res_plot} to demonstrate the contributions of each mineral; (c) Residual between best fit model and $Spitzer$ IRS spectrum; (d) Legend for the model plots, using shortened names from Table \ref{opacity}; Red: Cool disk component constituents, Blue: Warm disk component constituents.}
\figsetgrpend

\figsetgrpstart
\figsetgrpnum{A1.2}
\figsetgrptitle{CIDA 9}
\figsetplot{\gridline{(a) \fig{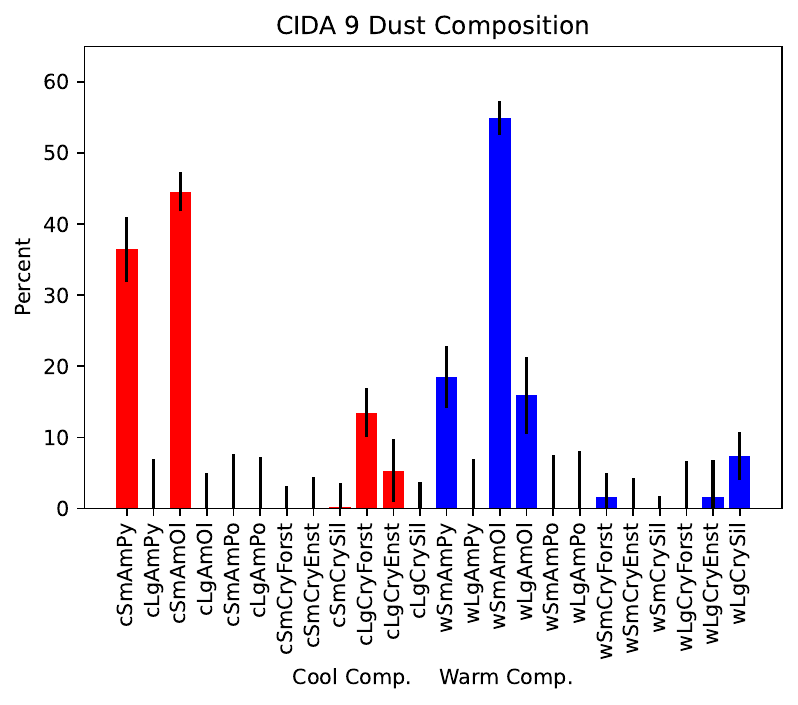}{0.5\textwidth}{}
              (b) \fig{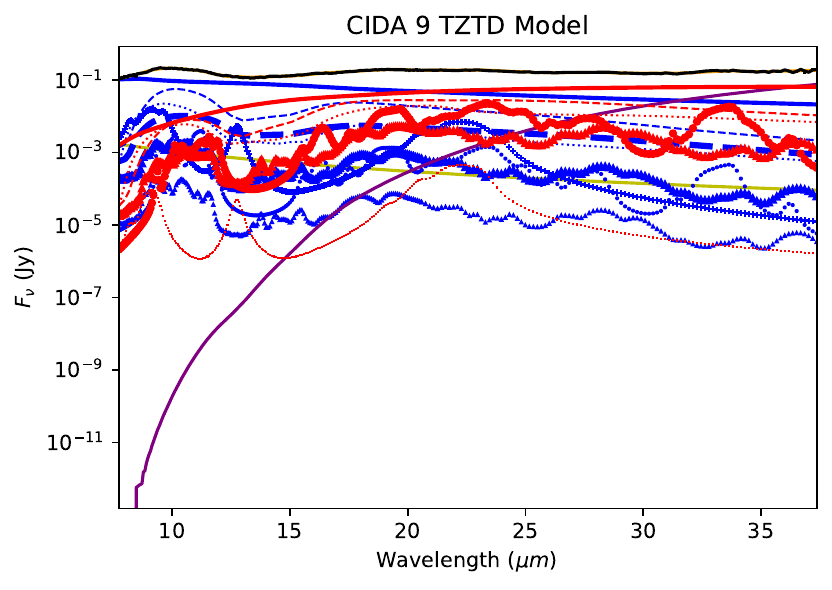}{0.5\textwidth}{}
              }
    \gridline{(c) \fig{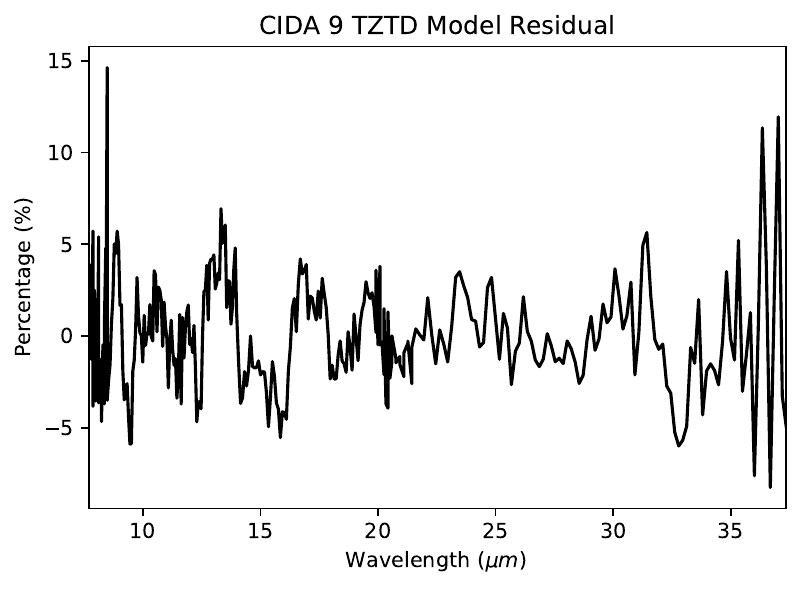}{0.5\textwidth}{}
              (d) \fig{tztd_legend.pdf}{0.45\textwidth}{}
              }}
\figsetgrpnote{As in Figure \ref{A1.1}, for CIDA 9}
\figsetgrpend

\figsetgrpstart
\figsetgrpnum{A1.3}
\figsetgrptitle{CQ Tau}
\figsetplot{\gridline{(a) \fig{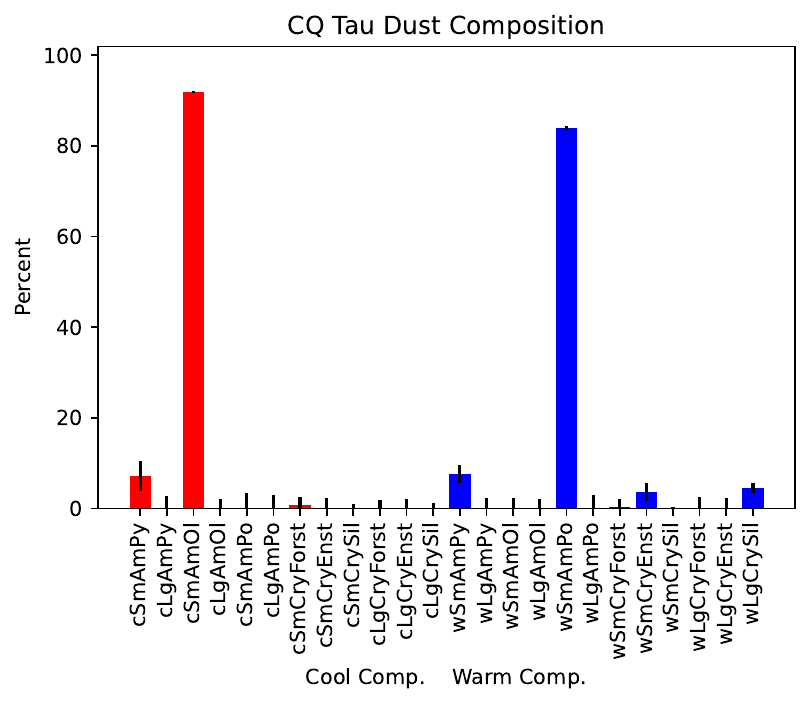}{0.5\textwidth}{}
              (b) \fig{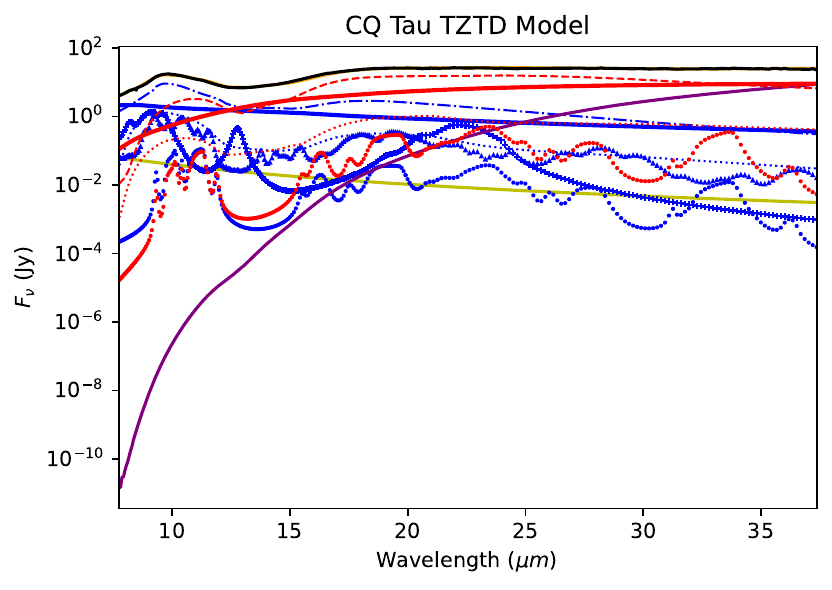}{0.5\textwidth}{}
              }
    \gridline{(c) \fig{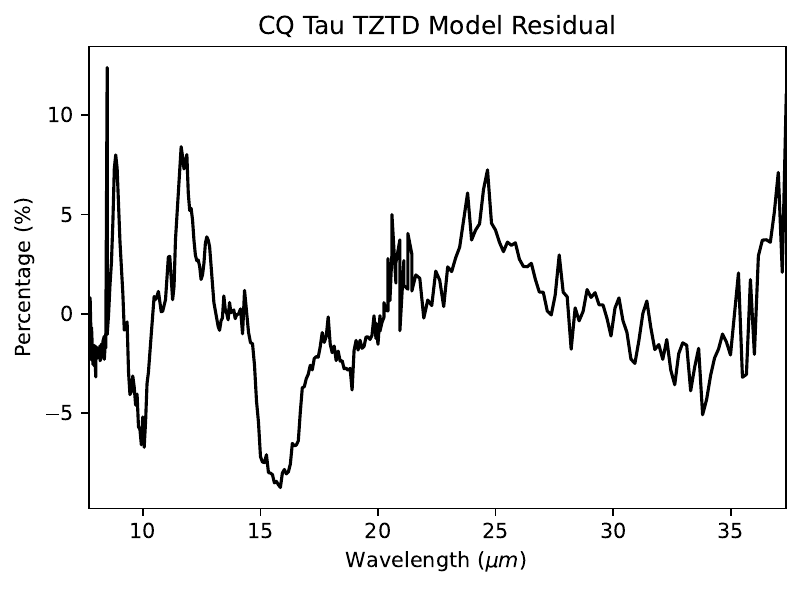}{0.5\textwidth}{}
              (d) \fig{tztd_legend.pdf}{0.45\textwidth}{}
              }}
\figsetgrpnote{As in Figure \ref{A1.1}, for CQ Tau}
\figsetgrpend

\figsetgrpstart
\figsetgrpnum{A1.4}
\figsetgrptitle{CS Cha}
\figsetplot{\gridline{(a) \fig{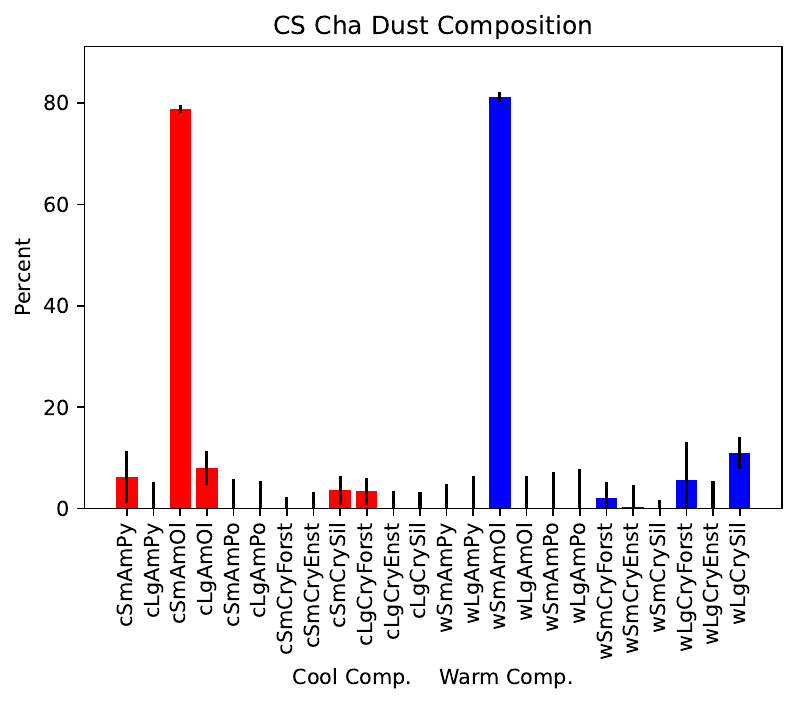}{0.5\textwidth}{}
              (b) \fig{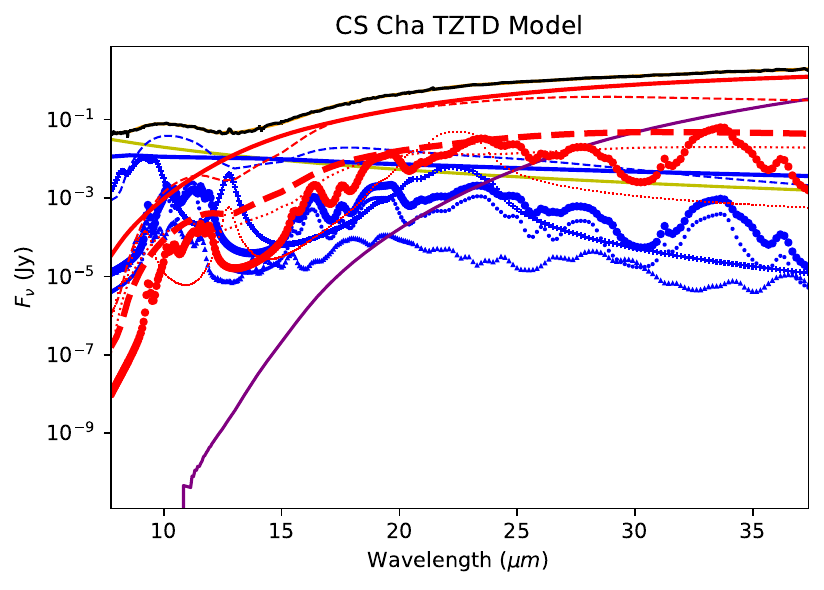}{0.5\textwidth}{}
              }
    \gridline{(c) \fig{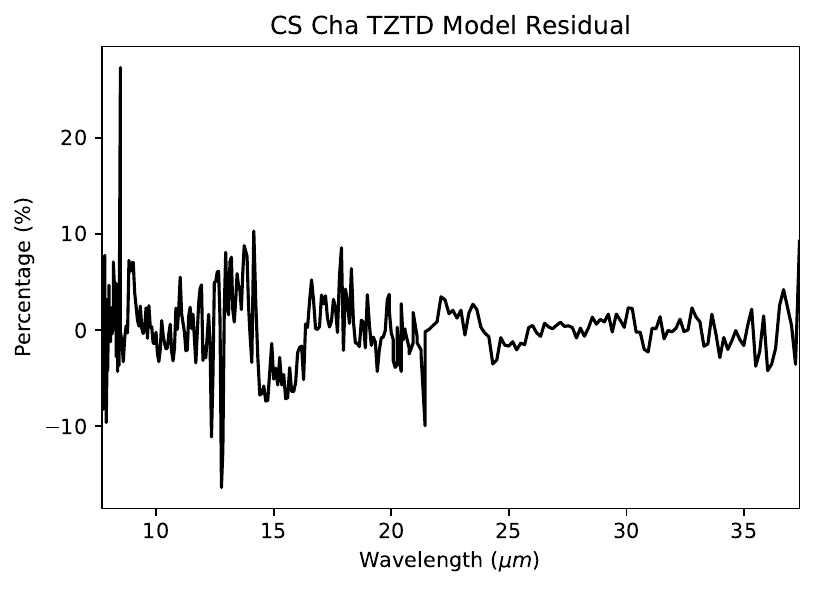}{0.5\textwidth}{}
              (d) \fig{tztd_legend.pdf}{0.45\textwidth}{}
              }}
\figsetgrpnote{As in Figure \ref{A1.1}, for CS Cha}
\figsetgrpend

\figsetgrpstart
\figsetgrpnum{A1.5}
\figsetgrptitle{DM Tau}
\figsetplot{\gridline{(a) \fig{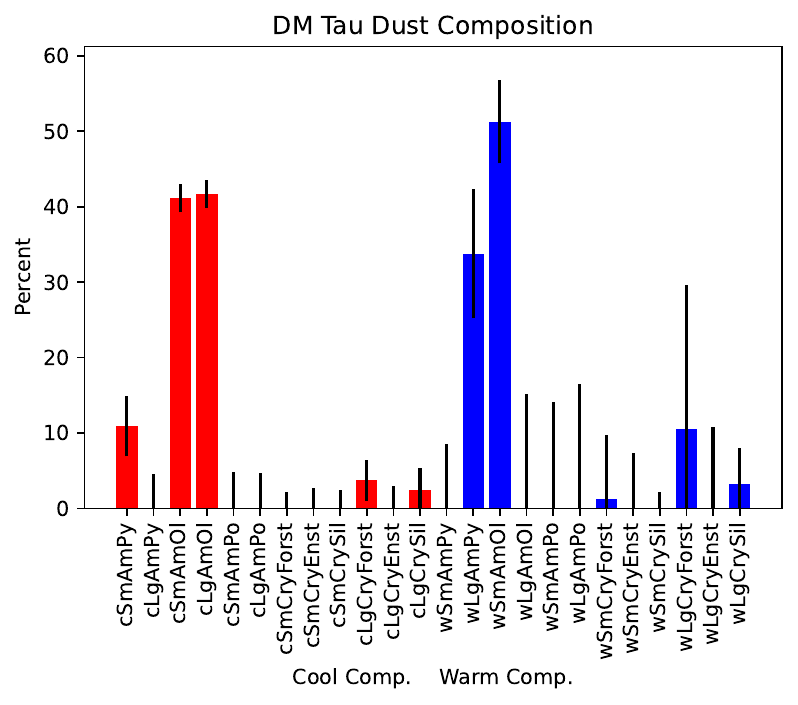}{0.5\textwidth}{}
              (b) \fig{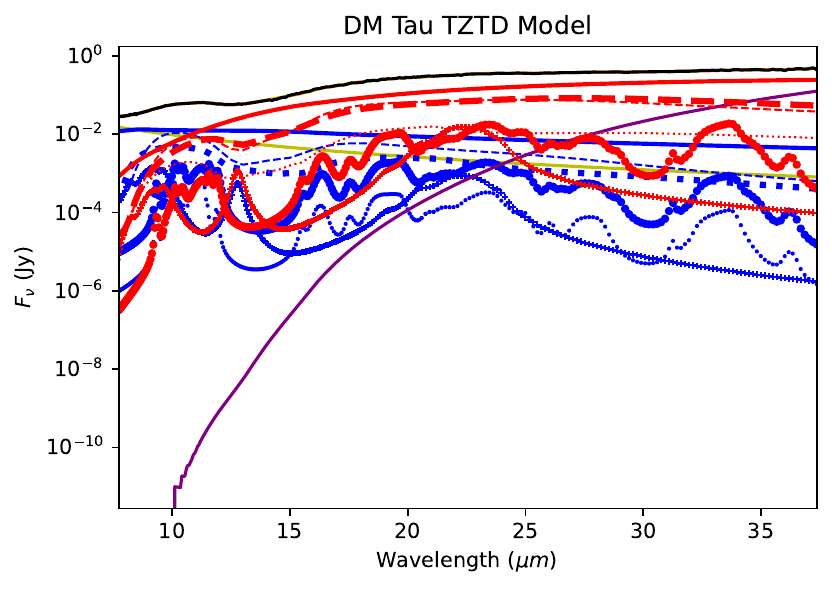}{0.5\textwidth}{}
              }
    \gridline{(c) \fig{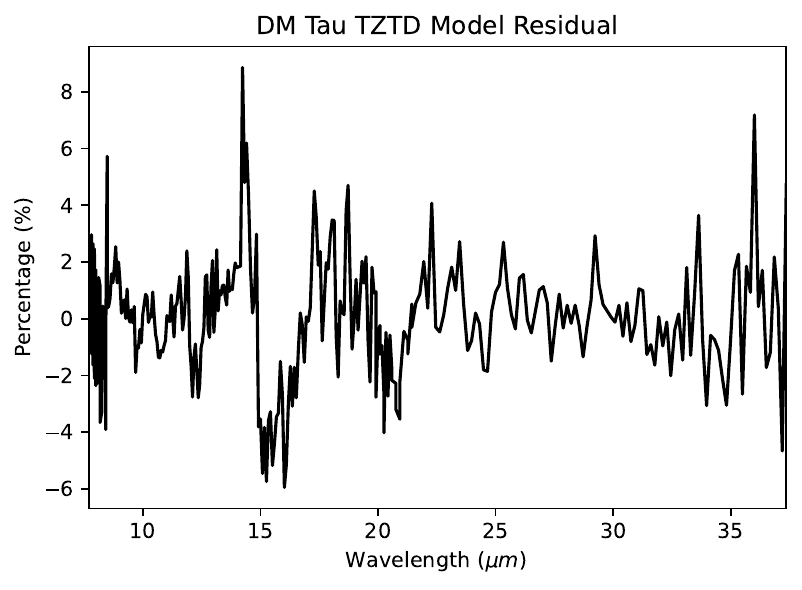}{0.5\textwidth}{}
              (d) \fig{tztd_legend.pdf}{0.45\textwidth}{}
              }}
\figsetgrpnote{As in Figure \ref{A1.1}, for DM Tau}
\figsetgrpend

\figsetgrpstart
\figsetgrpnum{A1.6}
\figsetgrptitle{DoAr 44}
\figsetplot{\gridline{(a) \fig{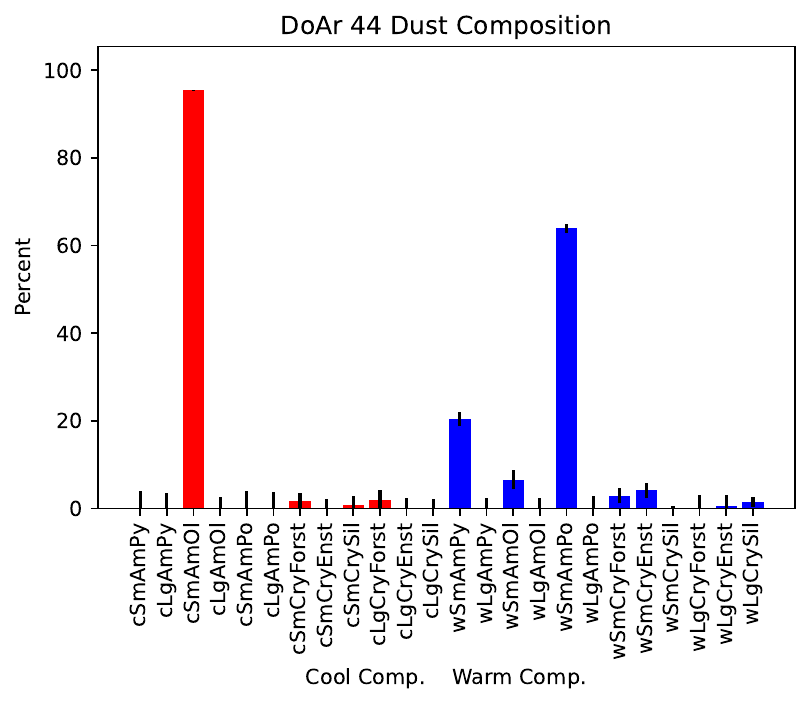}{0.5\textwidth}{}
              (b) \fig{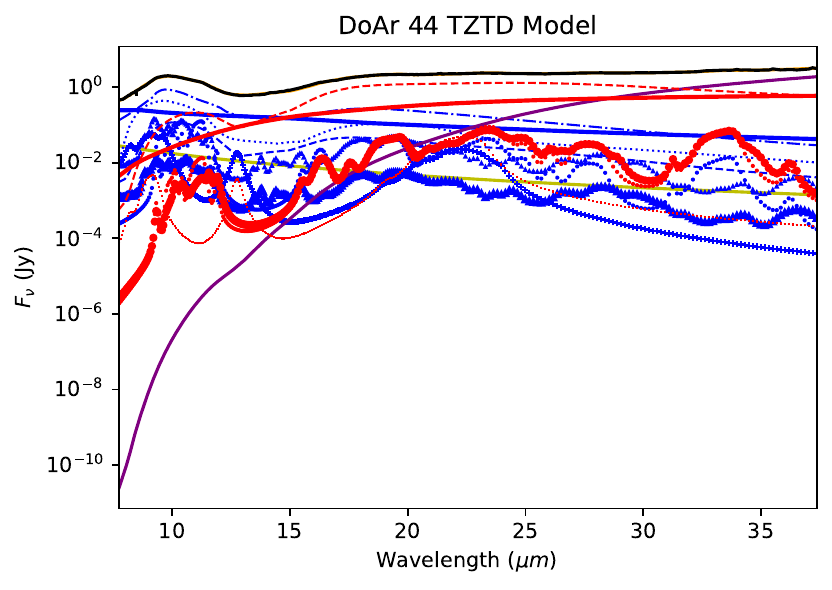}{0.5\textwidth}{}
              }
    \gridline{(c) \fig{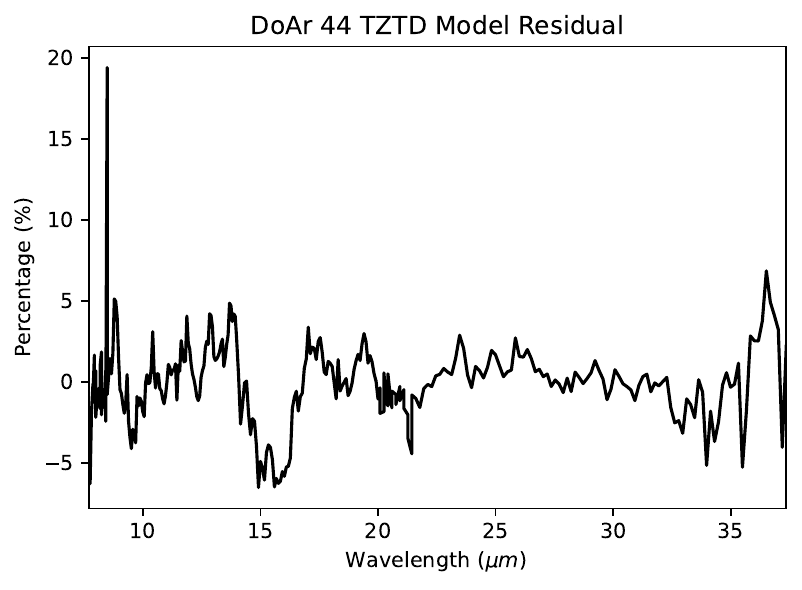}{0.5\textwidth}{}
              (d) \fig{tztd_legend.pdf}{0.45\textwidth}{}
              }}
\figsetgrpnote{As in Figure \ref{A1.1}, for DoAr 44}
\figsetgrpend

\figsetgrpstart
\figsetgrpnum{A1.7}
\figsetgrptitle{GG Tau AA Ab}
\figsetplot{\gridline{(a) \fig{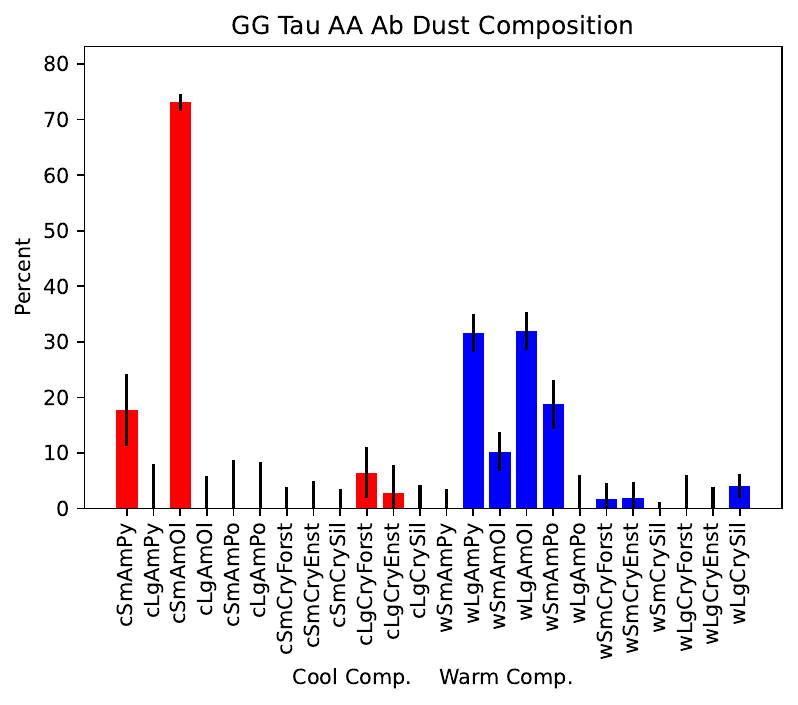}{0.5\textwidth}{}
              (b) \fig{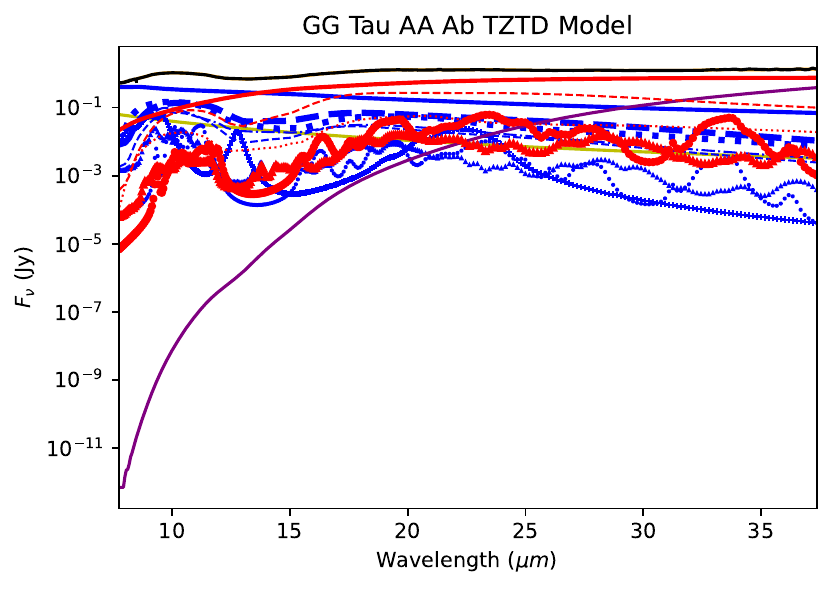}{0.5\textwidth}{}
              }
    \gridline{(c) \fig{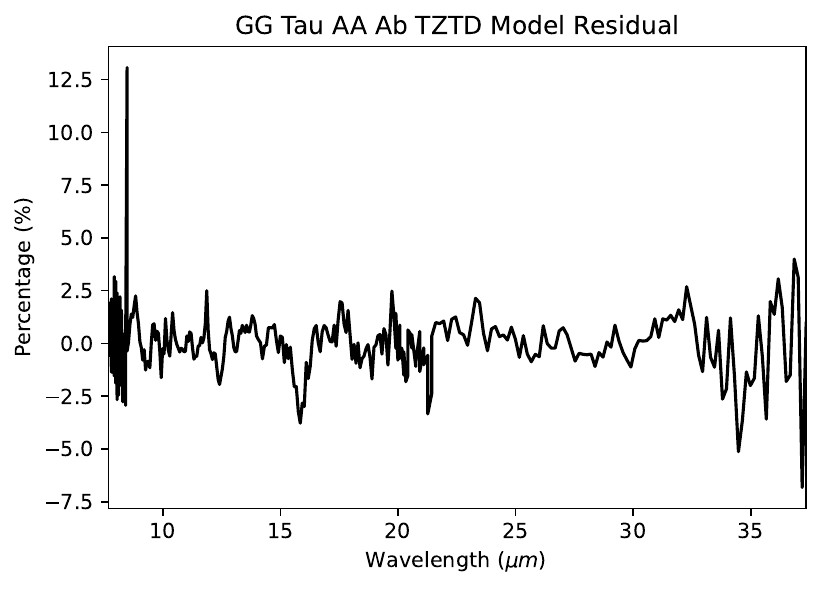}{0.5\textwidth}{}
              (d) \fig{tztd_legend.pdf}{0.45\textwidth}{}
              }}
\figsetgrpnote{As in Figure \ref{A1.1}, for GG Tau AA Ab}
\figsetgrpend

\figsetgrpstart
\figsetgrpnum{A1.8}
\figsetgrptitle{GM Aur}
\figsetplot{\gridline{(a) \fig{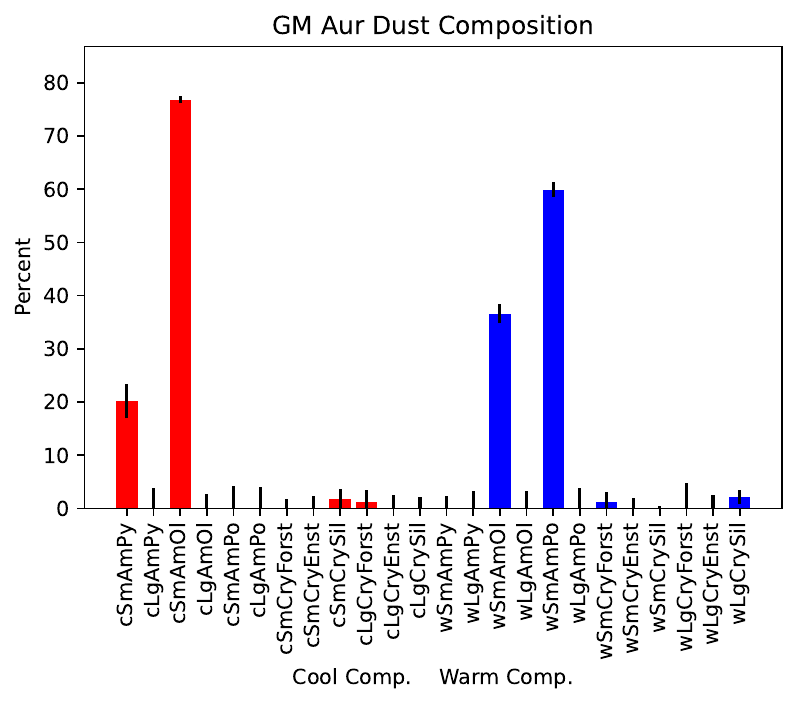}{0.5\textwidth}{}
              (b) \fig{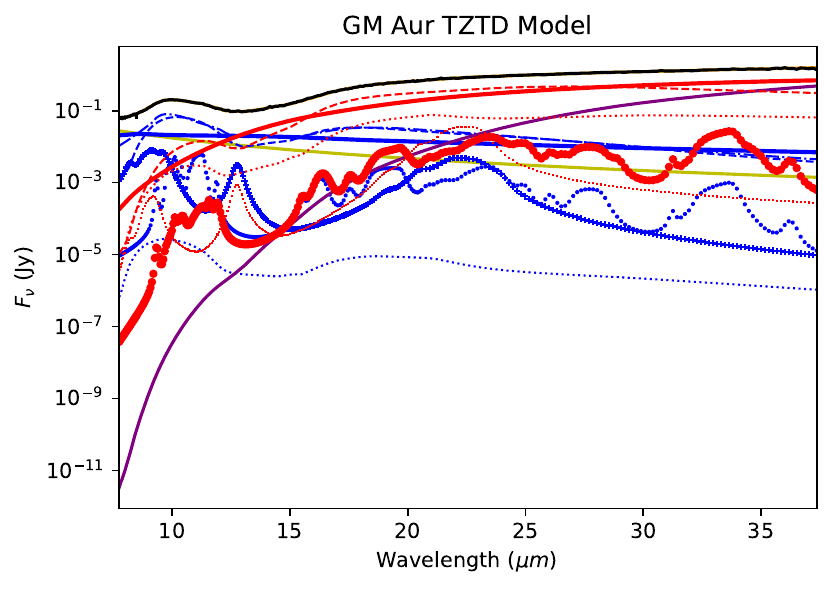}{0.5\textwidth}{}
              }
    \gridline{(c) \fig{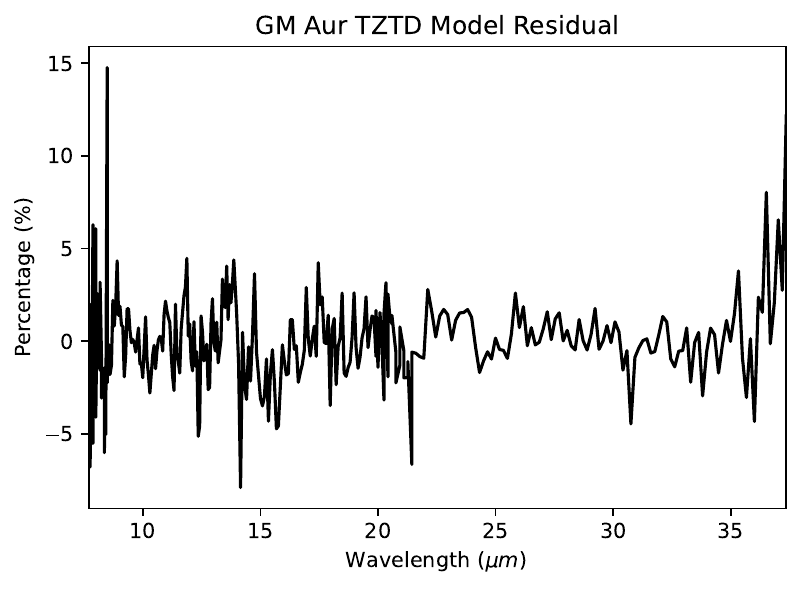}{0.5\textwidth}{}
              (d) \fig{tztd_legend.pdf}{0.45\textwidth}{}
              }}
\figsetgrpnote{As in Figure \ref{A1.1}, for GM Aur}
\figsetgrpend

\figsetgrpstart
\figsetgrpnum{A1.9}
\figsetgrptitle{IP Tau}
\figsetplot{\gridline{(a) \fig{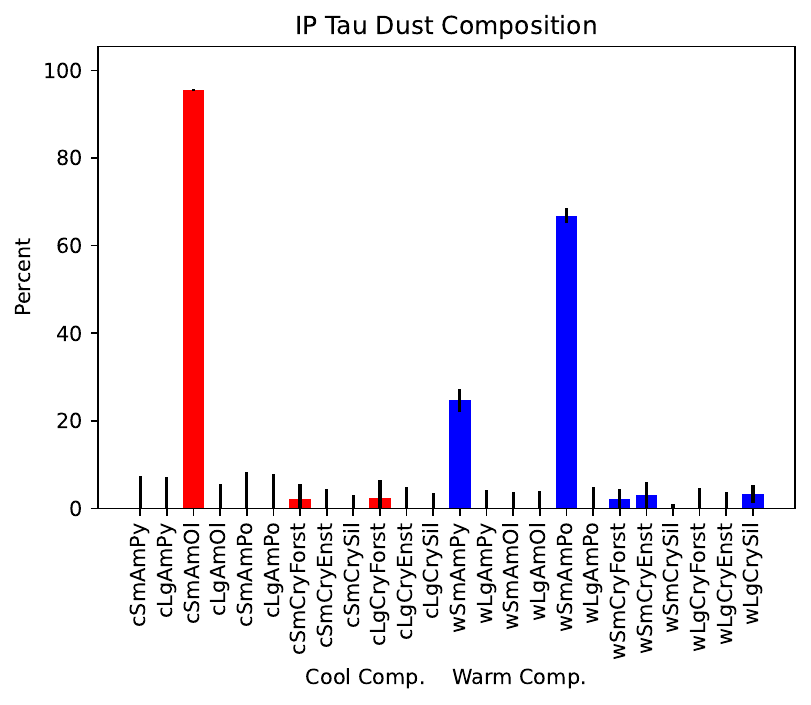}{0.5\textwidth}{}
              (b) \fig{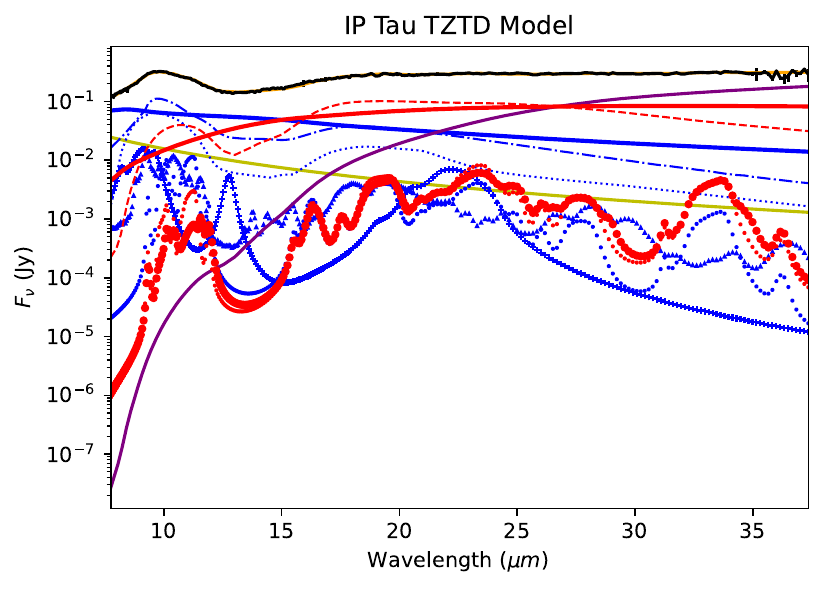}{0.5\textwidth}{}
              }
    \gridline{(c) \fig{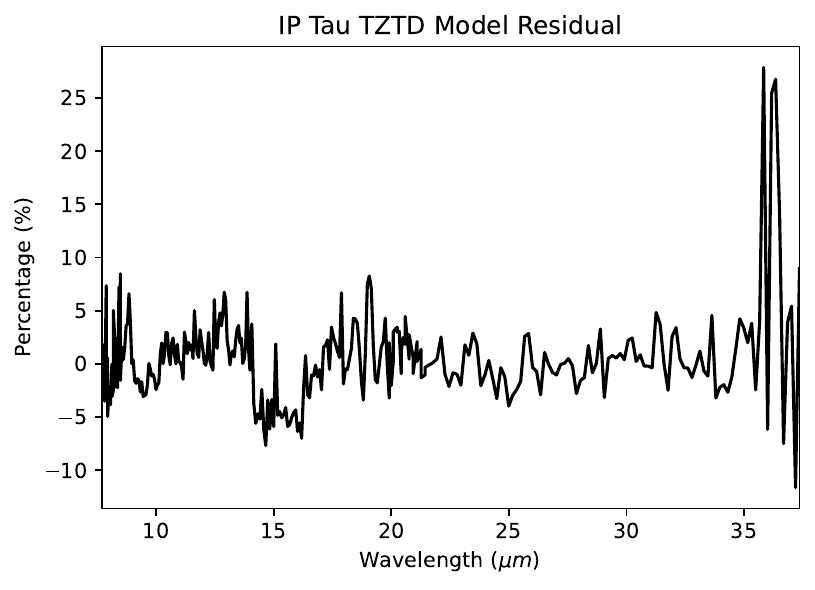}{0.5\textwidth}{}
              (d) \fig{tztd_legend.pdf}{0.45\textwidth}{}
              }}
\figsetgrpnote{As in Figure \ref{A1.1}, for IP Tau}
\figsetgrpend

\figsetgrpstart
\figsetgrpnum{A1.10}
\figsetgrptitle{J1604.3-2130}
\figsetplot{\gridline{(a) \fig{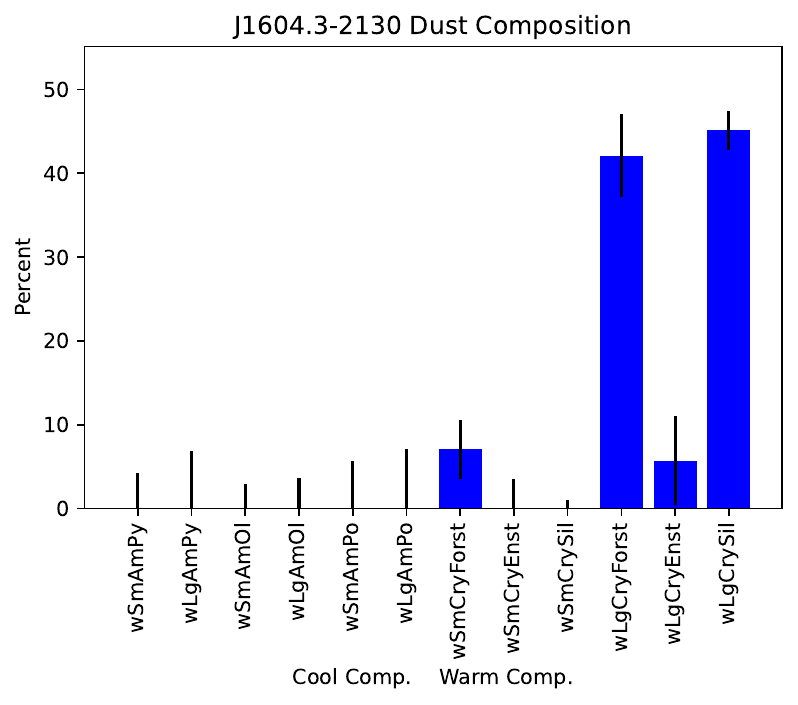}{0.5\textwidth}{}
              (b) \fig{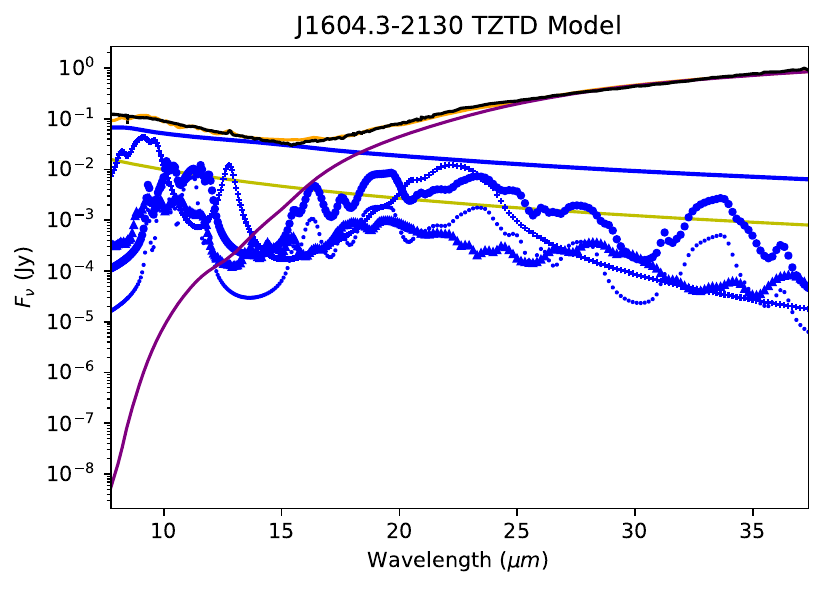}{0.5\textwidth}{}
              }
    \gridline{(c) \fig{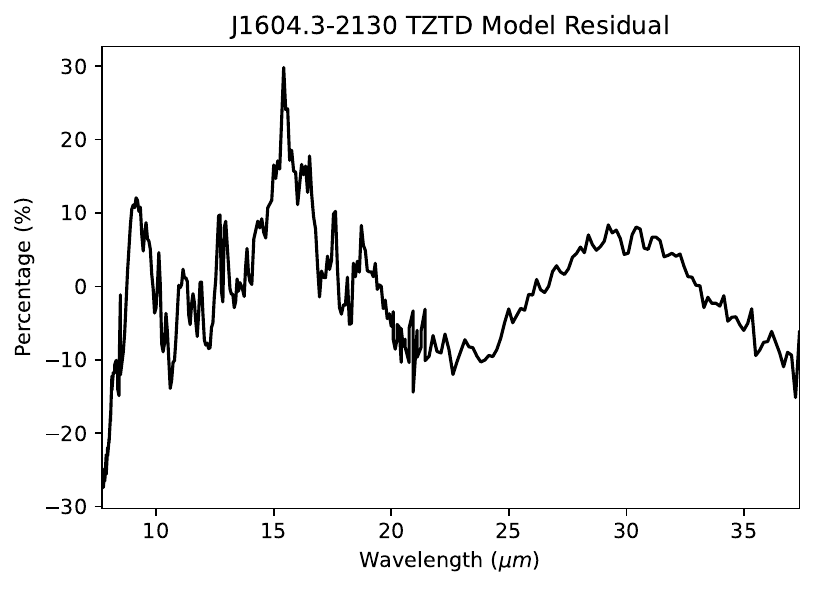}{0.5\textwidth}{}
              (d) \fig{tztd_legend.pdf}{0.45\textwidth}{}
              }}
\figsetgrpnote{As in Figure \ref{A1.1}, for J1604.3-2130}
\figsetgrpend

\figsetgrpstart
\figsetgrpnum{A1.11}
\figsetgrptitle{LkCa 15}
\figsetplot{\gridline{(a) \fig{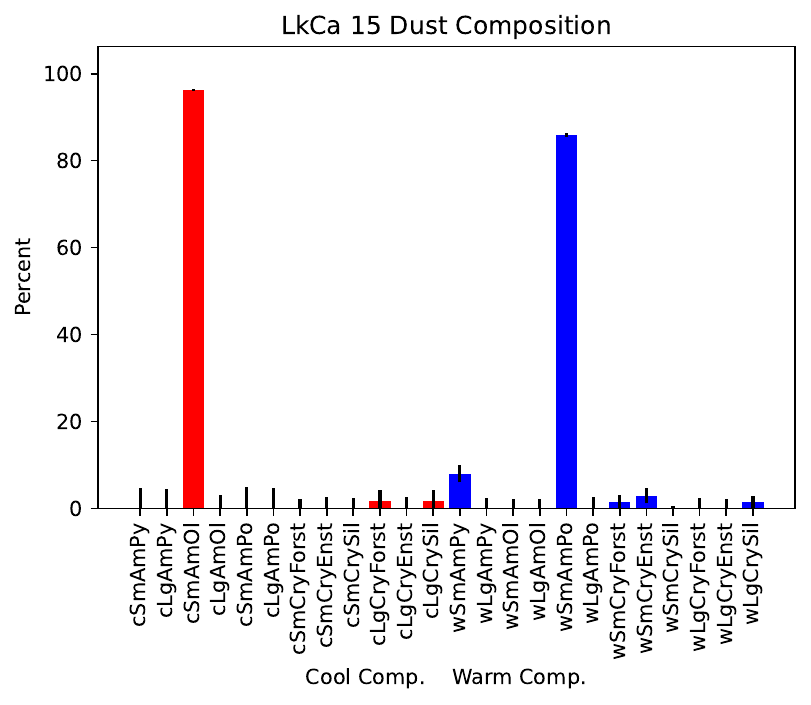}{0.5\textwidth}{}
              (b) \fig{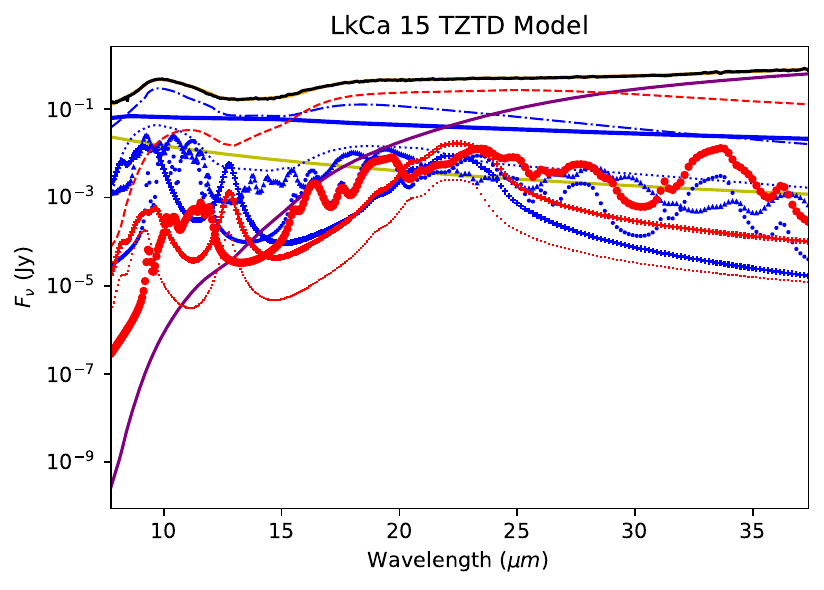}{0.5\textwidth}{}
              }
    \gridline{(c) \fig{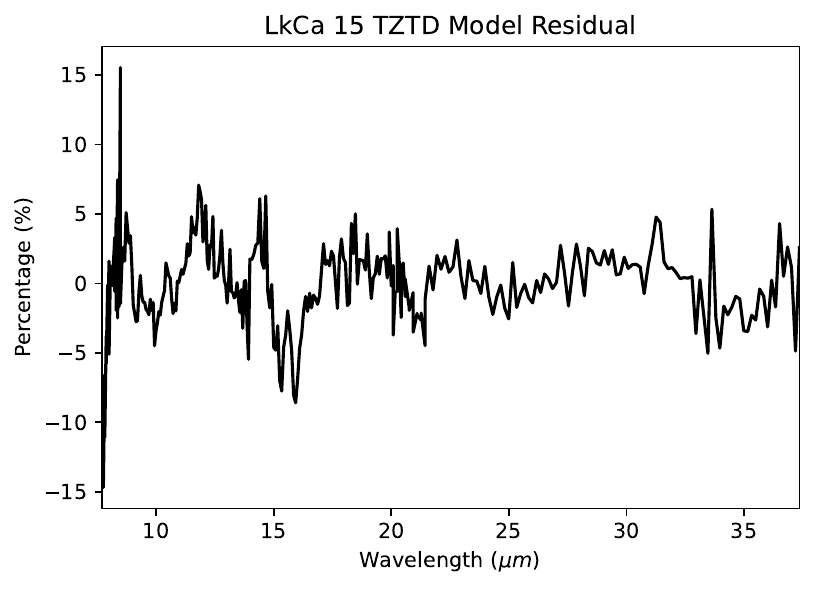}{0.5\textwidth}{}
              (d) \fig{tztd_legend.pdf}{0.45\textwidth}{}
              }}
\figsetgrpnote{As in Figure \ref{A1.1}, for LkCa 15}
\figsetgrpend

\figsetgrpstart
\figsetgrpnum{A1.12}
\figsetgrptitle{PDS 70}
\figsetplot{\gridline{(a) \fig{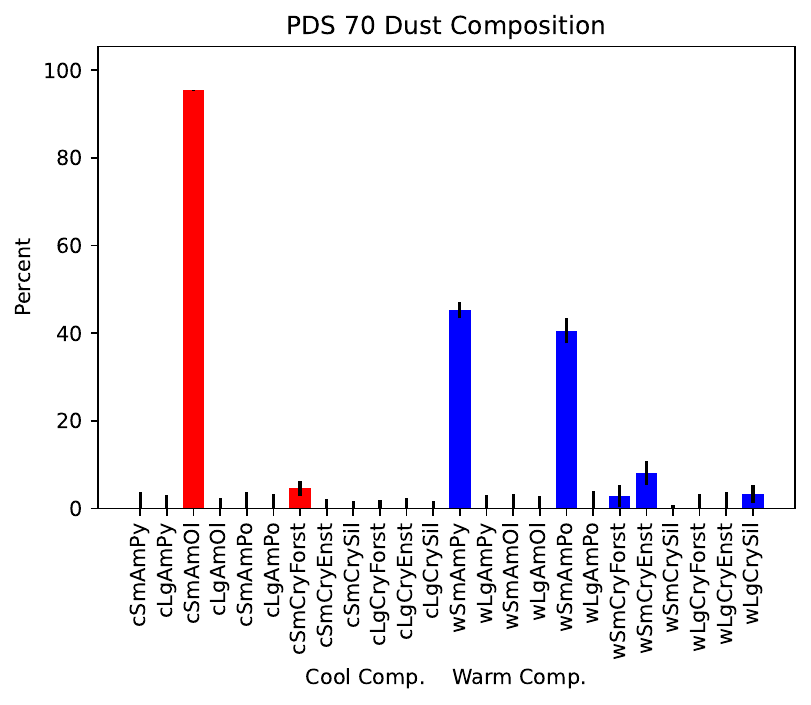}{0.5\textwidth}{}
              (b) \fig{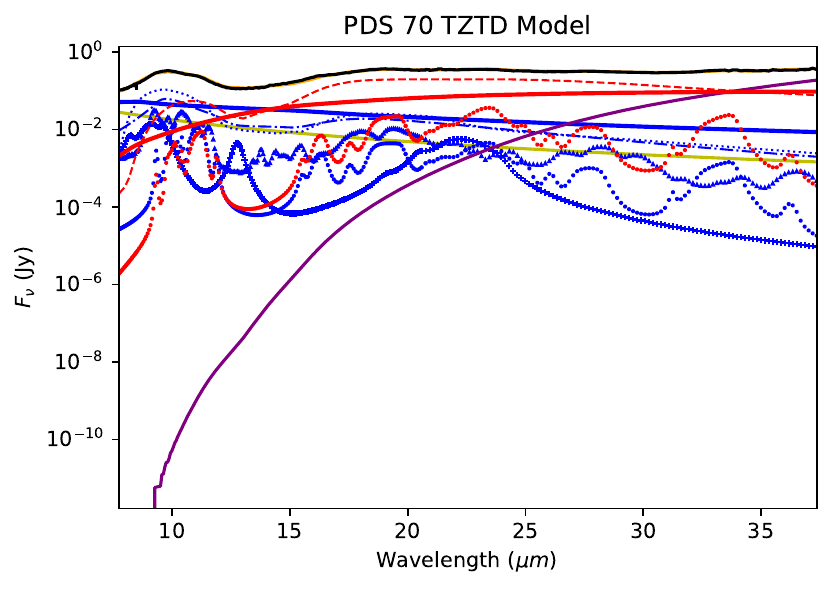}{0.5\textwidth}{}
              }
    \gridline{(c) \fig{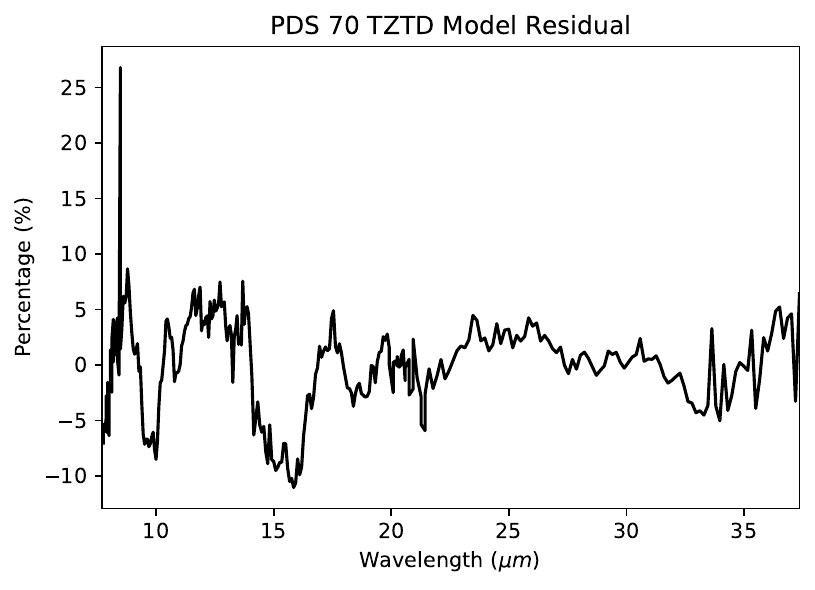}{0.5\textwidth}{}
              (d) \fig{tztd_legend.pdf}{0.45\textwidth}{}
              }}
\figsetgrpnote{As in Figure \ref{A1.1}, for PDS 70}
\figsetgrpend

\figsetgrpstart
\figsetgrpnum{A1.13}
\figsetgrptitle{RX J1842.9-3532}
\figsetplot{\gridline{(a) \fig{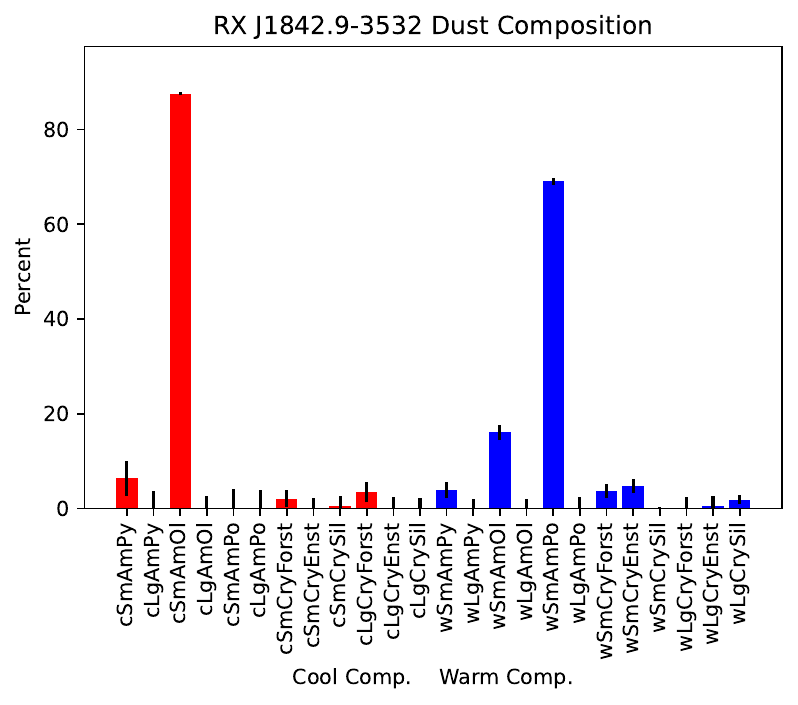}{0.5\textwidth}{}
              (b) \fig{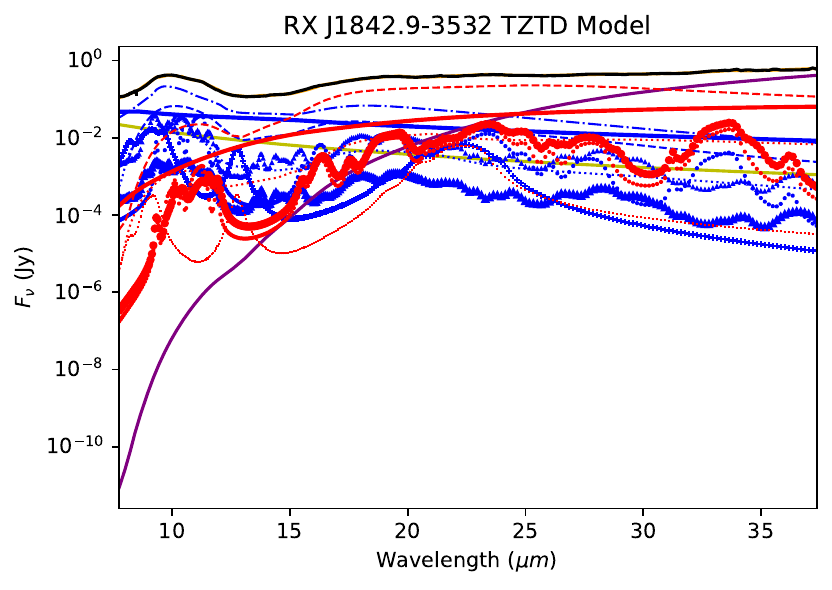}{0.5\textwidth}{}
              }
    \gridline{(c) \fig{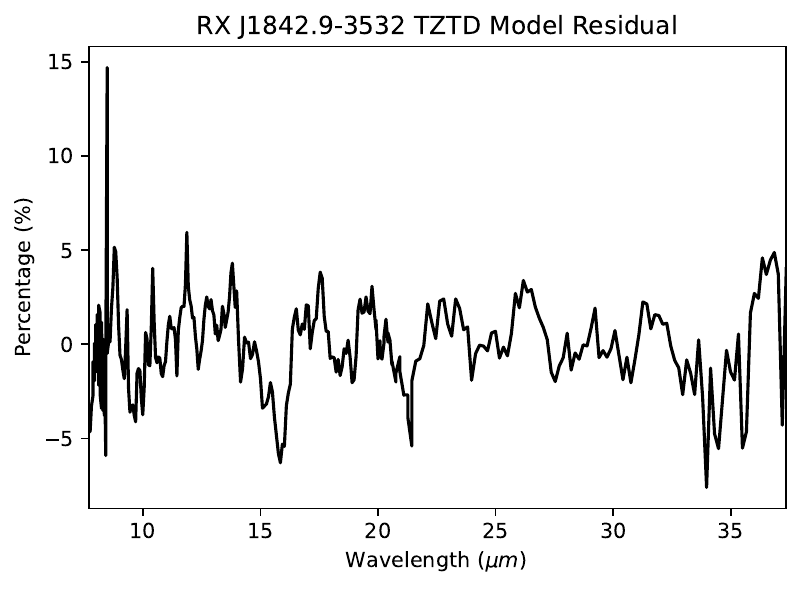}{0.5\textwidth}{}
              (d) \fig{tztd_legend.pdf}{0.45\textwidth}{}
              }}
\figsetgrpnote{As in Figure \ref{A1.1}, for RX J1842.9-3532}
\figsetgrpend

\figsetgrpstart
\figsetgrpnum{A1.14}
\figsetgrptitle{RX J1852.3-3700}
\figsetplot{\gridline{(a) \fig{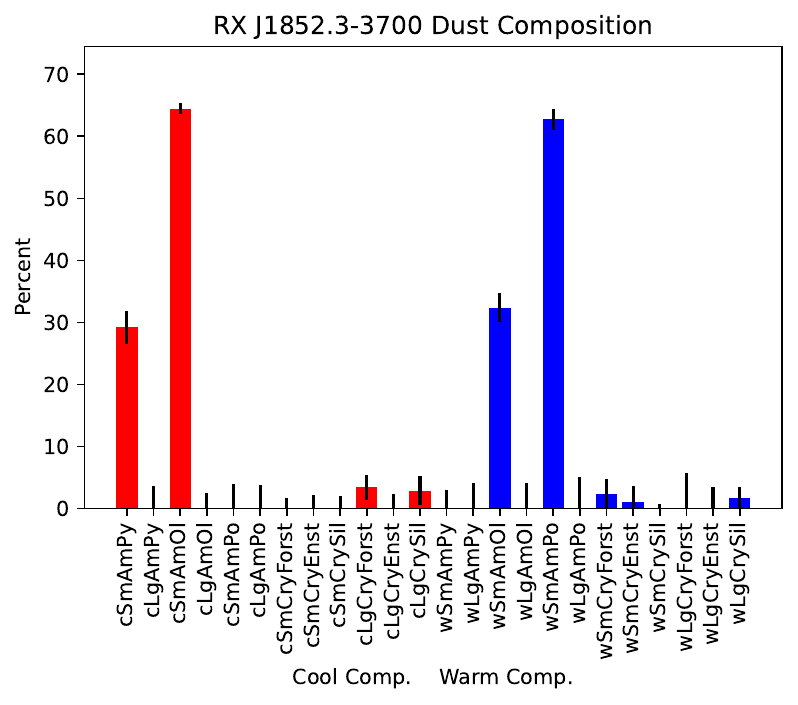}{0.5\textwidth}{}
              (b) \fig{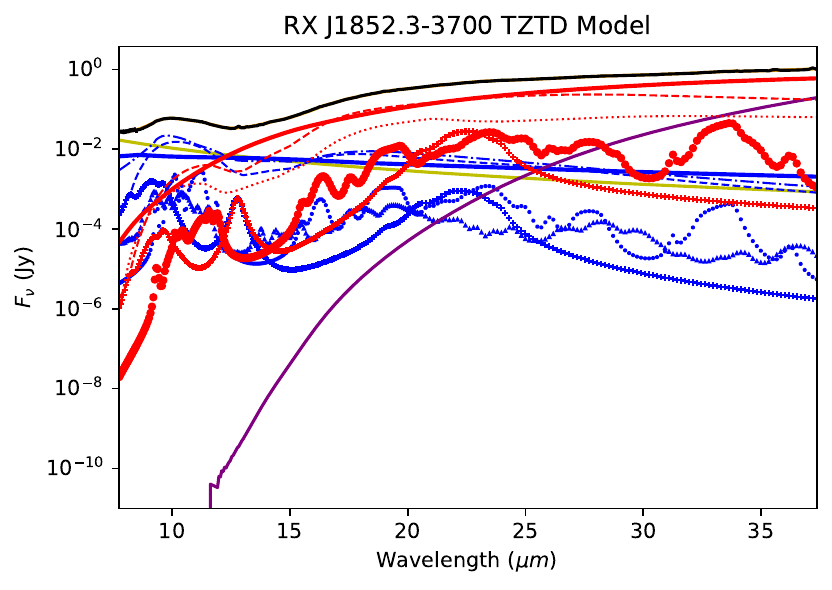}{0.5\textwidth}{}
              }
    \gridline{(c) \fig{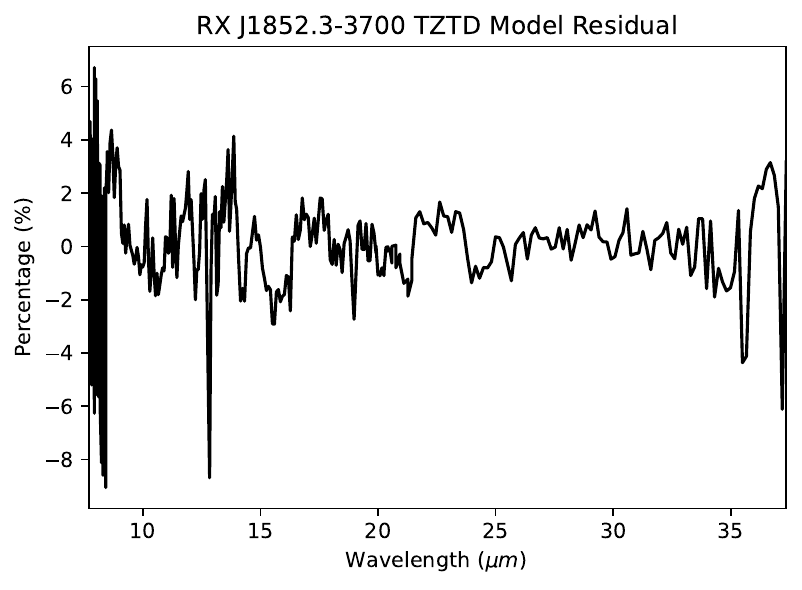}{0.5\textwidth}{}
              (d) \fig{tztd_legend.pdf}{0.45\textwidth}{}
              }}
\figsetgrpnote{As in Figure \ref{A1.1}, for RX J1852.3-3700}
\figsetgrpend

\figsetgrpstart
\figsetgrpnum{A1.15}
\figsetgrptitle{RY Tau}
\figsetplot{\gridline{(a) \fig{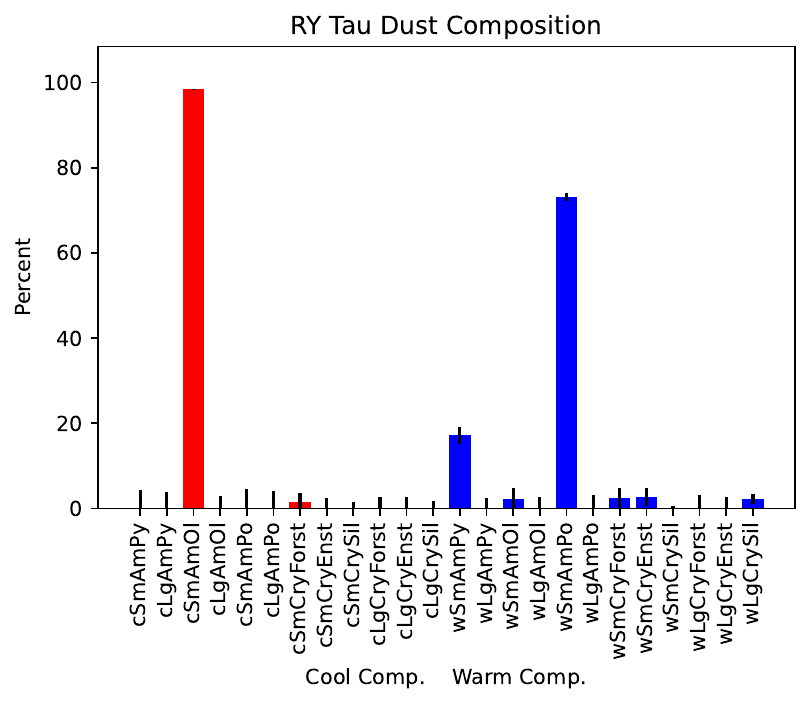}{0.5\textwidth}{}
              (b) \fig{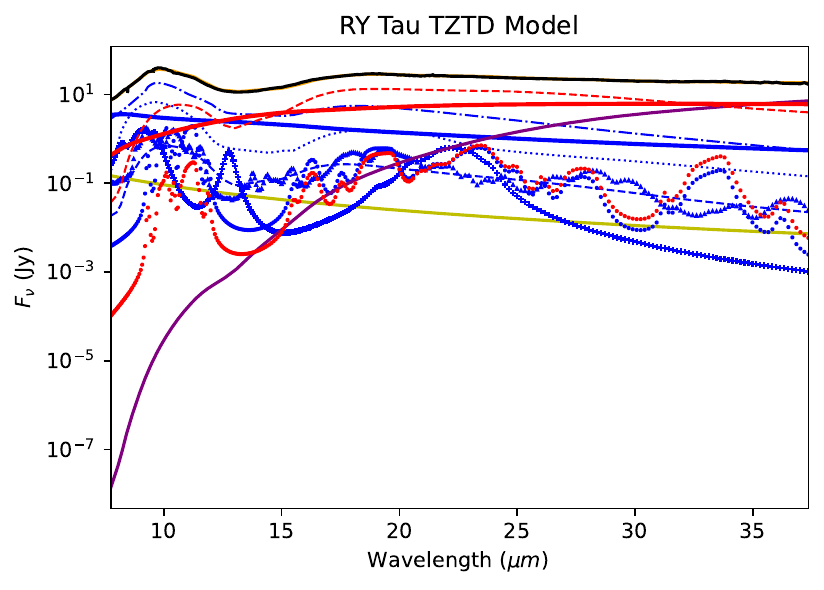}{0.5\textwidth}{}
              }
    \gridline{(c) \fig{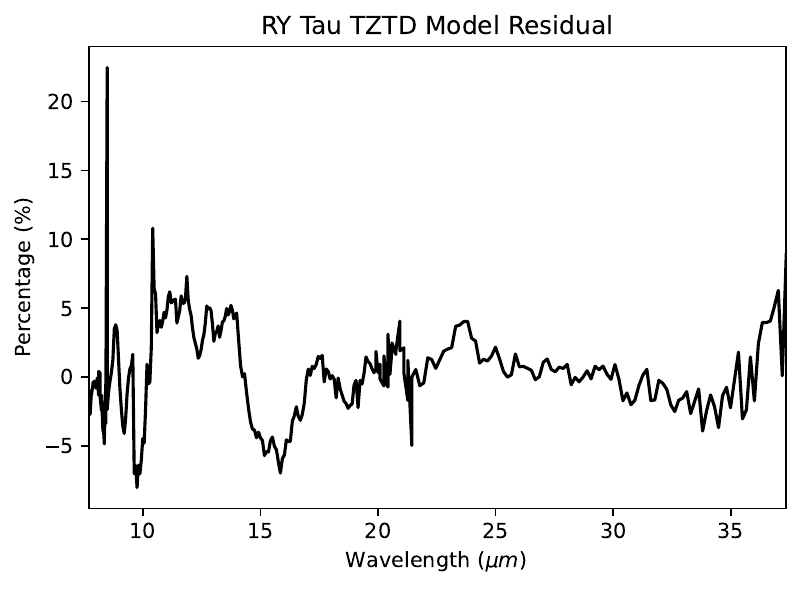}{0.5\textwidth}{}
              (d) \fig{tztd_legend.pdf}{0.45\textwidth}{}
              }}
\figsetgrpnote{As in Figure \ref{A1.1}, for RY Tau}
\figsetgrpend

\figsetgrpstart
\figsetgrpnum{A1.16}
\figsetgrptitle{SR 21}
\figsetplot{\gridline{(a) \fig{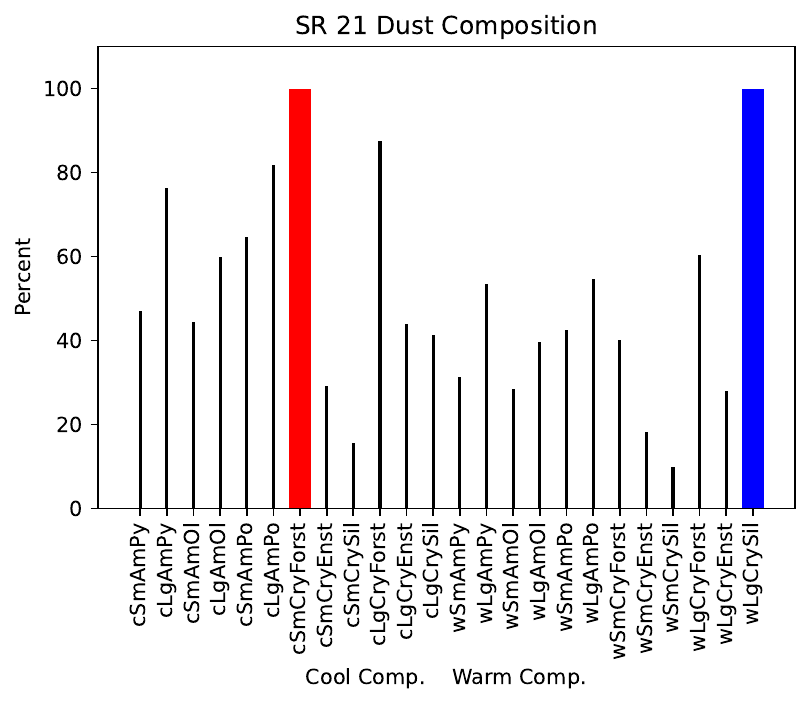}{0.5\textwidth}{}
              (b) \fig{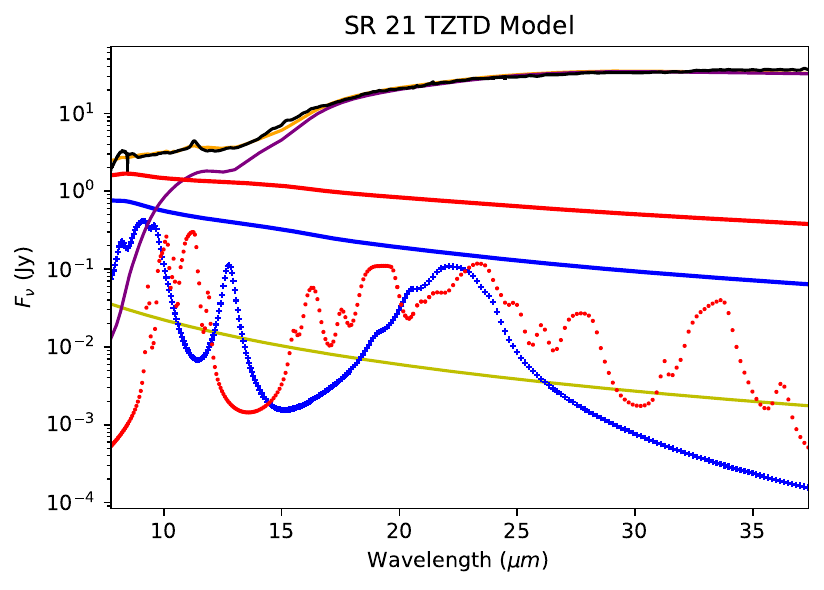}{0.5\textwidth}{}
              }
    \gridline{(c) \fig{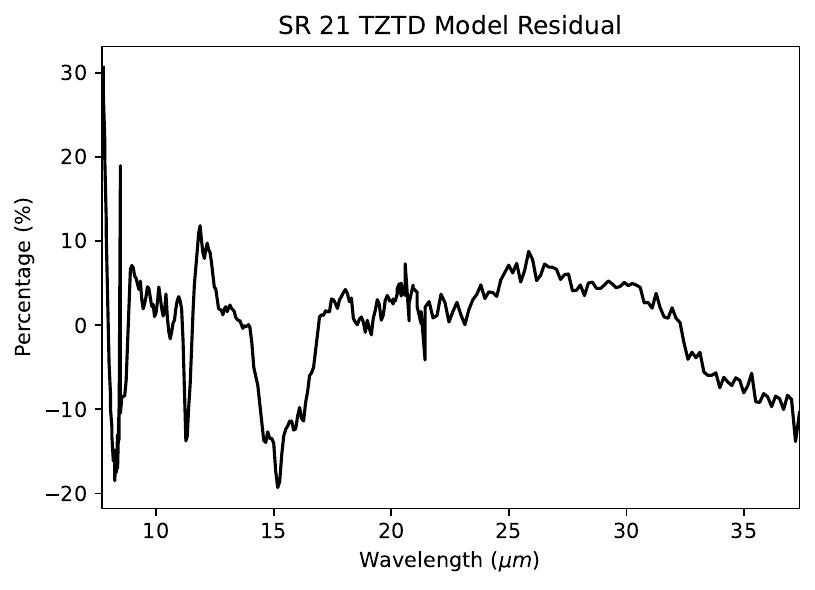}{0.5\textwidth}{}
              (d) \fig{tztd_legend.pdf}{0.45\textwidth}{}
              }}
\figsetgrpnote{As in Figure \ref{A1.1}, for SR 21}
\figsetgrpend

\figsetgrpstart
\figsetgrpnum{A1.17}
\figsetgrptitle{SR 24 S}
\figsetplot{\gridline{(a) \fig{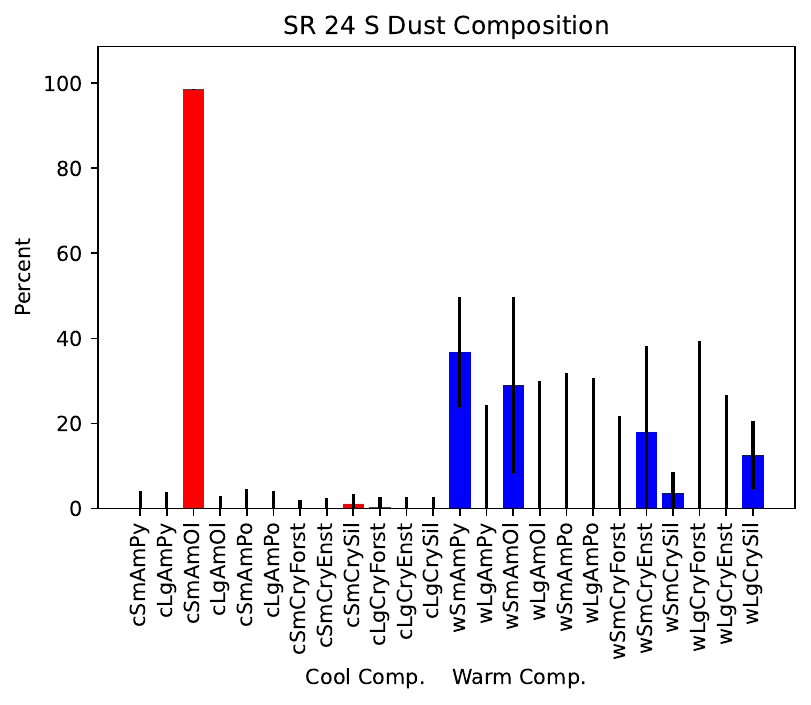}{0.5\textwidth}{}
              (b) \fig{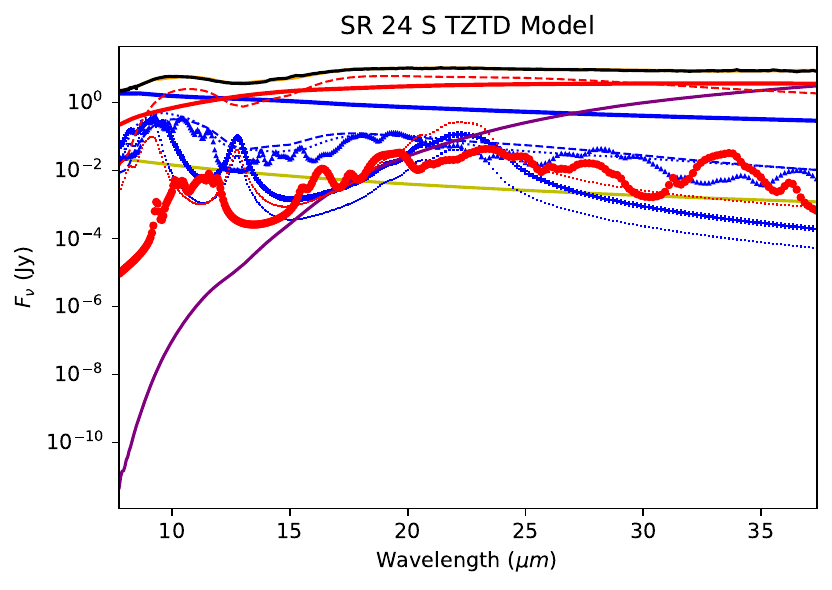}{0.5\textwidth}{}
              }
    \gridline{(c) \fig{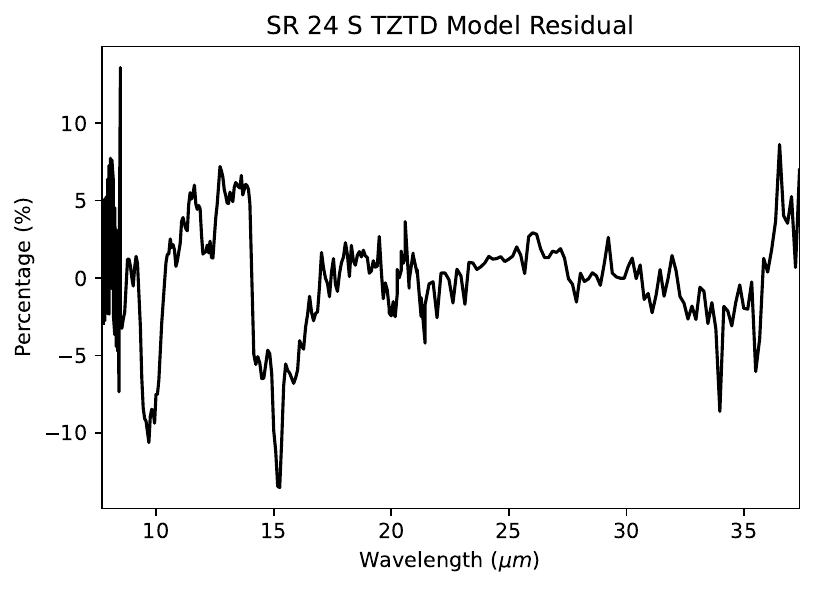}{0.5\textwidth}{}
              (d) \fig{tztd_legend.pdf}{0.45\textwidth}{}
              }}
\figsetgrpnote{As in Figure \ref{A1.1}, for SR 24 S}
\figsetgrpend

\figsetgrpstart
\figsetgrpnum{A1.18}
\figsetgrptitle{UX Tau A}
\figsetplot{\gridline{(a) \fig{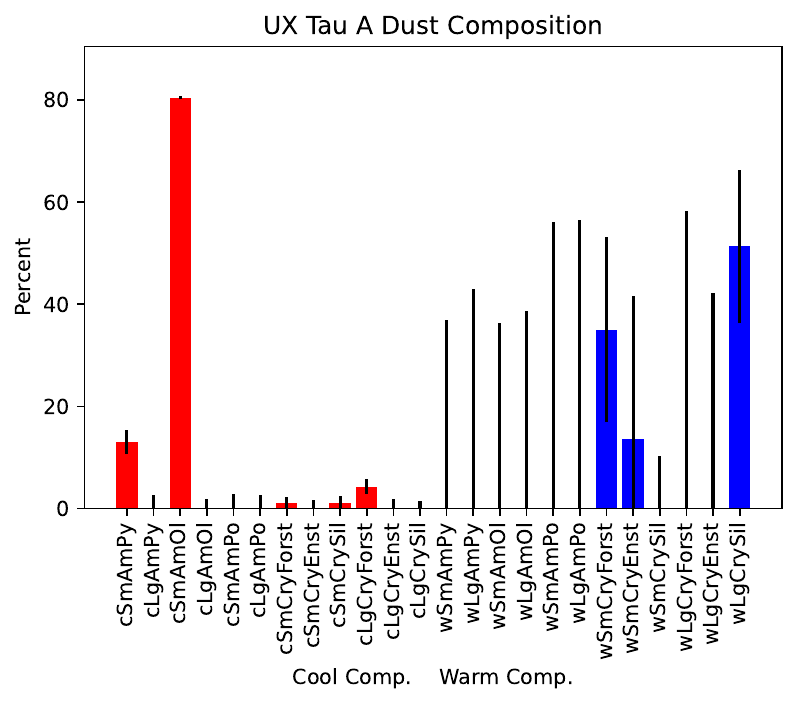}{0.5\textwidth}{}
              (b) \fig{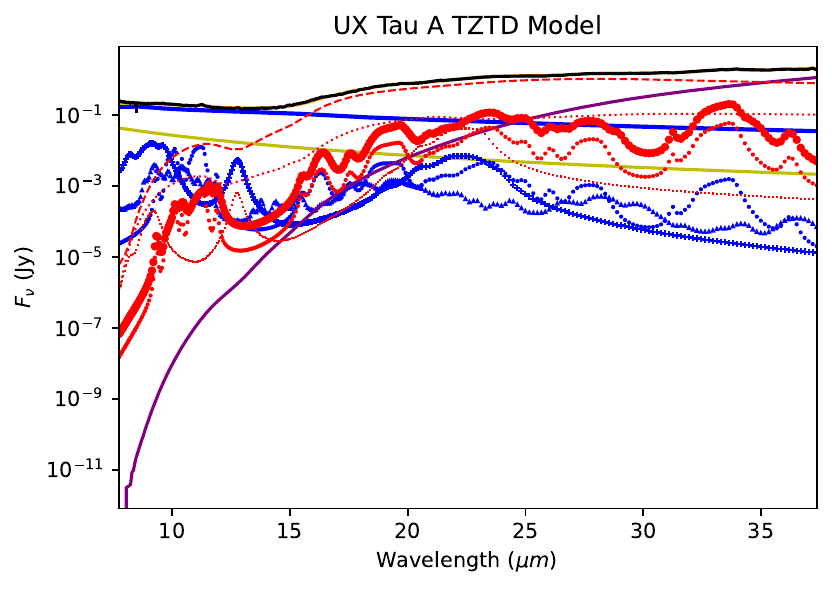}{0.5\textwidth}{}
              }
    \gridline{(c) \fig{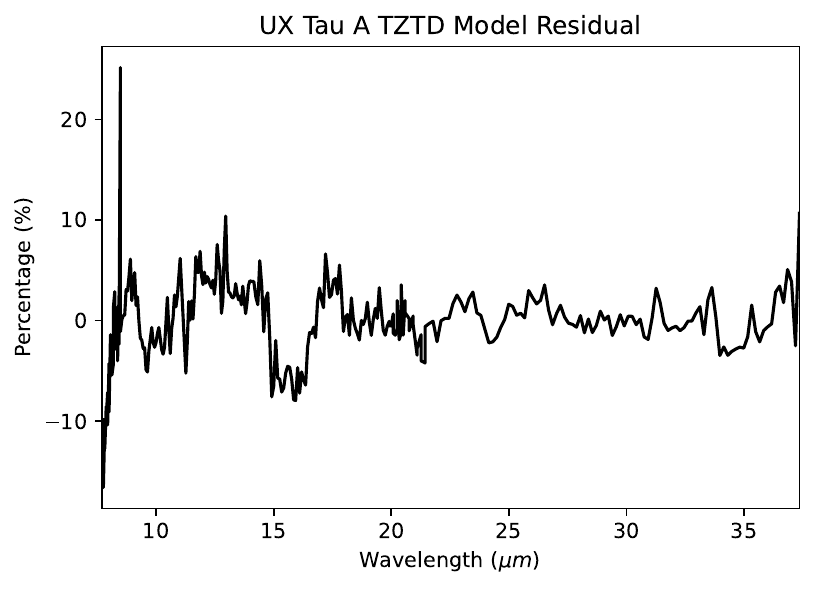}{0.5\textwidth}{}
              (d) \fig{tztd_legend.pdf}{0.45\textwidth}{}
              }}
\figsetgrpnote{As in Figure \ref{A1.1}, for UX Tau A}
\figsetgrpend

\figsetgrpstart
\figsetgrpnum{A1.19}
\figsetgrptitle{V1247 Ori}
\figsetplot{\gridline{(a) \fig{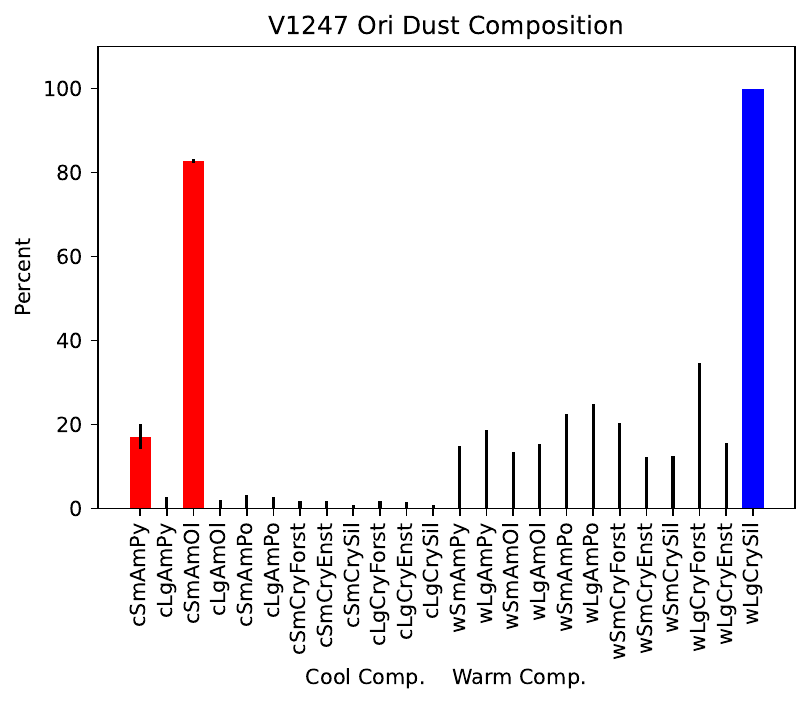}{0.5\textwidth}{}
              (b) \fig{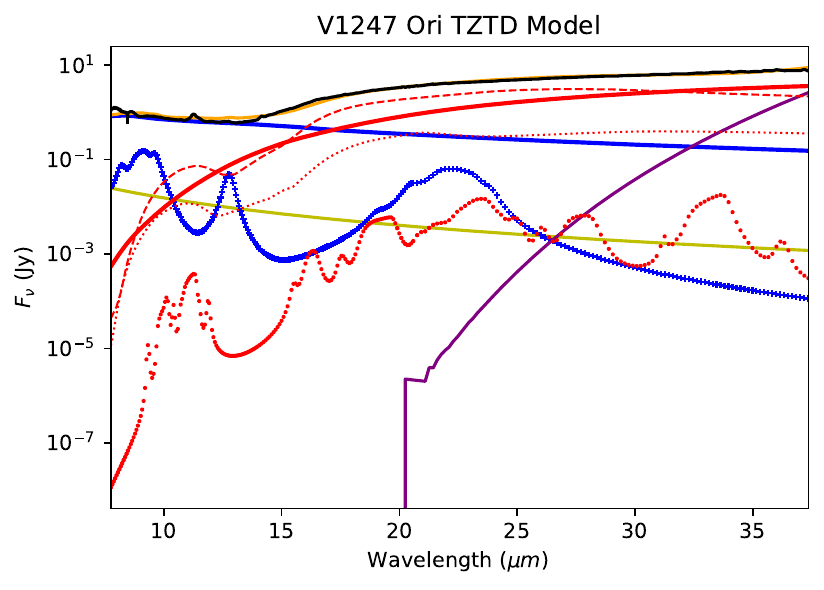}{0.5\textwidth}{}
              }
    \gridline{(c) \fig{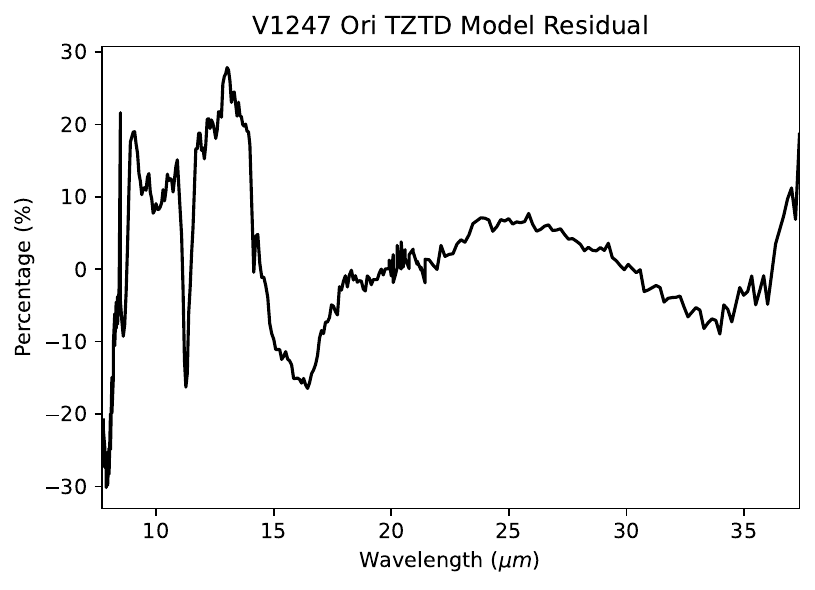}{0.5\textwidth}{}
              (d) \fig{tztd_legend.pdf}{0.45\textwidth}{}
              }}
\figsetgrpnote{As in Figure \ref{A1.1}, for V1247 Ori}
\figsetgrpend

\figsetgrpstart
\figsetgrpnum{A1.20}
\figsetgrptitle{WSB_60}
\figsetplot{\gridline{(a) \fig{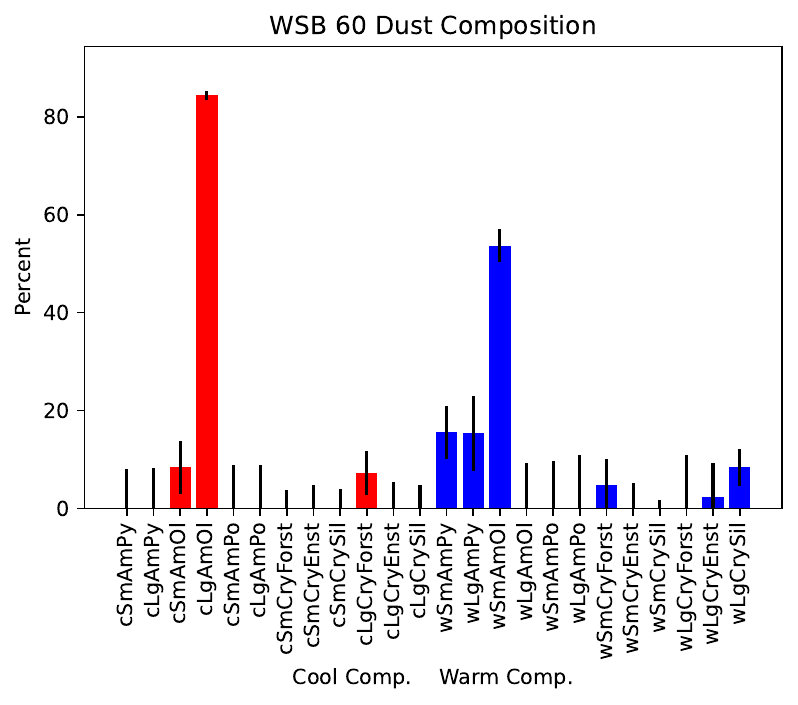}{0.5\textwidth}{}
              (b) \fig{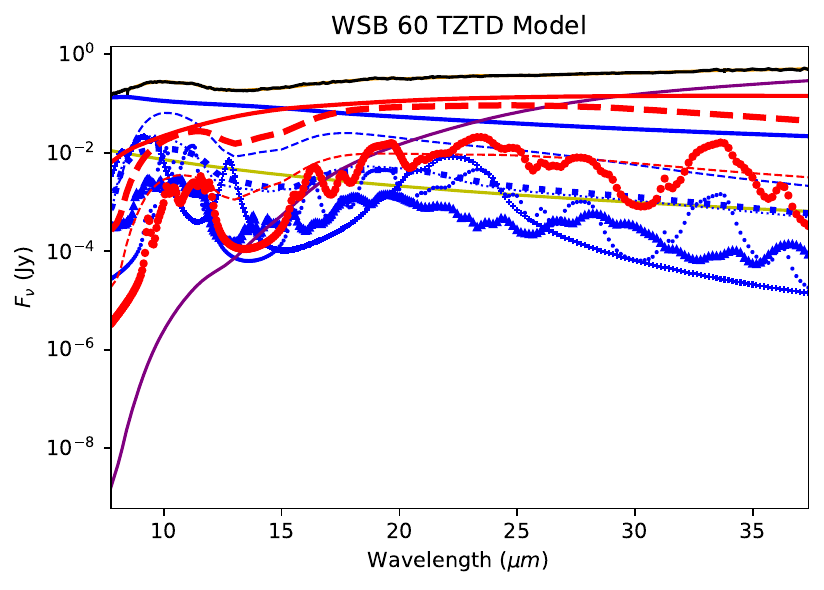}{0.5\textwidth}{}
              }
    \gridline{(c) \fig{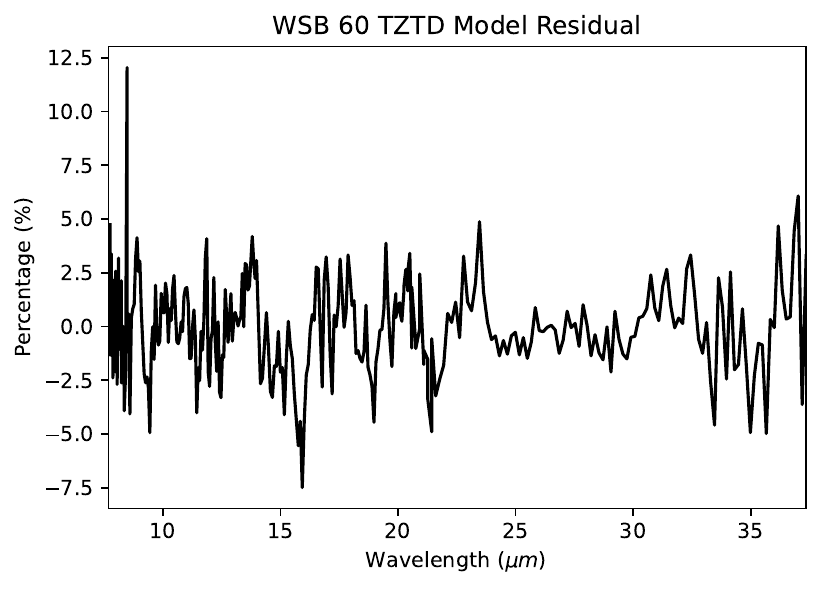}{0.5\textwidth}{}
              (d) \fig{tztd_legend.pdf}{0.45\textwidth}{}
              }}
\figsetgrpnote{As in Figure \ref{A1.1}, for WSB_60}
\figsetgrpend

\figsetend

\begin{figure*}[htb!]
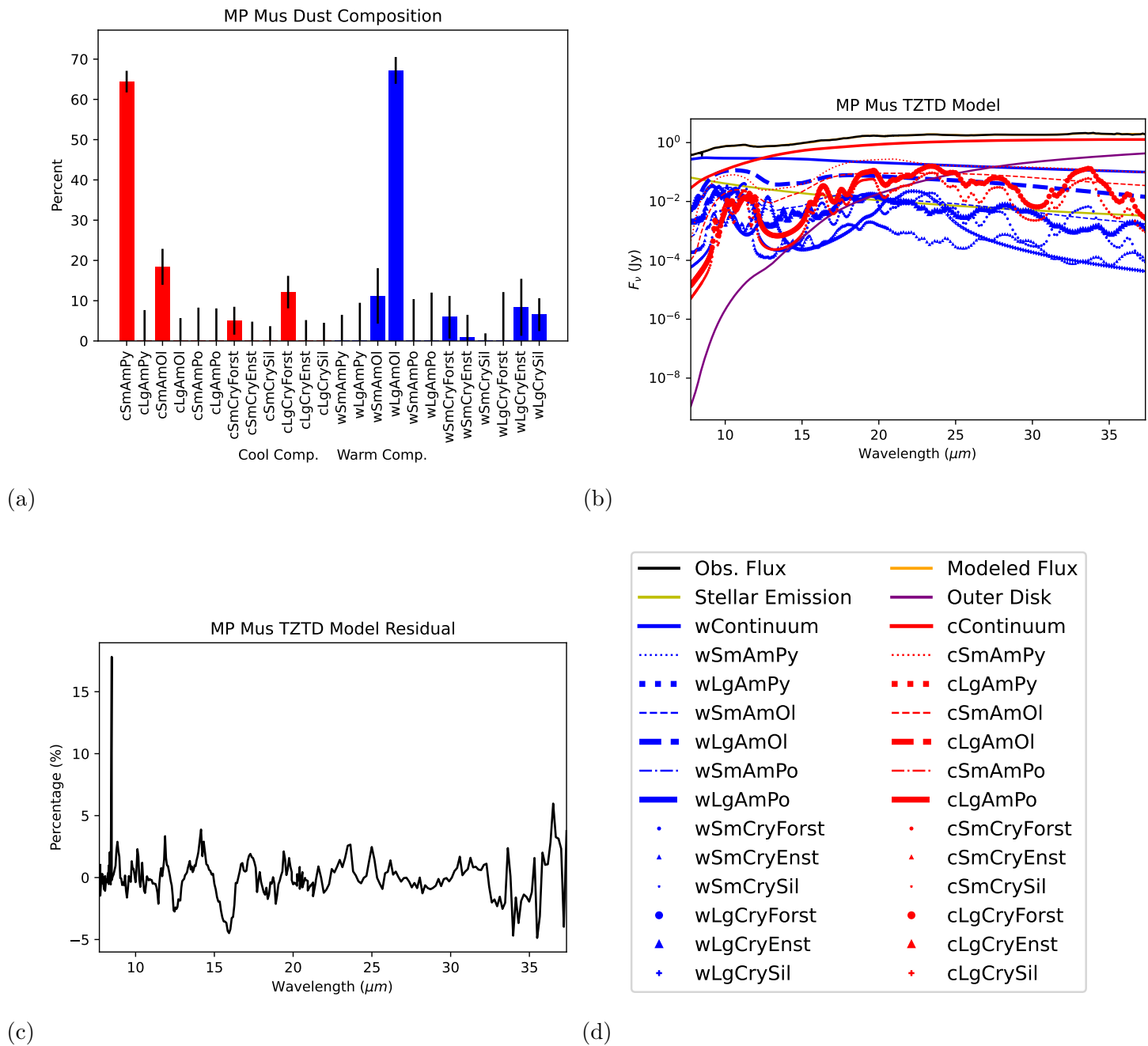

    \figurenum{A1}
    \gridline{(a) \fig{MP_Mus_tztd_comp.pdf}{0.5\textwidth}{}
              (b) \fig{MP_Mus_tztd_log.pdf}{0.5\textwidth}{}
              }
    \gridline{(c) \fig{MP_Mus_tztd_residual.pdf}{0.5\textwidth}{}
              (d) \fig{tztd_legend.pdf}{0.45\textwidth}{}
              }
    \caption{Results of TZTD empirical mineralogical analysis of MP Mus $Spitzer$ IRS spectrum; (a) Dust composition of optically thin portions of protoplanetary disk, using values from Table \ref{minres}, (b) Logarithmic scaling of corresponding model fit plot in Figure \ref{res_plot} to demonstrate the contributions of each mineral; (c) Residual between best fit model and $Spitzer$ IRS spectrum; (d) Legend for the model plots, using shortened names from Table \ref{opacity}; Red: Cool disk component constituents, Blue: Warm disk component constituents.}
\end{figure*}}

\end{document}